\documentclass[aps,superscriptaddress,twocolumn]{revtex4-2} 
\usepackage{hyperref,amsmath,amsfonts,siunitx,graphicx}
\graphicspath{{./grafs/}}
\newcommand{\etal}{\textit{et al.}}
\newcommand{\affdffct}{\affiliation{Departamento de F{\' \i}sica, 
    Faculdade de Ci{\^ e}ncias e Tecnologia, 
    Universidade Nova de Lisboa, 2829-516 Caparica, Portugal}}
\newcommand{\afcefema}{\affiliation{CeFEMA, 
    Av. Rovisco Pais, 1, 1049-001 Lisboa, Portugal}}
\newcommand{\acefitec}{\affiliation{CeFiTec,
    Faculdade de Ci{\^ e}ncias e Tecnologia, 
    Universidade Nova de Lisboa, 2829-516 Caparica, Portugal}} 
\newcommand{\affdfist}{\affiliation{Departamento de F{\' \i}sica,
    Instituto Superior T{\' e}cnico, Universidade de Lisboa,
    Av. Rovisco Pais, 1, 1049-001 Lisboa, Portugal}} 
\newcommand{\FracD}[1]{\ensuremath{\mathcal{D}^{#1}_{t^\ast}}}
\newcommand{\Tnet}{\ensuremath{\mathcal{T}_\mathrm{net}}}
\newcommand{\Tgrav}{\ensuremath{\mathcal{T}_\mathrm{grav}}}
\newcommand{\Tdiss}{\ensuremath{\mathcal{T}_\mathrm{diss}}}
\newcommand{\Tiner}{\ensuremath{\mathcal{T}_\mathrm{iner}}}
\newcommand{\adiss}{\ensuremath{\alpha_\mathrm{diss}}}
\newcommand{\adisss}{\ensuremath{\alpha_{\mathrm{diss}_s}}}
\newcommand{\adissp}{\ensuremath{\alpha_{\mathrm{diss}_p}}}
\newcommand{\ainer}{\ensuremath{\alpha_\mathrm{iner}}}
\newcommand{\signalop}{\mathop{\mathrm{sgn}}}
\begin{document}
\title{Physical pendulum model: Fractional differential equation and
  memory effects}
\author{L. N. Gon{\c c}alves}
\email{lng@fct.unl.pt}
\affdffct
\afcefema
\author{J. C. Fernandes}
\affdfist
\author{A. Ferraz}
\affdfist
\afcefema
\author{A. G. Silva}
\affdffct
\acefitec
\author{P. J. Sebasti{\~ a}o}
\affdfist
\afcefema
\date{\today}
\begin{abstract} 
A detailed analysis of three pendular motion models is presented. 
Inertial effects, self-oscillation, and memory, together with
non-constant moment of inertia, hysteresis and negative damping are
shown to be required for the comprehensive description of the free
pendulum oscillatory regime. The effects of very high initial
amplitudes, friction in the roller bearing axle, drag, and pendulum 
geometry are also analysed and discussed. The model that consists of a
fractional differential equation provides both the best explanation
of, and the best fits to, experimental high resolution and long-time
data gathered from standard action-camera videos.
\end{abstract}

\maketitle 
\begin{center}
  This article has been published by \\
  \textit{American Journal of Physics} \\
  at \url{https://doi.org/10.1119/10.0001660}
\end{center}
    
\section{Introduction} 

There's no classic like the physical pendulum.
It has been the subject of scientific enquiry since Galileo first
observed its isochrony 
\cite{Erlichson1999}.
It has been a source of technological development since Huygens' pendulum
clock 
\cite{Huygens1673}.
It has been a reference instrument since at least 1818
\cite{Kater1818,Jackson1961,Marson2012}
and it plays a fundamental role in the implementation of
gravitational wave observatories 
\cite{Blair1993,Mitrofanov1999,Uchiyama2000,Cagnoli2000}.
To this day, the pendulum continues to serve as a tool to understand
many diverse phenomena involving both oscillation and relaxation
\cite{Lima2008}
like parametric pumping (e.g. pendulum clock, swing, roller skating)
\cite{Sanmartin1984,Wirkus1998,Stilling2002,Post2007},
hysteresis
\cite{Greenwood2009,Obligado2013}, 
deterministic chaos 
\cite{Cuerno1992,Deserio2003},
charge density waves 
\cite{Azbel1984,Romeiras1987},
``macroscopic quantization''
\cite{Milburn1983,Yurke1986,Doubochinski2007,Shumaev2017},
bosonic Josephson junctions
\cite{Pigneur2018},
classical micro-canonical systems
\cite{Naudts2005,Baeten2011}
and dielectric relaxation 
\cite{Lukichev2014,Lukichev2015,Lukichev2016,Lukichev2019}.
The physical pendulum may even contribute to the understanding of some 
climate change effects like meteotsunamis
\cite{Rabinovich2009}.

The theoretical description of pendular motion has been the subject of
many studies, some of which gave rise to very sophisticated equations
of motion 
\cite{Fee2002,Peters2003,Fernandes2017,Kavyanpoor2017,Amabili2018}.
However, one that accurately matches long-time data, of a very high
amplitude physical pendulum, has been lacking. This probably stems
from lacking observations of the hysteretic nature of pendular
motion. This observation is now made and justifies the proposed
fractional model. This model fits experimental data gathered with a
modern action-camera and explains the observed hysteresis on the basis
of memory, non-constant moment of inertia and self-oscillation. 

Modern studies of the interdependency of amplitude on the period seem to
have started on the seventies of the XXth century 
\cite{Fulcher1976,Hall1977}.
Fourier transform analysis was used to advance those studies   
\cite{Zilio1982,Gil2008}.
A tentatively realistic model of the pendulum motion, considering
constant, linear and quadratic drag terms, was introduced by Squire  
\cite{Squire1986}
and studied in diverse combinations
\cite{Basano1991,Zonetti1999,Wang2002,Bacon2005,Simbach2005,Smith2012,Mungan2013}.  
More recently, Mathai \etal\ proposed a dry-friction damping term
dependent on the pendulum angle when studying an underwater pendulum 
\cite{Mathai2019}. 
The study of underwater pendula make apparent the effects of
the surrounding fluid 
\cite{Mathai2019,Eng2008,Bolster2010}.
In particular the emission of vortex rings at extreme angles was put
in evidence by Bolster \etal\ 
\cite{Bolster2010}.

Other experiments that provide insight into pendular motion
include the air-track
\cite{Whineray1991,Hinrichsen2018}, 
the drinking straw
\cite{Lorenceau2002,Smith2019},
and the free-fall 
\cite{Basano1989}. 

From the experimental point of view the study of pendular motion can
be conducted by measuring the angular position of the moving object or
by measuring its acceleration.
Recently, different studies used the latter approach
\cite{Fernandes2017,Hinrichsen2018,Alho2019,Larnder2019}.

It is worth mentioning that the study of pendular motion is a subject
in the extensive area of parameter identification of vibrating systems
\cite{Mann2009,Jaksic2011}.

This paper is organized as follows. 
The experimental setup is presented in section
\ref{sec:exper} together with some contextualizing data. 
The theoretical analysis regarding the equation of motion,
is introduced in section \ref{sec:eqofmotion}. 
Two initial conceptual models are presented and tested in
sections \ref{sec:classicmodel} and \ref{sec:lasermodel}. 
An introduction to time fractional derivatives and the concept of
memory follows. 
We move on describing the fractional differential equation of motion
and associated results.
We finish with a general conclusion. 

\section{Experimental}\label{sec:exper}

The physical pendulum used in this work is composed of: 
a $\SI{2.5}{\centi\metre}$ diameter roller bearing concentric with
an hollow acrylic disc with a diameter of $\SI{6}{\centi\metre}$; a
$\SI{52}{\centi\metre}$ long squared cross-section 
hollow bar; and a slightly longer threaded steel
$\SI{6}{\milli\metre}$ thick rod that is 
screwed both to the square bar and to the acrylic hollow disc by two
nuts. Rectangular
tiles, $\SI{18}{\centi\metre}$ wide and with lengths  
$l_1=(3\times n)\,\si{\centi\metre}$ where $n=1,\ldots,8$,  
made of cardboard, foam and tape were fixed at the end of the square
bar. Three slabs of composite cardboard, assembled to
form an alley, were used in half the measurement runs.  

\begin{figure*}
\centering
\begin{tabular}[b]{cc}
  \includegraphics[scale=0.4]{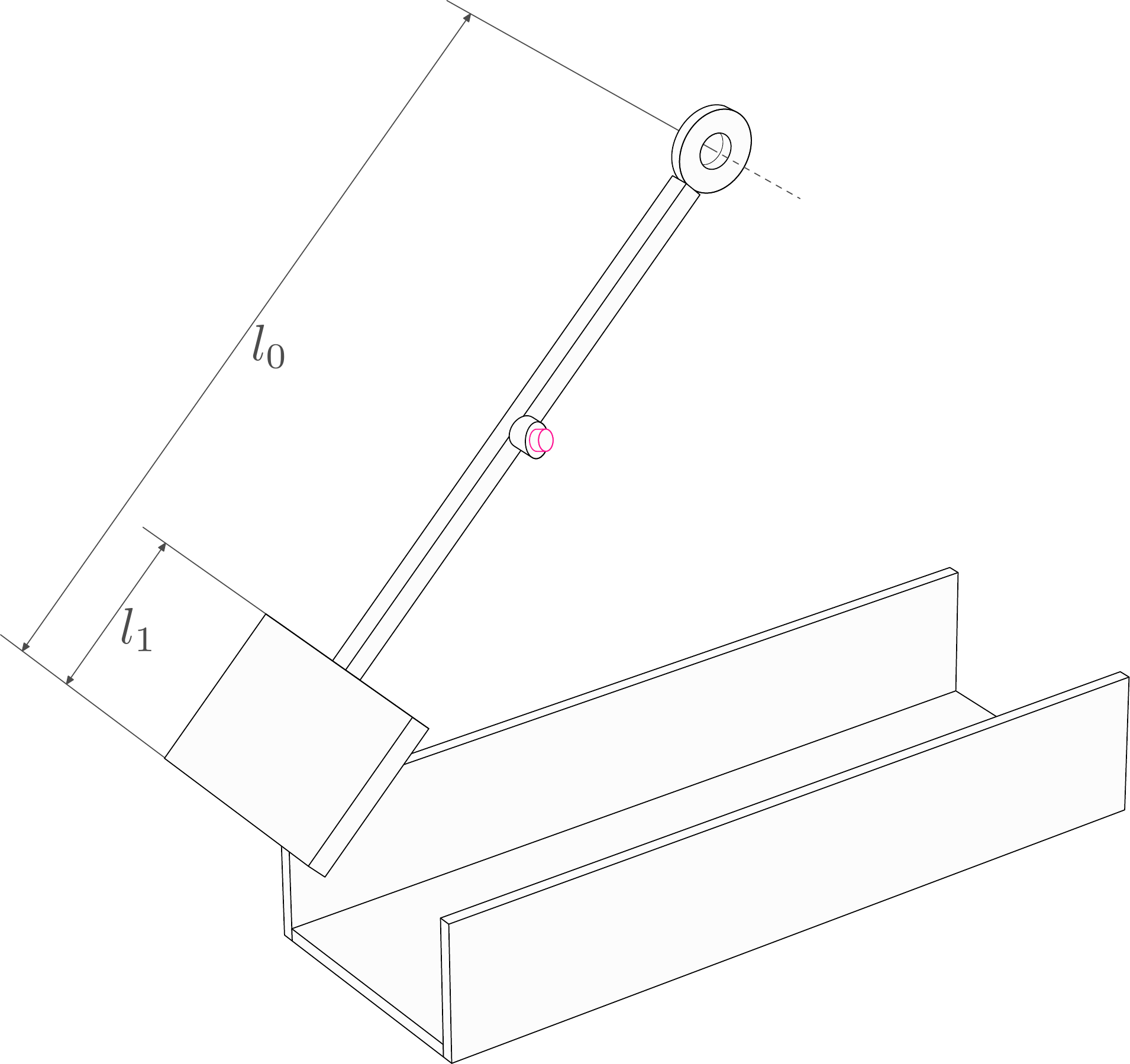} &
  \includegraphics[scale=0.4]{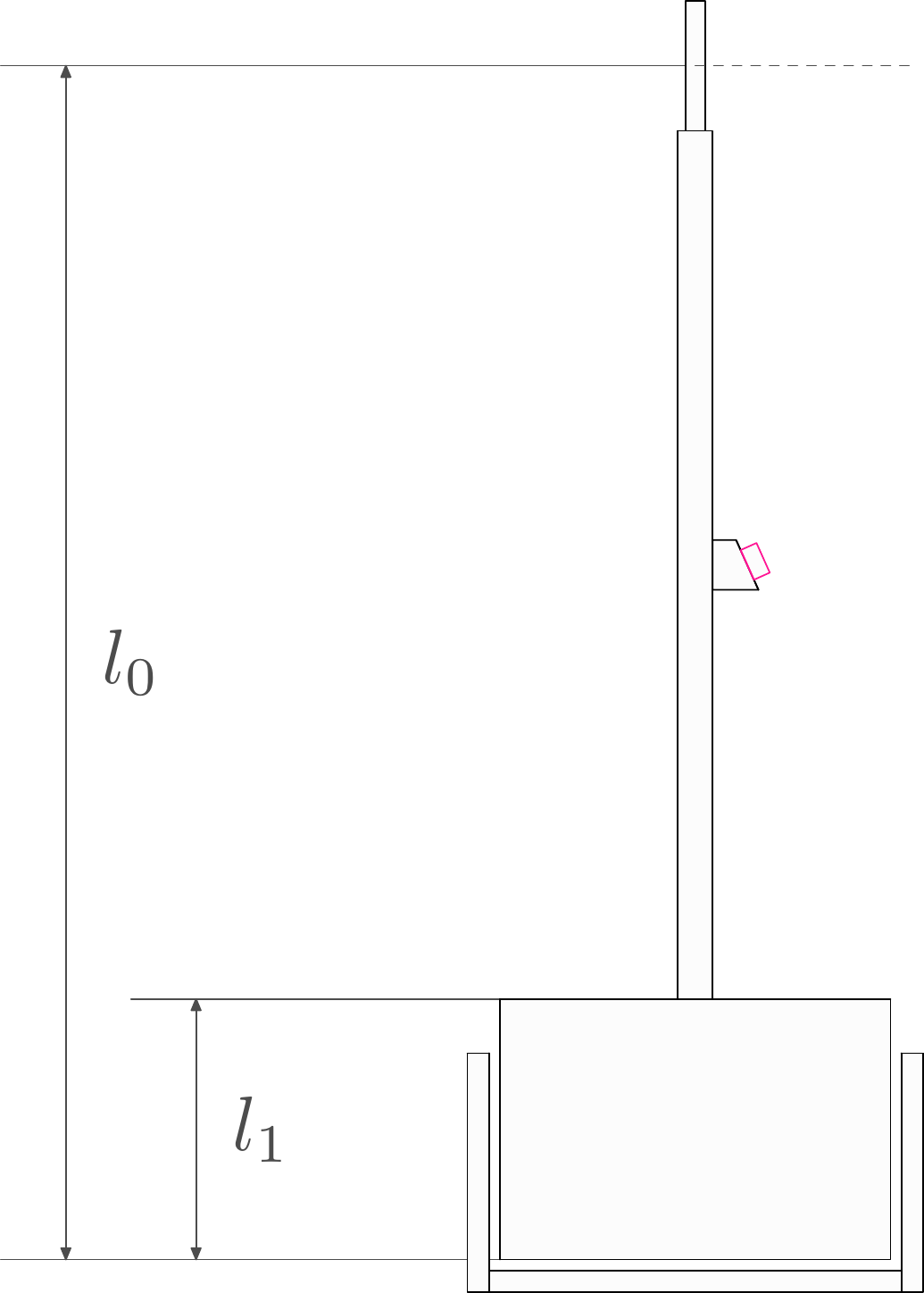} \\
  (a) & (b)
\end{tabular}
\caption{Schematics \protect \cite{featpost} of the apparatus: 
  (a) perspective view with
  $\theta=\SI{\pi/4}{\radian}$; (b) orthogonal view with $\theta=0$. 
  There may be a tile with an edge at the
  pendulum extreme (at a distance $l_0$ from the axle) and another edge
  at a distance $l_0-l_1$ (the height of the tile is $l_1$). Also,
  there may be an alley for the pendulum to pass through without
  touching. There is a clearance of approximately half centimeter
  between the pendulum and the alley. The schematics are drawn to
  scale showing both the pink tracking spot and a
  $l_1=\SI{12}{\centi\metre}$ tile.}  
\label{basicscheme}
\end{figure*}

Figure \ref{basicscheme} shows schematics of the apparatus.
Videos were recorded with a Sony HDR-AS100V action-camera using
a resolution of 1280x720 pixels at 120 frames per second (FPS). 
The camera was placed directly in front and
aligned with the pendulum axle at a distance of approximately 
$\SI{58}{\centi\metre}$. In this way the tracking spot was always within
the maximum possible camera view field. A pink\footnote{The color
  pink was chosen because it produced the best contrast.} circular
tracking spot was glued to the pendulum at approximately
$\SI{26}{\centi\metre}$ from the axle and facing the camera
lens (see Figure \ref{basicscheme}). 
This tracking spot orientation made possible to obtain a permanent
circular tracking template that contributed to avoid rotation artifacts. 
A carefull illumination of the apparatus proved necessary
to minimize motion blur.

The experimental $\theta(t)$ data was collected using 
the open source video analysis tool Tracker 
\cite{Tracker}.
The radial distortion of the video frames was corrected using
the ``Fisheye'' filter at $\SI{120}{\degree}$ and 48\% fixed pixels.
The tracking algorithm compares, for each frame, a previously defined
mark template with the current image within the tracking target area.
In view of the large number of video frames that are tracked on a
single video (may be larger than $10^4$), the Tracker's ``evolution
rate'' was set at 0\%. 
Also, due to heterogeneous lighting, the Tracker's ``automark'' was
reduced to 3 to obtain a manageable rejection rate.

A sequence of experimental runs was conducted according to the
following procedure. In each run the pendular motion was recorded from
an initial launch to its stopping.
Four kinds of measurement runs were analysed: launched (\texttt{l}) with 
initial speed; released from rest; 
with alley (\texttt{a}); and without alley. In Table
\ref{tab:kindofrun} the two 
letters codes used to identify each kind of run are presented.
For each kind of run two additional digits identify the tile length: 
no tile (\texttt{00}); $\SI{3}{\centi\metre}$ long tile (\texttt{03}); 
$\SI{6}{\centi\metre}$ long tile (\texttt{06}), etc.\ up to
$\SI{24}{\centi\metre}$ long tile (\texttt{24}) in steps of
$\SI{3}{\centi\metre}$. 
\begin{table}
  \caption{Run code infixes (e.g. \#\#\texttt{la}\# as explained in
    the text).}
  \centering
  \begin{tabular}{|l|c|c|}
    \hline
                 & with alley  & without alley \\
    \hline
    launched     & \texttt{la} & \texttt{ln}   \\
    \hline
    not launched & \texttt{na} & \texttt{nn}   \\
    \hline
  \end{tabular}
  \label{tab:kindofrun}
\end{table}
The complete run codes include also an end digit identifying the trial
number of each run kind. For example, \texttt{06na5} refers to
the fifth run using the $\SI{6}{\centi\metre}$ long tile without
launch but with alley.

In all runs the zero angle was defined by the pendulum's final
stopping position, the initial angle was always 
$\theta_\mathrm{ini}\geq\SI{\pi}{\radian}$ and the initial angular
velocity was always $\omega_\mathrm{ini}\leq 0$. 

Each run has its own specificity. The video camera
position and orientation affect pendulum angle measurement. The
position and orientation of the lamps may produce reflections that affect
some of the video-frames. The final equilibrium position may not be
exactly vertical. The roller bearing has some clearance and allows
sideways oscillations that may affect the main oscillation. 
When a tile enters or exits the alley its motion may be subject
to sudden perturbations. These specificities, while unavoidably
present, do not seem to significantly affect the experimental data. 

\begin{figure}[tbhp]
  \includegraphics[width=0.9\columnwidth]{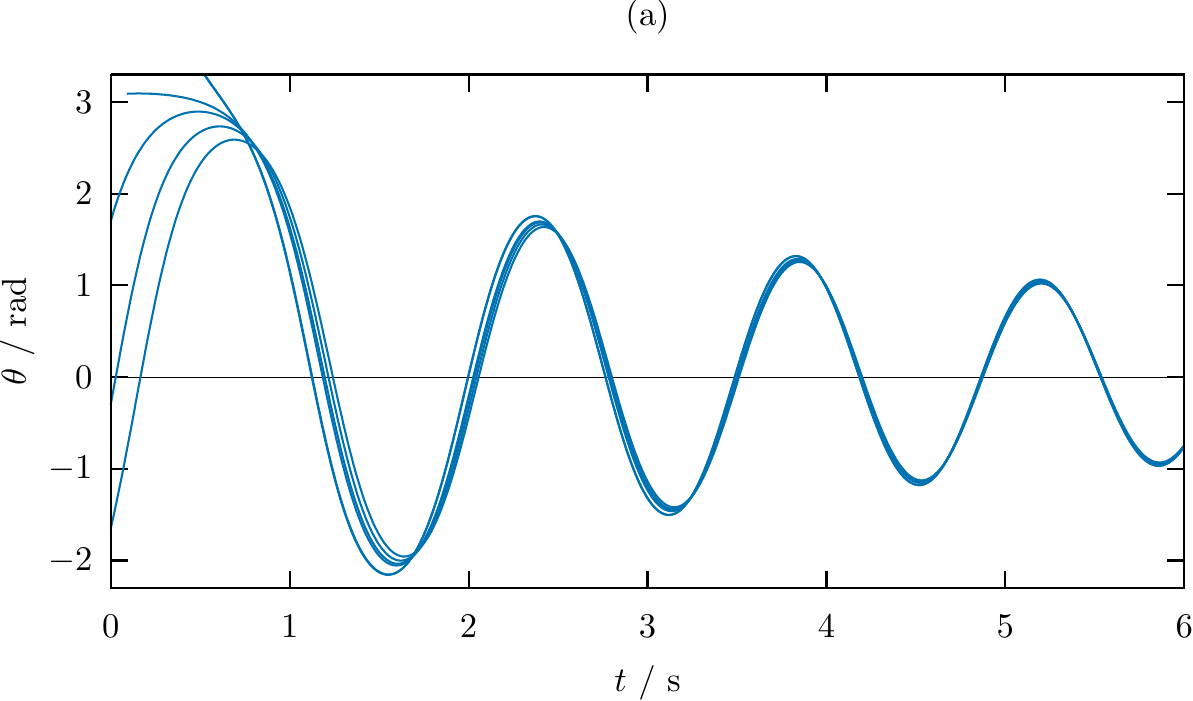} \\ 
  \caption{Superposition of five exploratory runs made with a 
    $l_1=\SI{12}{\centi\metre}$ tile, without alley and different
    initial conditions. One of the runs starts from rest at an
    initial angle close to $\SI{\pi}{\radian}$. For each run a time
    shift was introduced to allow a superposition of all runs at
    $t=\SI{10}{\second}$.}  
  \label{fig:inicondits}
\end{figure}

\begin{figure}[tbhp]
  \centering
  \includegraphics[width=0.9\columnwidth]{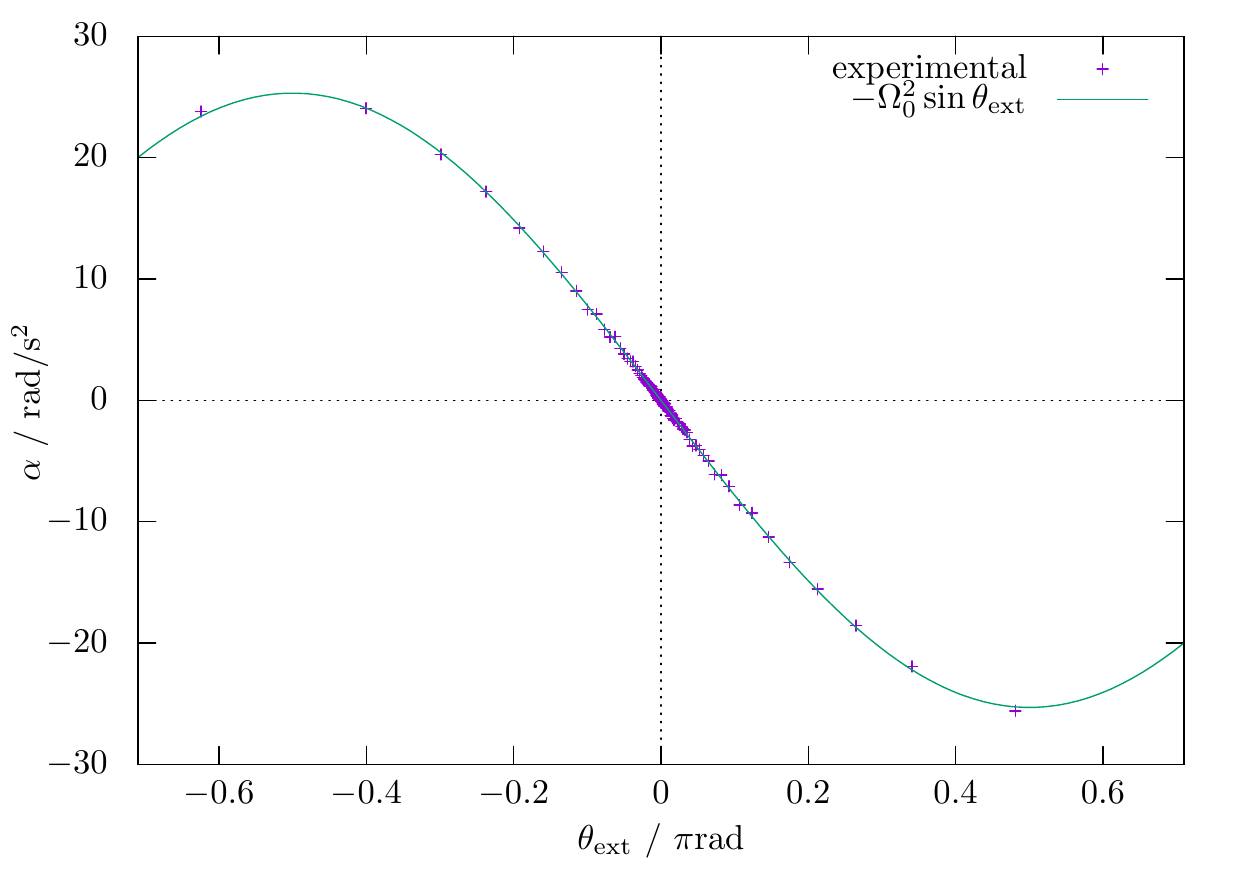} 
  \caption{Initial analysis of acceleration and angle for run 
    \texttt{24la1} (see text).}  
  \label{fig:ozsin}
\end{figure}

\begin{figure}[tbhp]
  \centering
  \includegraphics[width=0.9\columnwidth]{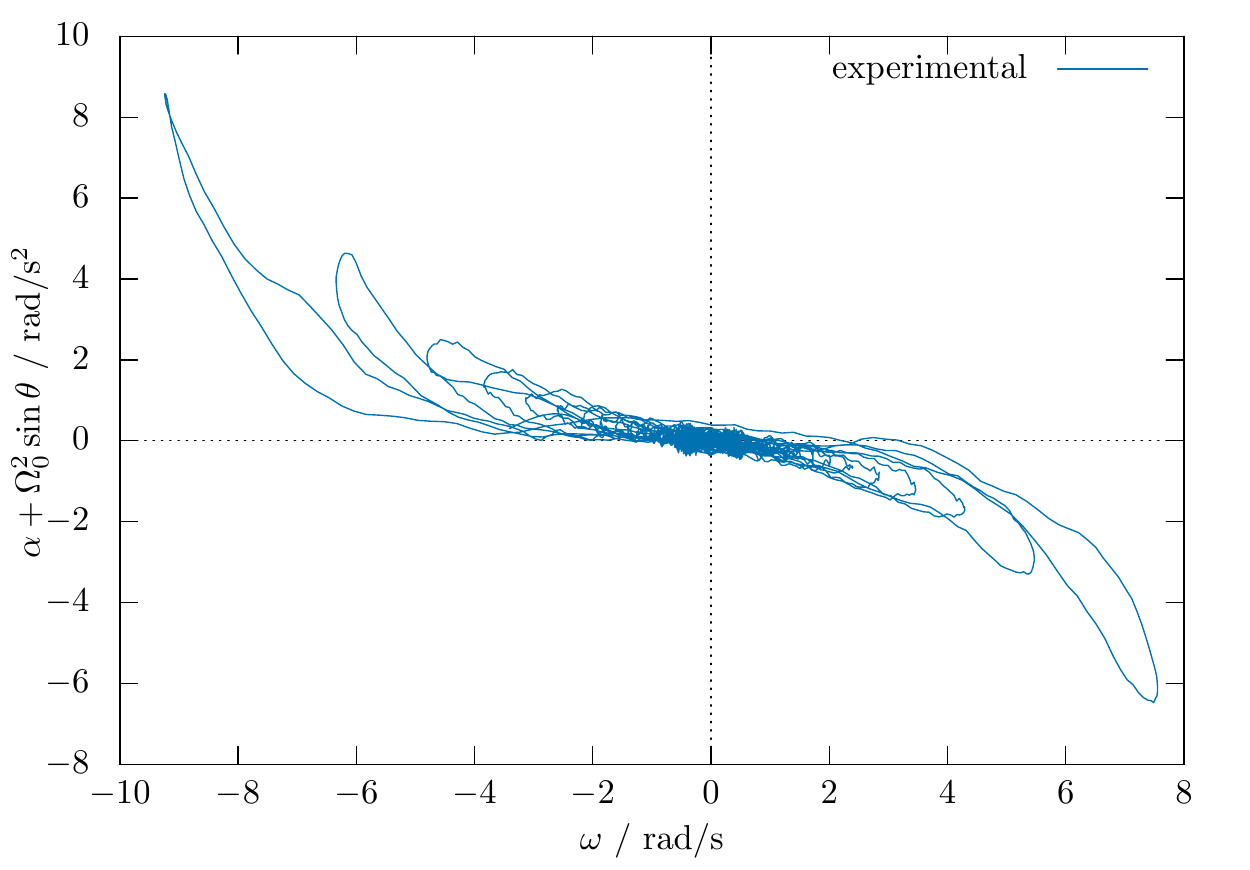} 
  \caption{Initial analysis of experimental data for run
    \texttt{24la1} (see text).}  
  \label{fig:lobolusdecaras}
\end{figure}

\begin{figure}[hbtp]
  \begin{center}
    \includegraphics[width=0.9\columnwidth]{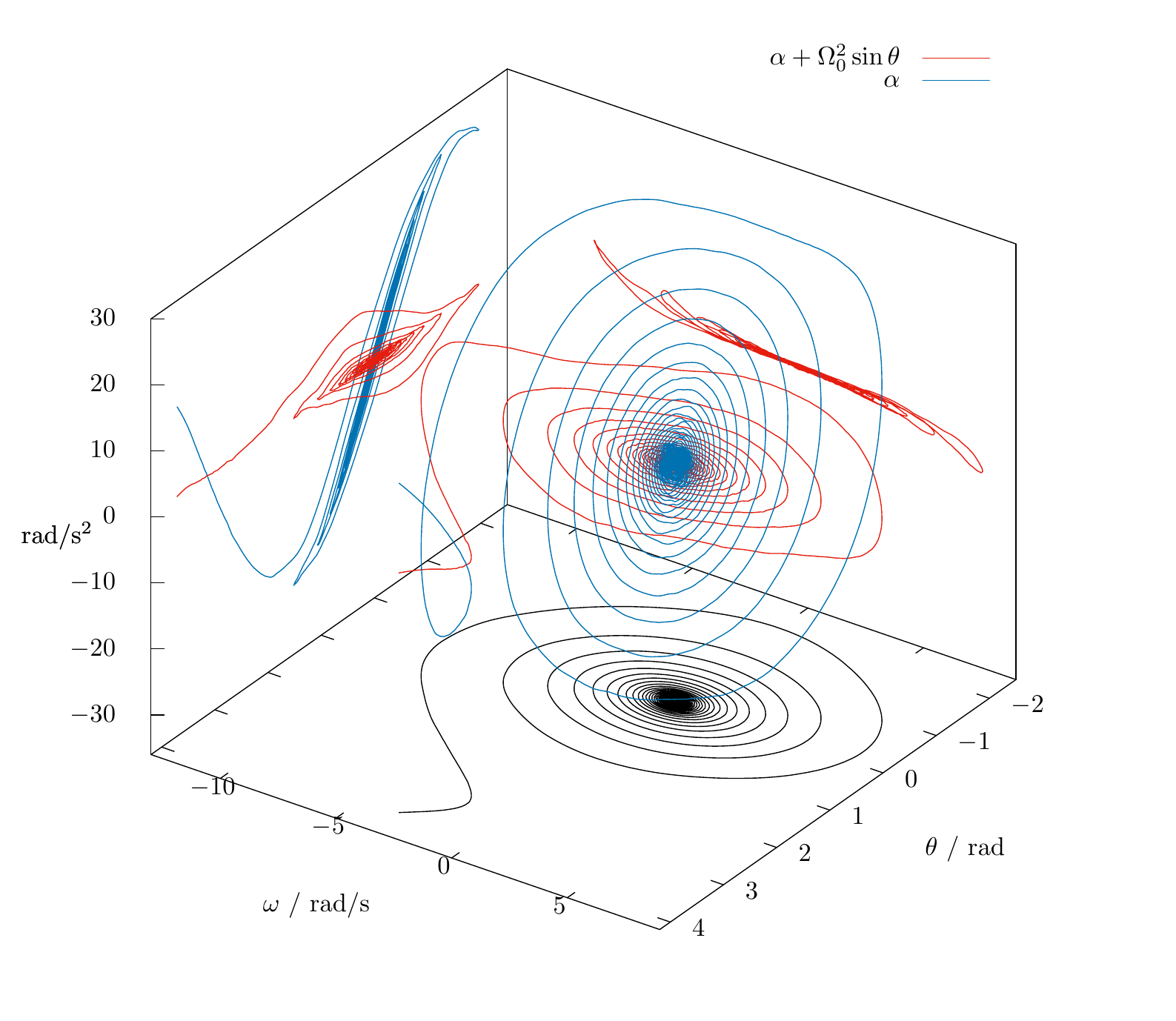}
  \end{center}
  \caption{Kinematic data for run \texttt{24la1}.}    
  \label{fig:S1}  
\end{figure}

Some exploratory $\theta(t)$ runs are presented In Figure
\ref{fig:inicondits} to illustrate how the pendular motion depends on 
different initial conditions. 
The angular acceleration $\alpha=\frac{d^2\theta}{dt^2}$ is plotted in
Figure \ref{fig:ozsin} as a function of the extreme angles
$\theta_\mathrm{ext}$ (for which the angular velocity
$\omega=\frac{d\theta}{dt}$ is zero). The values of both $\omega$ and
$\alpha$ are obtained directly from $\theta(t)$ data via
Savitzky-Golay filters 
\cite{Press1990}
using cubic polynomials and windows of 25 points (25 consecutive
video frames). It is clear that $\alpha(\theta_\mathrm{ext})$
is well fitted by a sine function with a coefficient of
proportionality $-\Omega_0^2$: 
\begin{equation}
  \label{eq:straightline}
  \alpha(\omega=0) \approx - \Omega_0^2\sin\theta_\mathrm{ext}.
\end{equation}
The acceleration may be evaluated relatively to
this function for any velocity:
\begin{equation}
  \label{eq:lobulosprovados}
  \Delta\alpha(\omega) = \alpha(\omega)+\Omega_0^2 \sin\theta(\omega).
\end{equation}
This is plotted in Figure \ref{fig:lobolusdecaras} which
puts in evidence the hysteretic nature of pendular motion. 
Note that Figures \ref{fig:ozsin} and \ref{fig:lobolusdecaras}
constitute mappings of the three-dimensional data plotted in 
Figure \ref{fig:S1}.

\section{General equation of motion}\label{sec:eqofmotion}

The net torque \Tnet\ acting on the pendulum is
\begin{equation}
  \label{eq:nettorque}
   \Tnet= \frac{dL}{dt}=\frac{dI}{dt}\omega + I\alpha
\end{equation}
where $L=I\omega$ is the angular momentum, $t$ is time, $I$ is the
moment of 
inertia, $\omega=\frac{d\theta}{dt}$ is the angular velocity, and
$\alpha=\frac{d\omega}{dt}$ is the angular acceleration. Under the
assumption of uniform air density and negligible pressure gradients,
the net torque is also the sum of gravitic and dissipative torques
$\Tnet=\Tgrav+\Tdiss$, so
\begin{equation}
  \label{eq:basiceq}
  I\alpha = \Tgrav+\Tdiss - \frac{dI}{dt}\omega.
\end{equation}
Usually, it is assumed that the pendulum is a strictly rigid body,
its moment of inertia is therefore constant and the
$-\frac{dI}{dt}\omega$ term vanishes. 
However, we may assume this term to be neither conservative nor
dissipative 
\begin{equation}
  \label{eq:tiner}
  -\frac{dI}{dt}\omega=\Tiner
\end{equation}
but inertial and dependent on the air surrounding the pendulum
as noted by Squire
\cite{Squire1986}:
\begin{quote}
  What does seem clear is that the acceleration of an oscillating body
  causes acceleration in the entrained air around it [\ldots].
\end{quote}
From this point of view it is reasonable to expect that the inertia does
not change significantly when either the pendulum is not moving
($\omega=0$) or its velocity is not changing ($\alpha=0$). On the
contrary, we expect inertia to change when velocity is changing
($\alpha\neq 0$). Furthermore, the greater the velocity, the greater
the amount of displaced air per unit time. Therefore, the changes of
inertia may be considered proportional to velocity, as a first 
approximation. There is also the possibility that the amount of air in
coherent motion with the pendulum may be directly correlated
with the moment of inertia. In this way the changes of inertia may
be considered proportional to acceleration, velocity, and moment of
inertia: 
\begin{equation}
  \label{eq:dIdt}
  \frac{dI}{dt}=\tau^2 I\omega\alpha
\end{equation}
where $\tau$ is an inertial characteristic time.
A dimensional analysis of the above expression provides some
additional insight on the underlying physics of this
phenomenon. In fact, $I\omega\alpha$ has dimensions of power,
therefore, equation (\ref{eq:dIdt}) states that the rate of change of
$I$ is proportional to a power and is positive 
when $\omega$ and $\alpha$ have the same sign.
Also, it is worth to note that $\tau^2 I\omega\alpha$ has dimensions
of ML$^2$T$^{-1}$=(ML$^{-1}$T$^{-1}$)L$^3$, which is viscosity times 
volume. In this way, the same rate of change of inertia can be obtained
either with low viscosity and big volume or with high viscosity and
small volume. 

The integration of equation (\ref{eq:dIdt}) is straightforward and
provides
\begin{equation}
  \label{eq:MOI}
  I=I_0 e^{\frac{(\tau\omega)^2}{2}}
\end{equation}
where $I_0$ is the moment of inertia at rest, leading to 
\begin{equation}
  \label{eq:angulmom}
  L=I_0 \omega e^{\frac{(\tau\omega)^2}{2}}.
\end{equation}
As a consequence of this, $L$ depends not only on $I_0\omega$ but
also on a speed dependent factor that ``inflates'' the angular momentum. 

Inserting equation (\ref{eq:dIdt}) in equation (\ref{eq:tiner}), one
finds  
\begin{equation}
  \label{eq:tiner2}
  \Tiner=-I(\tau\omega)^2\alpha 
\end{equation}
and inserting equation (\ref{eq:dIdt}) in equation (\ref{eq:basiceq}),
the result is 
\begin{equation}
  \label{eq:alpha2}
  \alpha=\frac{\Tgrav}{I}+\frac{\Tdiss}{I}-(\tau\omega)^2\alpha
\end{equation}
where the term $-(\tau\omega)^2 \alpha$ was already found by Basano
\etal\ 
\cite{Basano1989}. 

In conclusion 
\begin{equation}
  \label{eq:initialbasemodel}
  \alpha=\frac{\Tgrav+\Tdiss}{\left(1+\omega^2\tau^2\right)I}
\end{equation}
meaning that the pendulum is no longer considered 
a strictly rigid body because of the changing amount of air that is
dragged along and moves coherently with the rigid pendulum.
It is interesting to note that similar phenomena has been observed  
with water column oscillators
\cite{Lorenceau2002,Smith2019}. 

The gravitic torque can be expressed as
\begin{equation}
  \label{eq:Tgrav}
  \Tgrav = -Mgl_\mathrm{com}\sin\theta   
\end{equation}
where $M$ is the effective gravitic mass of the pendulum (discounting
buoiancy) and $l_\mathrm{com}$ is the distance between the rotation
axle and the center of mass (see Figure \ref{basicscheme}). 
It is important to note that the air surrounding
the pendulum contributes to inertia, equation (\ref{eq:dIdt}), but
neither contributes to gravitic mass nor to the center of mass because
it has null effective mass. 

Inserting equation (\ref{eq:Tgrav}) and equation (\ref{eq:MOI}) in
equation (\ref{eq:initialbasemodel}) we obtain  
\begin{equation}
  \label{eq:prevbasemodelbetter}
  \alpha=\frac{\frac{\Tdiss}{I_0}-\Omega_0^2\sin\theta}%
              {\left(1+\omega^2\tau^2\right)e^{\frac{(\tau\omega)^2}{2}}}.
\end{equation}
where 
\begin{equation}
  \label{eq:NatOmega0}
  \Omega_0=\sqrt{\frac{Mgl_\mathrm{com}}{I_0}}
\end{equation}
is, by definition, the natural angular frequency.

\begin{figure}
  \centering
  \includegraphics[width=0.9\columnwidth]{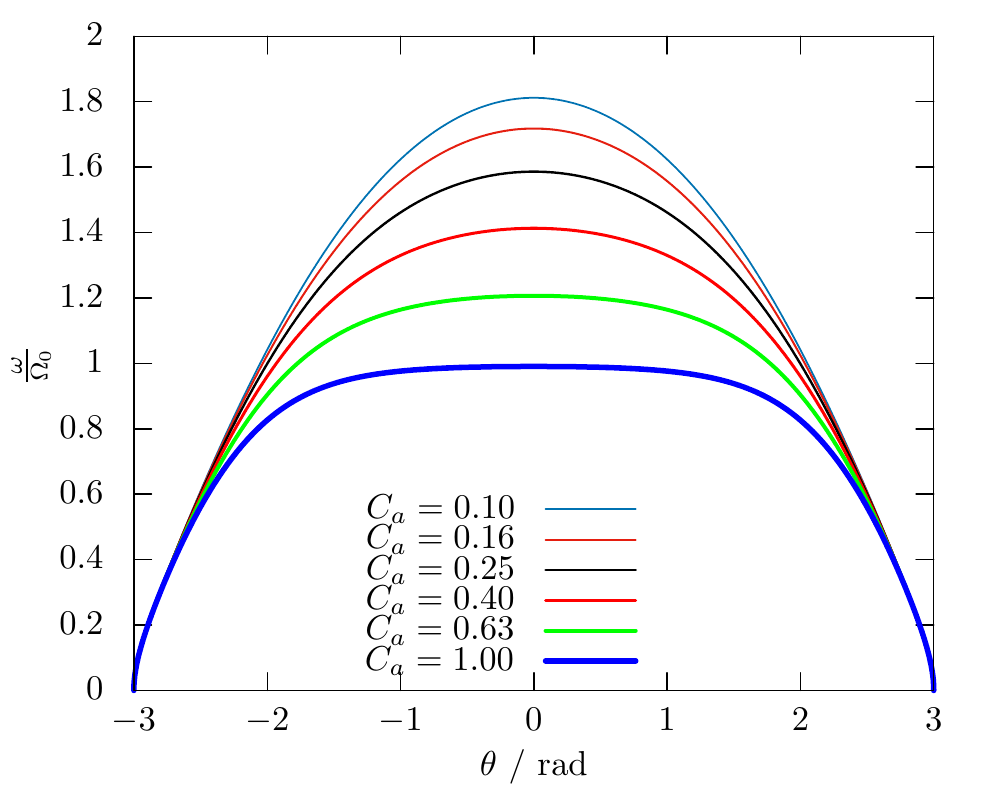}
  \caption{Plots of equation (\protect \ref{eq:pureinertialsolut}) 
    with $\theta_0=\SI{-3}{\radian}$.} 
  \label{fig:nullmodel}
\end{figure}

\begin{figure}
  \centering
  \includegraphics[width=0.9\columnwidth]{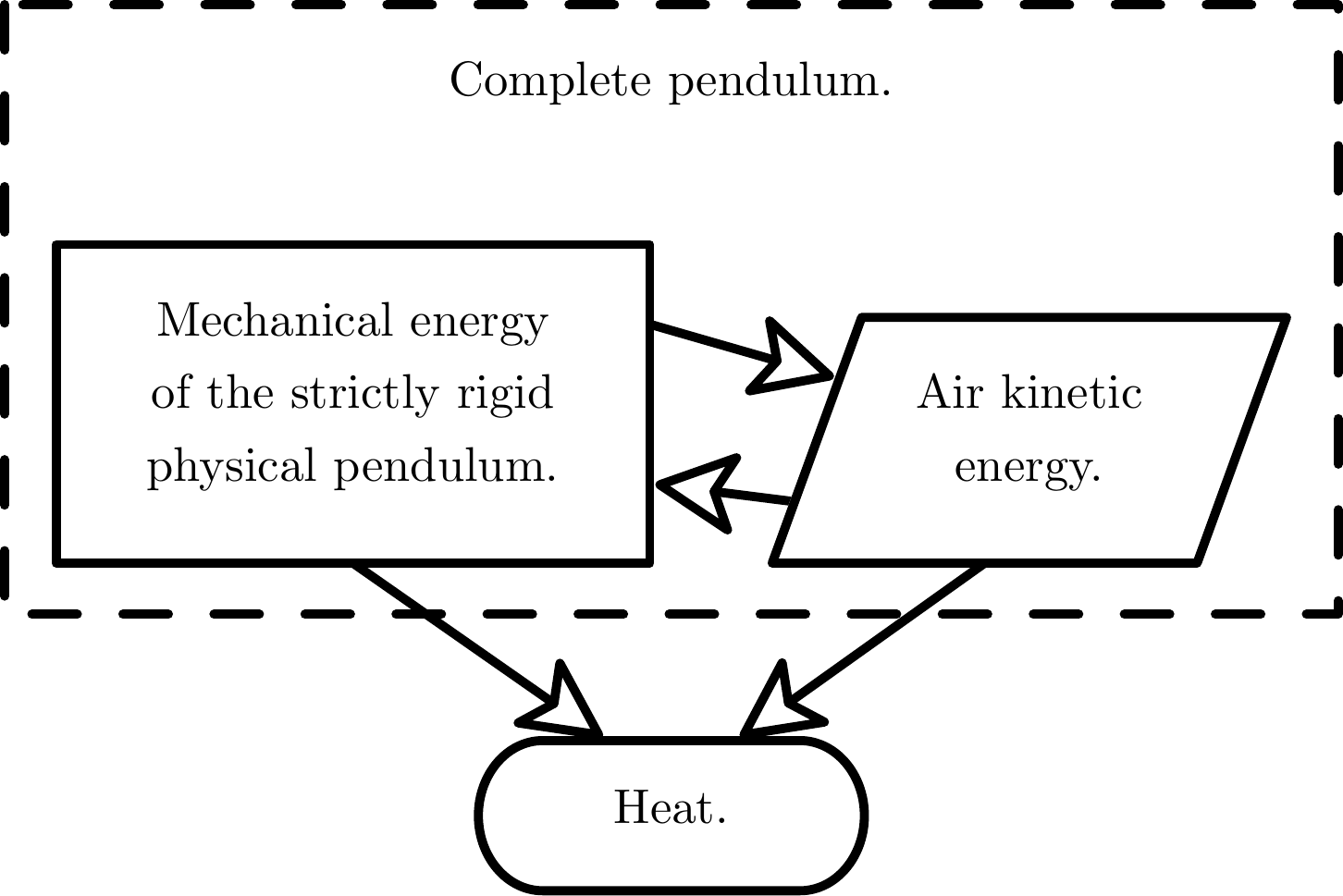} 
\caption{Schematics of the energy exchanges taking place during
  the physical pendulum motion. Part of the mechanical energy of the
  pendulum is converted into kinetic energy of the air. Some of this
  kinetic energy is recovered. The remaining part is dissipated as
  heat.}   
\label{dia:energyexchanges}
\end{figure}

As shall be confirmed in Figure \ref{plotfig:fourparamsclassic}, the 
denominator in equation (\ref{eq:prevbasemodelbetter}) may be twice
linearized in $x=(\tau\omega)^2$: 
\begin{equation}
  \label{eq:linearize}
  \frac{1}{\left(1+(\tau\omega)^2\right)e^{\frac{(\tau\omega)^2}{2}}}\approx
 \frac{1}{1+\frac{3}{2}(\tau\omega)^2}\approx 1-\frac{3}{2}(\tau\omega)^2
\end{equation}
Equation (\ref{eq:prevbasemodelbetter}) can, then, be written as
a general equation of motion:
\begin{equation}
  \label{eq:basemodel1}
  \alpha+\Omega_0^2\sin\theta=\adiss+\ainer
\end{equation}
in terms of both
\begin{equation}
  \label{eq:adissdef}
  \adiss=\frac{\Tdiss}{I_0}\left(1-\frac{3}{2}(\tau\omega)^2\right)
\end{equation}
and
\begin{equation}
  \label{eq:ainerdef}
\ainer=\frac{3}{2}(\tau\omega\Omega_0)^2\sin\theta=C_a\omega^2\sin\theta,
\end{equation}
where $C_a=\frac{3}{2}(\tau\Omega_0)^2$ is a non-dimensional
coefficient that measures the effect of ``added mass''  
\cite{Ermanyuk2000,Neill2007,Messer2010,Pantaleone2011,Raza2012,Konstantinidis2013}. 
Equation (\ref{eq:ainerdef}) is similar to the inertial nonlinearity
term introduced by Kavyanpoor and Shokrollahi
\cite{Kavyanpoor2017}
in a generalized Duffing oscillator equation of motion.
Suppose now that $\adiss=0$ in equation (\ref{eq:basemodel1}).
Then we have
\begin{equation}
  \label{eq:nullmodel}
  \alpha+(\Omega^2_0-C_a\omega^2)\sin\theta=0.
\end{equation}
Note that the ``added mass'' reduces the angular frequency of the
pendulum. This reduction may be named ``inertial redshift'' in
analogy with what is known as ``damping redshift'' 
\cite{Peters2004}.
Equation (\ref{eq:nullmodel}) provides the following solution for a
pendulum that initiates its movement from rest at an angle $\theta_0$.
\begin{equation}
  \label{eq:pureinertialsolut}
  \frac{\omega}{\Omega_0}=\sqrt{\frac{1-e^{2(\cos\theta_0-\cos\theta)C_a}}{C_a}}
\end{equation}
This solution is plotted in Figure \ref{fig:nullmodel} and proves that
\Tiner\ in fact is not dissipative because the solution is an even
function in $\theta$ (the speed returns to the same value after a full
swing). 
Figure \ref{fig:nullmodel} reflects the effect of equation
(\ref{eq:dIdt}): when the speed is increasing the moment of inertia is
also increasing. The very curious
consequence of this moment of inertia evolution is that when the speed
increases, the ratio acceleration to velocity shows a flattening
behaviour. 
In fact, this is equivalent: (i) to a storage of kinetic
energy in the air when the speed is increasing; and (ii) to a release
of kinetic energy from the air to the strictly rigid part of the
pendulum when the speed is decreasing. 
This stored kinetic energy is converted into potential energy because 
the air that moves together with the strictly rigid part of the
pendulum provides a ``gentle push''. This ``gentle push'' constitutes
a kind of parametric pumping 
\cite{Sanmartin1984,Wirkus1998}.
Figure \ref{dia:energyexchanges}
contains a schematic diagram of the energy exchanges taking place.

\subsection{Mechanical energy}\label{sec:mechener}

In order to calculate the mechanical energy,  
we start by rewriting 
equation (\ref{eq:prevbasemodelbetter}) with the first linearization
of equation (\ref{eq:linearize}) in the form of a specific
torques equation as follows:  
\begin{equation}
  \label{eq:specifictorques}
  \left(1+\frac{3}{2}(\tau\omega)^2\right)\alpha 
          +\Omega_0^2\sin\theta=\frac{\Tdiss}{I_0}.
\end{equation}

On multiplication by $I_0d\theta=I_0\omega\,dt$, 
and using equation (\ref{eq:NatOmega0})
the specific torques
are converted to differential energies:
\begin{equation}
  \label{eq:specificpowers}
  -Mgl_\mathrm{com}\sin\theta\,d\theta -\Tdiss d\theta = 
             \left(1+\frac{3}{2}(\tau\omega)^2\right)I_0\omega\,d\omega.
\end{equation}
Upon integration and using the conveniently chosen
integration constant $E-Mgl_\mathrm{com}$, we obtain an 
energy equation: 
\begin{equation}
  \label{eq:energies}
  Mgl_\mathrm{com}(1-\cos\theta)+E_\mathrm{diss}
  +\frac{1}{2}I_0\left(1+\frac{3}{4}(\tau\omega)^2\right)\omega^2 = E
\end{equation}
where $E$ is the total energy of the system and $E_\mathrm{diss}$ is
the dissipated energy (this makes the mechanical energy 
$E_\mathrm{mech}=E-E_\mathrm{diss}$).
Given that the gravitic potential energy may be defined relatively to the
lowest equilibrium point as
\begin{equation}
  \label{eq:graviticenergy0}
  E_\mathrm{p}=Mgl_\mathrm{com}(1-\cos\theta)
\end{equation}
and presenting the maximum 
\begin{equation}
  \label{eq:graviticenergy}
  E_{\mathrm{p}_\mathrm{MAX}}=2Mgl_\mathrm{com}=2I_0\Omega_0^2,
\end{equation}
then the remaing term is the kinetic energy
\begin{equation}
  \label{eq:kineticenergy}
  E_\mathrm{k}=\frac{1}{2}I_0
             \left(1+\frac{3}{4}(\tau\omega)^2\right)\omega^2=
                \frac{1}{2}I_0 \omega^2 + \frac{3}{8}I_0\tau^2\omega^4.
\end{equation}
This means that the kinetic energy of the physical pendulum has one 
term associated with the moving rigid body
\begin{equation}
  \label{eq:kineticrigid}
  E_\mathrm{rig} = \frac{1}{2}I_0 \omega^2
\end{equation}
and another term associated with the moving surrounding air 
\begin{equation}
  \label{eq:kineticair}
  E_\mathrm{air} = \frac{3}{8}I_0\tau^2 \omega^4.
\end{equation}
This is similar to situations involving a component of
pseudowork-energy balance in dissipative systems 
\cite{Penchina1978,Copeland1982,Sherwood1983,Arons1999,Guemez2013,Guemez2016,Guemez2018}.

Using equation (\ref{eq:graviticenergy}), equation (\ref{eq:energies})
may be written in the following normalized form 
$\mathcal{E} = \frac{E}{E_{\mathrm{p}_\mathrm{MAX}}}$
\begin{equation}
  \label{eq:nondimmechener}
  \mathcal{E} =
          \left( \sin\frac{\theta}{2} \right)^2 +
          \left( \frac{\omega}{2\Omega_0} \right)^2 + 
          \left( \sqrt{3}\tau\omega\frac{\omega}{2\Omega_0} \right)^2
          +\frac{E_\mathrm{diss}}{E_{\mathrm{p}_\mathrm{MAX}}}.
\end{equation}
From the above equation the normalized mechanical energy 
(written like a three-dimensional Pythagorean Theorem) is:
\begin{equation}
  \label{eq:amplitudes}
  \mathcal{A}^2=
  \mathcal{A}_\theta^2+{\mathcal{A}_\omega}^2+\mathcal{A}_\mathrm{air}^2. 
\end{equation}
where 
$\mathcal{A}=\sqrt{\mathcal{E}_\mathrm{mech}}=\sqrt{\mathcal{E}-E_\mathrm{diss}/E_{\mathrm{p}_\mathrm{MAX}}}$,
$\mathcal{A}_\theta=\sin(\theta/2)$, 
$\mathcal{A}_\omega=\omega/(2\Omega_0)$ and 
$\mathcal{A}_\mathrm{air}=\sqrt{3}\tau\omega\mathcal{A}_\omega$.
In view of the fact that $\mathcal{A}_\mathrm{air}^2$ is 
likely to be very small, and also that equation (\ref{eq:dIdt}) is
just an approximation, we choose to consider
\begin{equation}
  \label{eq:amplitudesobservable}
  \mathcal{A}^2\approx \mathcal{A}_\theta^2+{\mathcal{A}_\omega}^2.
\end{equation}
It is interesting to note that the above equation allows for a critical
angular velocity 
\begin{equation}
  \label{eq:velocritica}
  \omega_c = 2\Omega_0
\end{equation}
for which the maximum kinetic energy exceeds the maximum potential
energy thus separating the oscillatory regime from the rotational motion
\cite{Naudts2005,Lima2010}.
Also, equation (\ref{eq:amplitudesobservable}) provides a way to
define the instantaneous phase $\phi$ of the pendular motion as
\begin{equation}
  \label{eq:phasephi}
  \tan\phi=\frac{\mathcal{A}_\theta}{\mathcal{A}_\omega}.
\end{equation}
As a consequence, the phase speed $\frac{d\phi}{dt}$ is
\begin{equation}
  \label{eq:phasespeed}
  \Phi=
  \frac{\frac{d\mathcal{A}_\theta}{dt}\mathcal{A}_\omega
       -\mathcal{A}_\theta\frac{d\mathcal{A}_\omega}{dt}}%
        {\mathcal{A}_\theta^2+{\mathcal{A}_\omega}^2}=
  \frac{\left(\cos\frac{\theta}{2}\right)\frac{\omega^2}{4\Omega_0}
         -\left(\sin\frac{\theta}{2}\right)\frac{\alpha}{2\Omega_0}}%
  {\left(\sin\frac{\theta}{2}\right)^2+\left(\frac{\omega}{2\Omega_0}\right)^2}.
\end{equation}

\subsection{Dissipation, inertia and negative damping}

In addition to the viscous or Stokes drag, usually used in damped
harmonic motion, Coulomb dry-friction and turbulent or Newton drag 
are also included in the standard dissipative acceleration expression  
\cite{Squire1986,Nelson1986,Takahashi1999,Arora2006,Guo2011,Smith2012,Dahmen2014,Klein2017}:
\begin{equation}
  \label{eq:standarddiss}
  \adisss =-C_1\omega-\frac{C_0+C_2\omega^2}{\signalop(\omega)}
\end{equation}
where  
\begin{equation}
  \label{eq:sgn}
  \signalop(x) = \left\{\begin{array}{l} 
      1 \Leftarrow x\geq 0 \\ 
      -1 \Leftarrow x<0
\end{array}\right. 
\end{equation}

Negative damping can be perceived as positive forcing 
\cite{Graef1972,Fulcher2006,Stoop2006,Jenkins2013}
that happens
whenever $\alpha + \Omega_0^2 \sin\theta$ has the same sign of
$\omega$, that is, the conditions observed in the odd quadrants of 
Figure \ref{fig:lobolusdecaras} (highlighted in Figure
\ref{plotfig:dissipativehysteresis}). 

For strictly dissipative torques, 
$\alpha + \Omega_0^2 \sin\theta = \adiss$, negative damping can't be
observed. However, in view of the inertial torque, 
the condition for negative damping is
\begin{equation}
  \label{eq:negativedampingdefinition}
  (\adiss+\ainer)\omega > 0.
\end{equation}
Given the \adiss\ and \ainer\ expressions, it is clear both that, for 
positive angular velocity, the sum $\adiss+\ainer$ can only be
positive for positive quadratic coefficient,
i.e. $C_a\sin\theta-C_2>0$, and that, for angular velocities close to
zero, negative damping is never observed due to the non-null Coulomb
friction. 

It is important to note that this result is counter-intuitive since
the moving air, surrounding the pendulum, should contribute with
positive forcing at near-zero speeds.
From this point of view, the classical description of the dissipative
and inertial torques doubtly will account for the detailed description
of the pendular motion in particular for angular velocities close to
zero. In fact, in Figure \ref{fig:lobolusdecaras}, one can observe
typical characteristics of hysteresis and, close to null speed, positive
forcing.

\section{Classic model}\label{sec:classicmodel}

The standard dissipative acceleration $\adisss$, equation
(\ref{eq:standarddiss}), is a parabolic function of angular velocity
$\omega$  
\cite{Kostov2008}.
Some authors have used power-laws
\cite{Crawford1975,Ravindra1994,Baltanas2001,Mickens2003,Jaksic2011,Elliott2015,Lukichev2016,Plastino2018b}.
In order to test if such models provide better fits we considered the
following power-law
\begin{equation}
  \label{eq:powerlawdiss}
  \adissp =-
\frac{C_0+C_3\left|\frac{\omega}{\Omega_0}\right|^{p}}{\signalop(\omega)}.
\end{equation}
We verified that, although $\adissp(\omega)$ contains a null slope
at zero speed, it does in fact allow better fits. 
Note that this feature bypasses all considerations concerning the
mathematical expression of low speed friction 
\cite{Flores2008,Muvengei2012}.

So, the classic model used is
\begin{align}
  \alpha+\Omega_0^2\sin\theta & = \ainer+\adissp \nonumber \\
  & = C_a\omega^2\sin\theta
   -\frac{C_0+C_3\left|\frac{\omega}{\Omega_0}\right|^{p}}{\signalop(\omega)}.
       \label{eq:classicmodel}  
\end{align}

\subsection{Classic model results}\label{sec:classicresults}

We used Fitteia
\cite{Sebastiao2013}
to fit the classic model \S\ref{sec:classicmodel} 
and the OPA model \S\ref{sec:lasermodel}. Fitteia is a powerfull
fitting and plotting online platform that fulfills
most of the requirements suggested in 
\cite{Grosse2014}.

\begin{figure*}[p]
\centering
\includegraphics[width=0.85\textwidth]{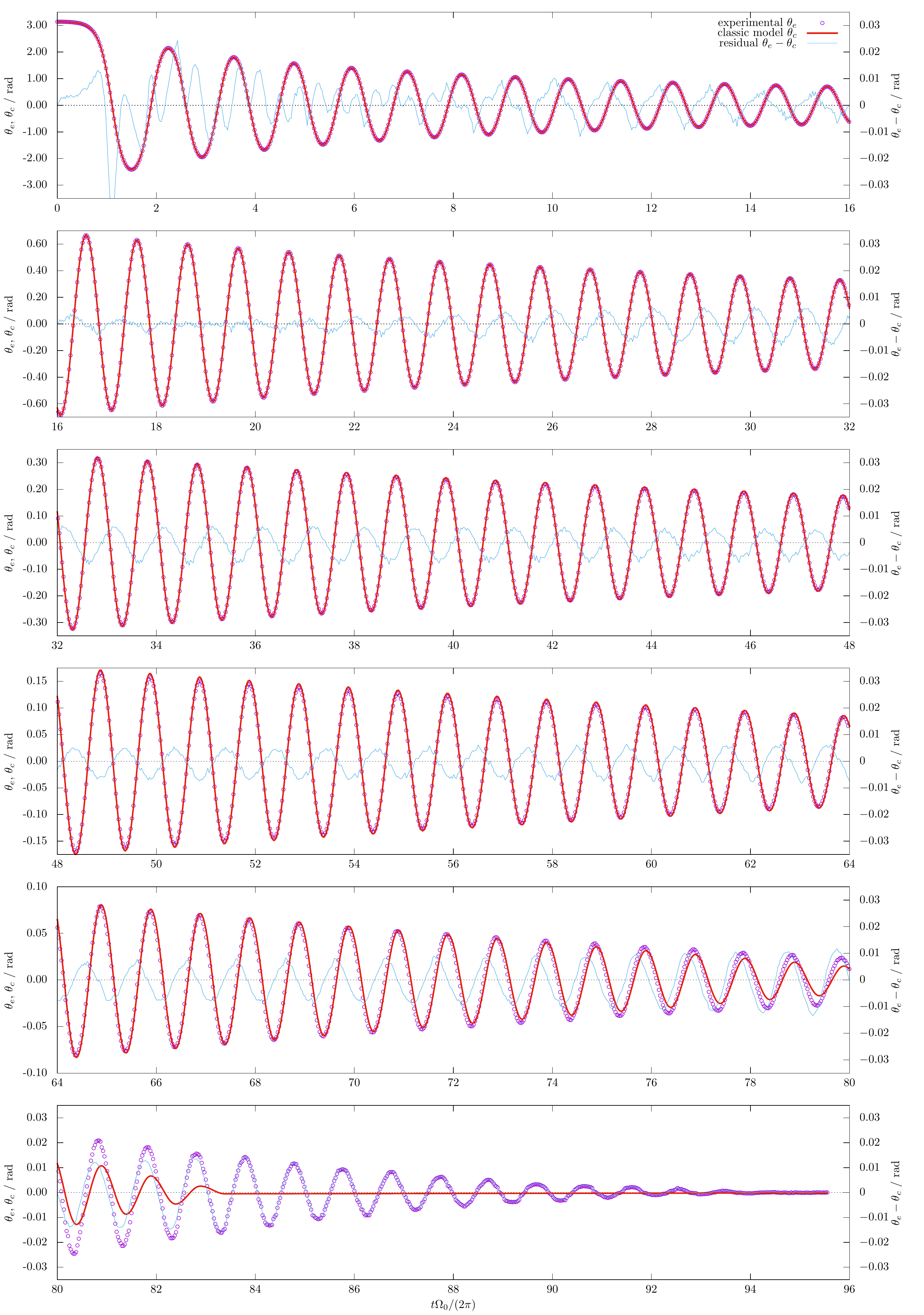}
\caption{Results for $\theta$  corresponding to run \texttt{03na1} fitted
  by the classic model. The complete run is shown to reveal the
  persistent spurious dephasing between the data and the model.}
\label{plotfig:spuriousosc}
\end{figure*}

\begin{figure}[hbtp]
  \begin{center}
    \includegraphics[width=0.9\columnwidth]{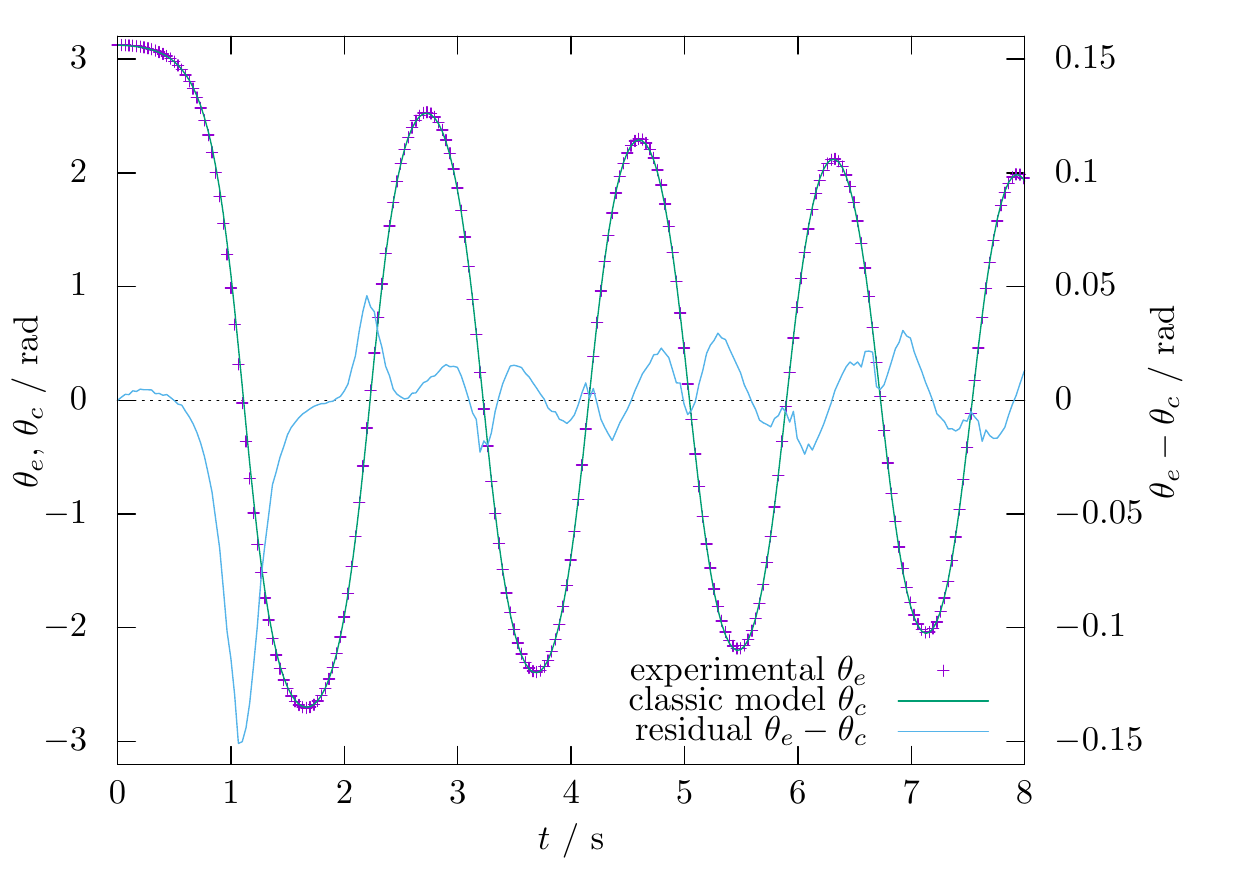}
  \end{center}
  \caption{Results for $\theta$  corresponding to run \texttt{00nn1}
  fitted by the classic model. Only the first few cycles 
  are shown to reveal the single 5\% residual peak near 
  $t=\SI{1}{\second}$.}
  \label{fig:S2}  
\end{figure}

\begin{figure}[hbtp]
  \begin{center}
    \includegraphics[width=0.9\columnwidth]{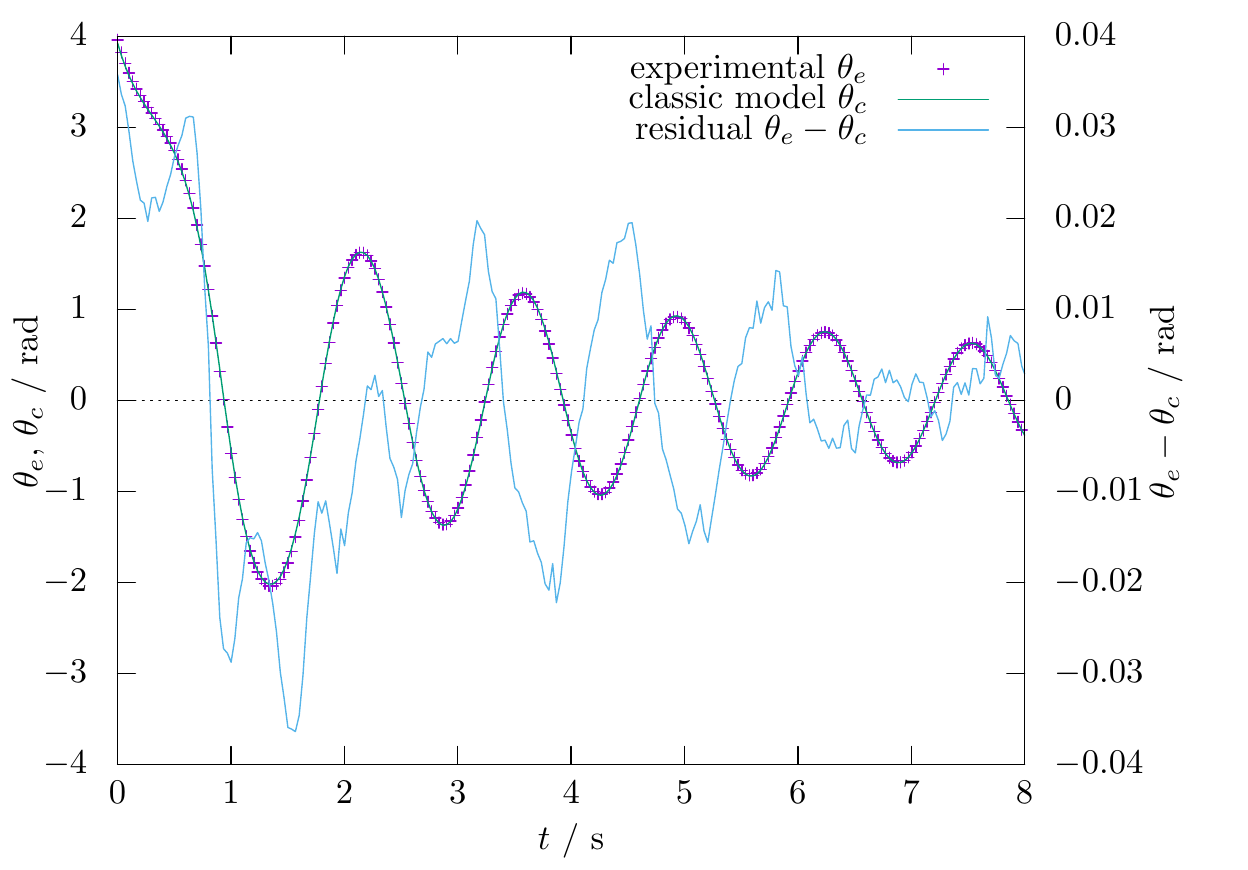}
  \end{center}
  \caption{Results for $\theta$  corresponding to run \texttt{18la1}
  fitted by the classic model. Only the first few cycles 
  are shown to reveal the high accuracy of the fit.}
  \label{fig:S3}  
 \end{figure}

The classic model, equation (\ref{eq:classicmodel}), matches quite
accurately all our experimental data. An example is given in Figure 
\ref{plotfig:spuriousosc} (also see Figures \ref{fig:S2} and \ref{fig:S3}).

\begin{figure}
\centering
\includegraphics[width=0.9\columnwidth]{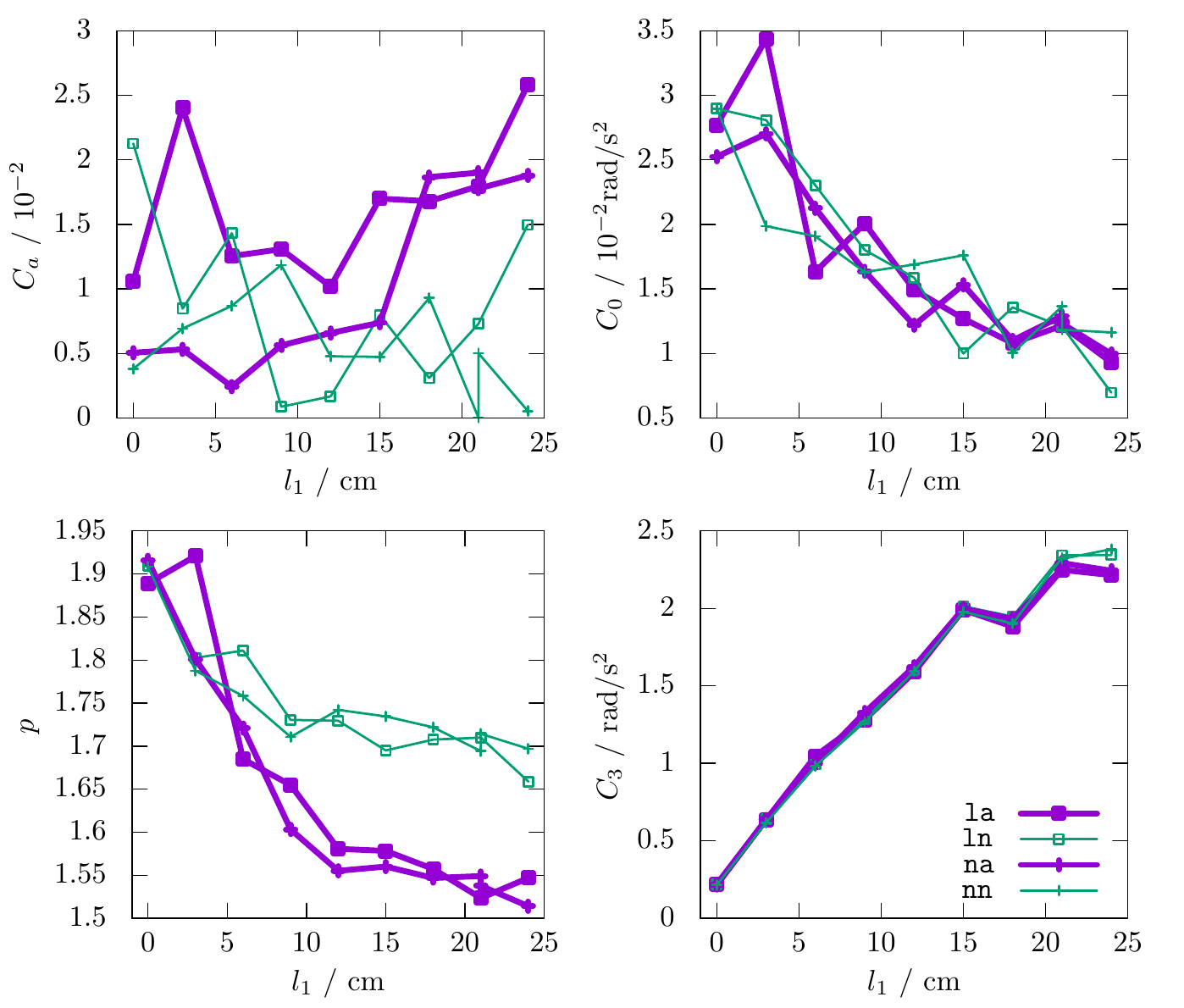}
\caption{Results from the classic model for parameters
  $C_a$, $C_0$, $p$ and $C_3$. The obtained values of $C_a$ prove
  that, in fact, equation (\protect \ref{eq:linearize}) is valid because 
  $(\tau\omega)^2=\frac{2C_a\omega^2}{3\Omega_0^2}
                 =\frac{8C_a}{3}\mathcal{A}_\omega^2 <
                 \frac{8C_a}{3} < 0.08 $. The results from the classic
  model for $\Omega_0$ were very much confirmed by the fractional
  model and can therefore be observed in 
  Figure \protect \ref{plotfig:fittexOmegaZer}.}
\label{plotfig:fourparamsclassic}
\end{figure}

We expected the classic model to provide clear results for $C_a$,
the inertial or added-mass parameter that we introduced, but that's
not so much the case. The complete set of obtained $C_a$ values is
presented in Figure \ref{plotfig:fourparamsclassic}.
We expected $C_a$ to increase when $l_1$ increases but this is only
apparent for alley runs (\texttt{la} and \texttt{na}). Also only the
three highest tiles show different $C_a$ between alley and no alley
runs. 

The results for $C_0$ are more consistent (also shown in Figure
\ref{plotfig:fourparamsclassic}).
We think that the decrease of $C_0$ on increasing $l_1$ is caused by
the larger tiles making very small and very slow oscillations disappear.

The most interesting results are those of $p$ (also shown in Figure
\ref{plotfig:fourparamsclassic}).
It is clear that there is a difference between runs with alley and
runs without alley. This difference can only be detected for the
larger tiles and shows that the alley reduces turbulence. 

The results for $C_3$ are very clear (also shown in Figure
\ref{plotfig:fourparamsclassic}) showing almost linear increase with
$l_1$ and independence of the alley. 
Note that if we had chosen the standard dissipative acceleration
\adisss, equation (\ref{eq:standarddiss}), it would be $C_2$ to
measure the importance of turbulence
\cite{Guo2011} 
and not $p$ in equation (\ref{eq:powerlawdiss}).

We conclude that the classic model is generally very satisfatory as it
can generally identify which runs used the alley and which runs did
not. However, the classic model
sometimes deviates a little from the first few swings of the pendulum
and,
furthermore, the classic model seems unable to eliminate a persistent 
dephasing or spurious oscillation visible in the residuals of the
fits as, for instance, in Figure \ref{plotfig:spuriousosc}.
Spurious oscillations are a long-standing problem in engineering
and are usually associated with delayed action and self-oscillations
\cite{Minorsky1942,Jenkins2013}.
\section{OPA model}\label{sec:lasermodel}

\begin{figure}
\centering
\begin{tabular}{c}
    \includegraphics[width=0.9\columnwidth]{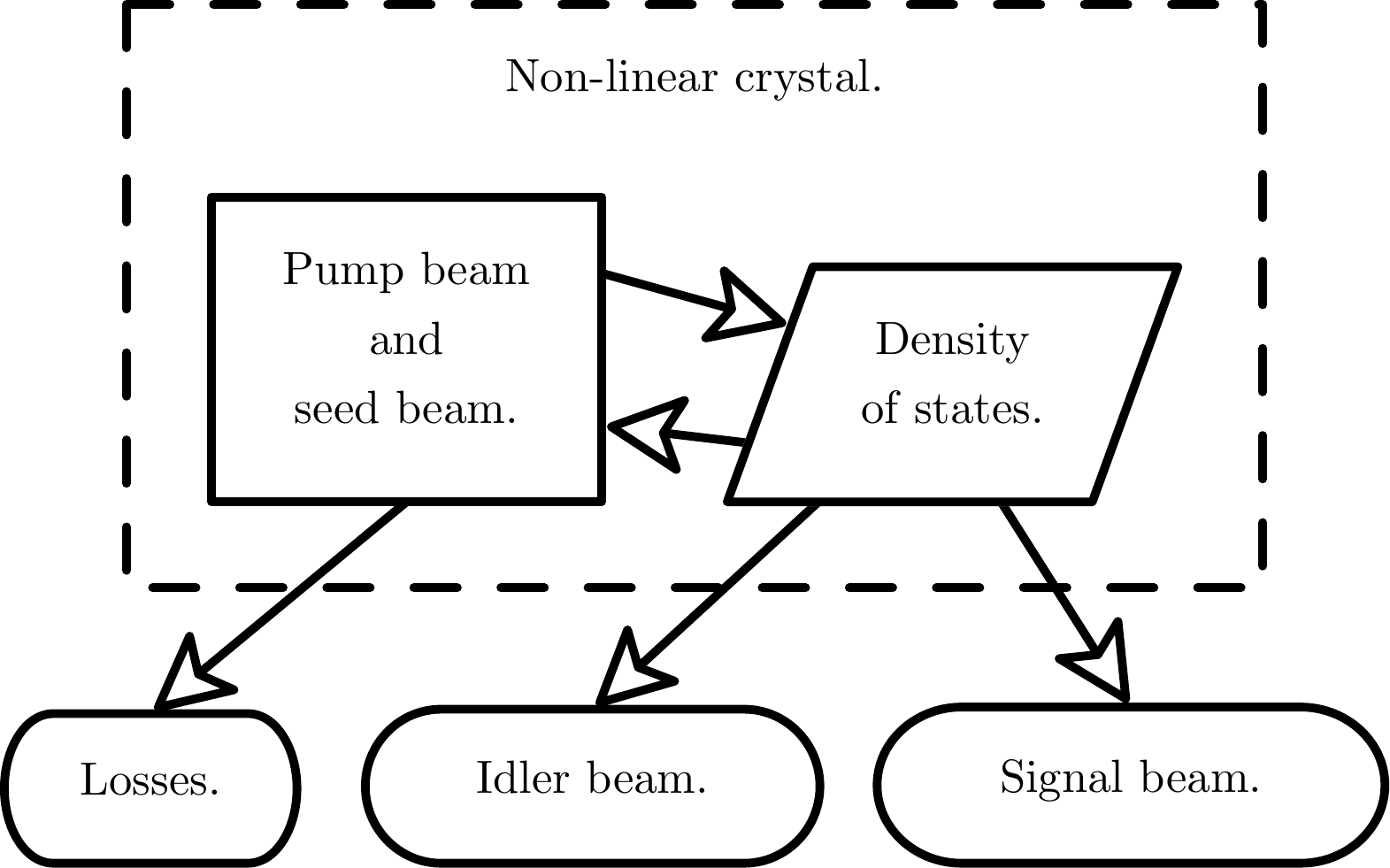} 
\end{tabular}
\caption{Schematics of the energy exchanges taking place during optical
  parametric amplification. The process of parametric light
  amplification is a process occurring in the presence of a non-linear
  crystal. Most of the input energy, which is given by the pump beam,
  is used/converted to an increase of density of states and therefore
  increasing the intensity of the signal beam. Some pump energy is
  recovered and lost and the remaining part in used in the idler
  beam. Compare with Figure \protect \ref{dia:energyexchanges}.}  
\label{dia:energyexchanges2}
\end{figure}

Pulse stretching and compression are two of the most crucial stages in
chirped pulse amplifiers. Although chirped pulse conventional
amplification (CPA) systems have enabled the development of
high-energy few-cycles pulses, an alternative technique for the
generation of high-energy ultrashort 
pulses results from a combination of the CPA' stages with optical
parametric amplification (OPA), which is a nonlinear optical phenomena 
\cite{Cerullo2003,Shen2006,Boyd2008},
has been used.
The conjugation of CPA with a nonlinear three-wave-mixing
process, occurring within an adequate non-linear crystal where a
stronger and higher frequency input wave (pump pulse) 
\emph{exchanges energy} with a weaker and lower frequency
input wave (seed pulse), generates an output signal pulse and also an
auxiliary wave (idler pulse) due to energy and moment conservation
(see Figure \ref{dia:energyexchanges2}).

\begin{figure}
\centering
\includegraphics[width=0.9\columnwidth]{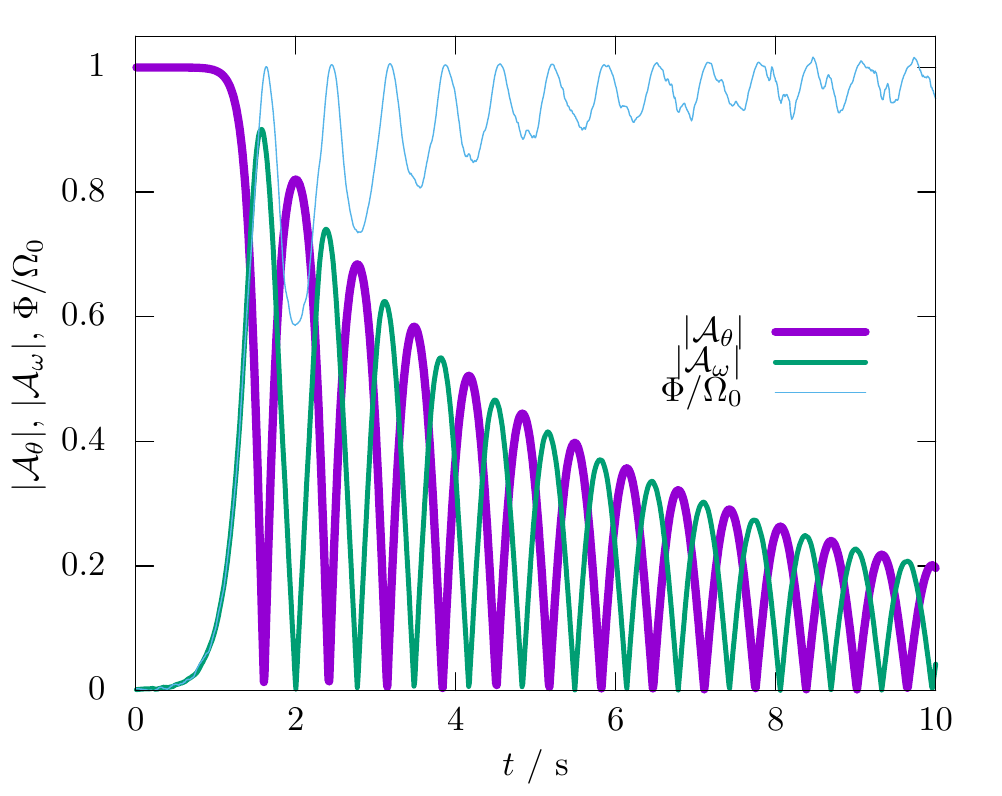}
\caption{Results for $\mathcal{A}_{\theta}$ and for
  $\mathcal{A}_{\omega}$ corresponding to run \texttt{24na1} fitted
  by the classic model. The obtained parameter values of $\Omega_0$
  and $C_a$ were used to calculate $\mathcal{A}_{\omega}$ and 
  the phase speed $\Phi$ (from equation 
  \protect \ref{eq:phasespeed}). Note the noise propagated into the
  calculation of $\Phi$.}    
\label{plotfig:schluck}
\end{figure}

In this context we have found an interesting analogy between the
physical pendulum relaxation (\S\ref{sec:mechener}) and
the amplification of chirped laser pulses
\cite{Lehmann2013,Schluck2015}.
Specially interesting is the similarity between our Figure
\ref{plotfig:schluck} and the numerical
solutions of the equations relating the pump $A_p$ and the seed $A_s$
amplitudes:  
\begin{align}
  \frac{dA_p}{d\zeta} & = \frac{\zeta BA_s}{4} \label{eq:laseroB1} \\
  \frac{dA_s}{d\zeta} & =-\frac{2A_s+2B^\ast A_p}{\zeta} 
    \label{eq:laseroB2} \\
  \frac{dB}{d\zeta}   & =-\frac{\zeta A_p A_s^\ast}{2} \label{eq:laseroB3}
\end{align}
where $B$ is density of states and $\zeta$ is a self-similar
coordinate. Within this analogy the amplitudes of the pump
pulse and the seed pulse correspond respectively to the amplitudes 
$\mathcal{A}_{\theta}$ and $\mathcal{A}_{\omega}$
of the energies involved in the movement of the pendulum,
suggesting an explanation/understanding of the motion of the
pendulum as it happens with OPA.   
This analogy can be made mathematically explicit:
\begin{align}
  A_p   & \equiv \mathcal{A}_\theta\label{eq:lasertransB1} \\
  A_s   & \equiv j\mathcal{A}_\omega\label{eq:lasertransB2} \\
  B     & \equiv -j\Phi^\prime \label{eq:lasertransB3} \\
  \zeta & \equiv \sqrt{t} \label{eq:lasertransB4}
\end{align}
where $j=\sqrt{-1}$ and $\Phi^\prime$ is a phase speed similar but not
equal to the phase speed defined in equation (\ref{eq:phasespeed}).
Equations (\ref{eq:laseroB1}), (\ref{eq:laseroB2}) and
(\ref{eq:laseroB3}) are, therefore, analogous to 
\begin{align}
  \frac{d \mathcal{A}_\theta}{dt} & =
                            \Phi^\prime \mathcal{A}_\omega 
                                    \label{eq:laserB1} \\
  \frac{d \mathcal{A}_\omega}{dt} & =
                            -\frac{C_4 \mathcal{A}_\omega 
                                 +\Phi^\prime \mathcal{A}_\theta}{t}
                                    \label{eq:laserB2} \\
  \frac{d \Phi^\prime}{dt} & =
                         -\Omega_0^2 \mathcal{A}_\theta \mathcal{A}_\omega 
                             \label{eq:laserB3}
\end{align}
where we have included the coefficient $C_4$
for generality. It's quite curious that equations
(\ref{eq:laserB1}), (\ref{eq:laserB2}) and (\ref{eq:laserB3}) can be
understood as equations of either energy transfer or amplitude
exchange where phase speed $\Phi^\prime$ plays the role of exchange rate
(analogous to density of states $B$). The explicit appearance of
time $t$ in one of the equations and their bad fit of the data led us 
to try a variety of similar sets of equations. We finally arrived at
the following compromise between smallest modification and best fit: 
\begin{align}
  \frac{d \mathcal{A}_\theta}{dt} & = \Phi^\prime \mathcal{A}_\omega 
                                     \label{eq:pump} \\
  \frac{d \mathcal{A}_\omega}{dt} & =
       -\frac{C_4 \mathcal{A}_\omega}{t}-\Phi^\prime
       \mathcal{A}_\theta -\signalop(\mathcal{A}_\omega)C_5
                                    \label{eq:seed} \\
  \frac{d \Phi^\prime}{dt} & =
        -\Omega_0^2 \mathcal{A}_\theta \mathcal{A}_\omega
                             \label{eq:phaspeed}
\end{align}
where we see that there is
an additional coefficient ($C_5$) to account for Coulomb friction.
Given that $\omega=\frac{d\theta}{dt}$ is equivalent to 
\begin{equation}
  \label{eq:Arelation}
  \frac{d\mathcal{A}_\theta}{dt}=
       \cos\frac{\theta}{2}\mathcal{A}_\omega\Omega_0,
\end{equation}
equation (\ref{eq:pump}) implies that
\begin{equation}
  \label{eq:opaphasespeed}
  \Phi^\prime = \Omega_0\cos\frac{\theta}{2}.
\end{equation}
This makes equation (\ref{eq:seed}) exactly equivalent to
\begin{equation}
  \label{eq:newDNHM}
  \frac{d\omega}{dt}+\frac{2\Omega_0 C_5}{\signalop(\omega)}+
                     \frac{C_4}{t}\omega+\Omega_0^2\sin\theta=0
\end{equation}
which is the equation of motion of a pendulum damped by both Coulomb
friction and a laminar drag that changes with time. This $C_4\omega/t$  
term provides the required hysteretic behaviour but, once again, 
there is no provision for negative damping.

The analogy between the non-linearities of the pendulum and non-linear
optics was also noted in reference
\cite{Christian2017}. 
Regarding analogies between a forced harmonic oscillator and
non-linear optics see 
\cite{George1983,Boscolo2014}.

\subsection{OPA model results}\label{sec:laseresults}

\begin{figure*}
\centering
\includegraphics[width=0.9\textwidth]{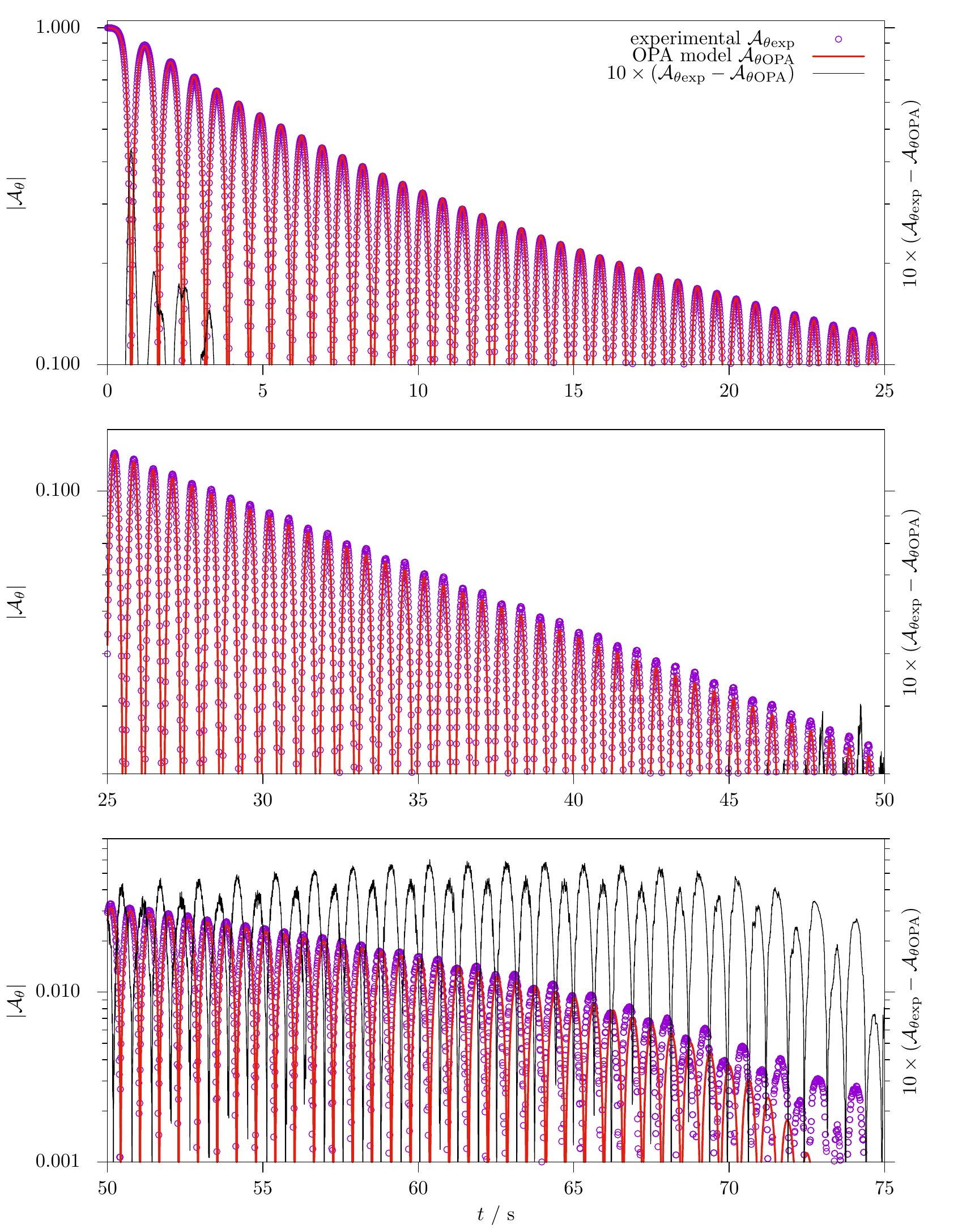}
\caption{Results for $\mathcal{A}_{\theta}$ corresponding to run
  \texttt{09la1} fitted by the OPA model. Fitted parameter values are 
  $\Omega_0=\SI{5.05}{\radian\per\second}$, $C_4=3.25$, 
  $C_5=\SI{1.57e-3}{\per\second}$, and
  $t_0=\SI{-8.85}{\second}$ (this is the zero of model time in the
  experimental time scale).}  
\label{fig:laserplots}
\end{figure*}

Equations (\ref{eq:pump}), (\ref{eq:seed}), and (\ref{eq:phaspeed})
allow for quite good fits. 
Figure \ref{fig:laserplots} presents one example.
Many of OPA model fits show a 5\% residual peak near
$t=\SI{1}{\second}$ similarly to the classic model 
(as in Figure \ref{plotfig:spuriousosc}).  

The explicit inverse dependence on time ($t^{-1}$) in
equation (\ref{eq:seed})
obviously imposes an hyperbolic amplitude decay matching the
experimental data for initial times. 

As the OPA model can't describe negative
damping, we resume its study and keep the idea that the experimental
data is consistent with energy storage and converted energy release
(parametric amplification that is similar to either self-oscillations
or parametric pumping). 
So, now we have two models that fit our data but not in a completely
satisfatory way. We continue investigating better models.
\section{Fractional derivatives}

One common example, often used to introduce the concept of memory, is 
what happens when the air and dust entrained by a moving car, on a
desert sandy road, ends up overcoming and falling on the 
wind-shield when the car slows down to a full stop. This intuitive event,
representing a memory effect of the whole system, can also be observed 
in underwater pendula
\cite{Bolster2010}. 

Memory effects are an integration of past events (history) and, in the
case of a physical pendulum, cannot be
modelled exclusively by an added mass that depends on the
instantaneous only $\theta$ and $\omega$ (see Figure
\ref{plotfig:dissipativehysteresis}). 
Memory effects are traditionaly described by the Basset history force
\cite{Basset1888,Hamilton1973,Herringe1976,Thomas1992,Mainardi1995,Chang1998,Candelier2004,Hinsberg2011,Baleanu2013,Daitche2015,Annamalai2017,Maris2019}
\begin{equation}
  \label{eq:bassetforce}
  F_H=\frac{3}{2}\rho d_s^2\sqrt{\pi\nu}
      \int_0^t a(t^\prime) \frac{d t^\prime}{\sqrt{t-t^\prime}}
\end{equation}
where $\rho$ is the fluid specific mass, $d_s$ is the diameter of a
sphere, $\nu$ is the kinematic viscosity and $a$ is the translational
acceleration of the sphere in the stationary fluid.
The expression of the Basset force is equivalent to a fractional
semi-derivative 
\cite{Tatom1988,Mainardi1997,Bombardelli2008,Lukerchenko2010}
\begin{equation}
  \label{eq:bassetforceassemider}
  F_H \propto \mathcal{D}_t^\frac{1}{2} v,
\end{equation}
where $v$ is the translational velocity of the sphere in the
stationary fluid. The above semi-derivative can be
generalized  
\begin{equation}
  \label{eq:bassetforceasfracder}
  F_H \propto \mathcal{D}_t^\beta v
\end{equation}
establishing a connection between the fractional derivative order 
and the permanence of memory  
\cite{Du2013}.
The above constitutes a sound basis for the introduction of a fractional
derivative in the physical pendulum equation of motion. Nevertheless,
there is a diversity of reasons, listed below, that provide additional
support.   
\begin{enumerate}
\item Fractional derivatives integrate all causes of memory effects
  and may, therefore, be used either as a replacement of those causes 
\cite{Ozgen2013} 
or as a completion of a rough model
\cite{Olejnik2018}.
\item Fractional models can describe negative damping  
  \cite{Sakakibara1997}, 
  mechanical energy increases, and/or odd-even symmetry breakings 
  \cite{Seredynska2005,Yin}.
\item Significative reductions in the number of model parameters was
  achieved in viscoelasticity 
  \cite{Bagley1983,Torvik1984,Gaul1991,Metzler2003},
  thermal systems
  \cite{Aribi2014},
  acoustics
  \cite{Falaize2014},
  electronics
  \cite{Quintana2006},
  and biology
  \cite{Magin2010}.
\item The equivalence between differential
  equations of non-integer order with a constant coefficient and
  differential equations of integer order with a varying coefficient,
  like equation (\ref{eq:newDNHM}), has been conjectured 
  \cite{Mainardi2018,Li2018}.
\item Additional memory terms were found necessary for the BBO
  equation  
  \cite{Odar1964,Catalano1985,Parmar2012,Lambertz2012}
  both when extended to compressible fluids  
  \cite{Parmar2011,Annamalai2017}
  and when considering axisymmetric bodies
  \cite{Lawrence1986}.
\item The Basset history force was found non-negligible in
  oscillatory motion
  \cite{Abbad2004}.
\item The term $-(\tau\omega)^2 \alpha$ in equation
  (\ref{eq:alpha2}) was identified as a 
  \emph{``summary of the history integral effect''} 
  \cite{Basano1989}.
\end{enumerate}
Taking all these reasons into consideration, we introduce in section
\S\ref{sec:fractionalmodel} a fractional differential equation of
motion for the physical pendulum. 

We shall use a time-independent-order fractional derivative but note
that some authors prefer variable-order
\cite{Coimbra2003,Pedro2005}.

A few aspects of fractional derivatives, related to their widespread
use, are mentioned next. 

\begin{figure*}
\centering
\includegraphics[width=0.9\textwidth]{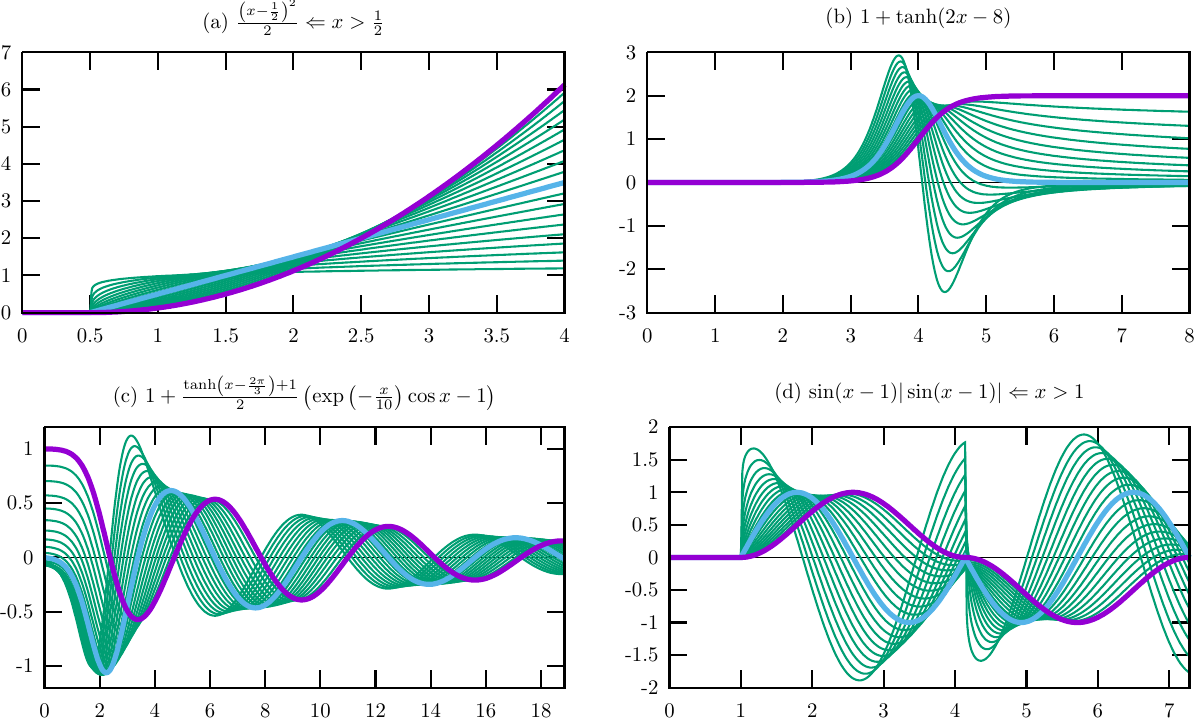} 
\caption{Plots of four functions in purple, first derivatives
  in blue and left Gr\"unwald-Letnikov fractional derivatives of order
  $\beta=0.1,\; 0.2,\ldots,1.9$ in green as calculated by
  equation (\protect \ref{eq:grunletexperder}) 
  with $h=(x\mbox{ range})/700$ and $N=500$
  for (a) and (b) or $N=100$ for (c) and (d).} 
\label{grun-let-examps}
\end{figure*}

The fractional derivative provides a functional interpolation between
closest integer derivatives. Figure \ref{grun-let-examps} contains
examples.
 
The fractional derivative of a trigonometric function is
proportional to a dephasing of the original function and the amount of
dephasing is itself proportional to the order of the fractional
derivative. 

Fractional derivatives have been studied for a long time and were
defined in many ways but became an
instrument of physicists only recently and therefore their physical
meaning has been the subject of several discussions
\cite{Bagley1983b,Moshrefi-Torbati1998,Podlubny2008},
particularly in respect of projectile motion
\cite{GomezAguilar2018,Ebaid2011}
and damped harmonic motion
\cite{Rekhviashvili2019}.

Recently, an early introduction of fractional calculus syllabus was
proposed  
\cite{Khubalkar2018}.

\subsection{Calculating fractional derivatives}

Our calculations are based on the Gr\"unwald-Letnikov definition of
fractional derivative 
\cite{Herrmann2014} 
which is considered a fractional derivative in a strict sense 
\cite{Ortigueira2015c}.
As time fractional derivatives must be calculated from past data only
(ignoring future data) for causality to be kept, a time fractional
derivative must be a left derivative. Consider a function $f(t)$, where
$t$ is time, which is known only at a discrete set
of steps $t_i=t_{i-1}+h, \; i=0,\ldots , N$ so that
$f_i=f(t=t_i)$. Consider also that the time fractional derivative of
order $\beta$ corresponds to the operator $\mathcal{D}^\beta_t$. Then
we use 
\begin{equation}
  \label{eq:grunletexperder}
  \left(\mathcal{D}^\beta_t f\right)_i = 
     \frac{1}{h^\beta}\sum_{k=0}^N W_k f_{i-k}
\end{equation}
where 
\[ W_k=\left(1-\frac{\beta+1}{k}\right) W_{k-1}, \quad W_0=1, \]
$h=\SI{1/120}{\second}$, coinciding with
the videos' frame rate, and $N=100$ (we take one hundred steps into
the past to calculate the fractional derivative). This means that
$Nh\approx T$ where $T$ is the oscillation period. 

\subsection{Solving fractional differential equations}

As for solving integer differential equations, 
fractional differential equations require adequate numerical methods.
Following some preliminary tests with the compact numerical method
proposed by Seredy{\' n}ska and Hanyga
\cite{Seredynska2000}
to solve fractional differential equations for nonlinear oscillators,
a version of the algorithm proposed by Spanos and Evangelatos 
\cite{Spanos2010}
was implemented because it generally follows the principles
of predictor-corrector methods 
\cite{Diethelm2002}.

A general pendulum equation of motion may be written as
(\S\ref{sec:fractionalmodel}) 
\begin{equation}
  \label{eq:beginningfraceqofmot}
  \alpha+\Omega_0^2\sin\theta = \alpha_\mathrm{dl}
        +\mathcal{D}^\beta_t \alpha_\mathrm{dh}
\end{equation}
where $\alpha_\mathrm{dl}$ is a known algebric function of the angular
velocity describing dissipative acceleration at low speed and 
$\alpha_\mathrm{dh}$ is a known algebric function of the angular
velocity describing dissipative acceleration at high speed but
ignoring memory effects.
Supposing known initial angle $\theta_0$, angular velocity $\omega_0$
and angular acceleration $\alpha_0$, the method starts by making a
prediction about some values of the next step using a Taylor series
expansion
\begin{align}
  \theta_{i+1}= & \theta_i + \left(\omega_i+\alpha_i\frac{h}{2}\right)h \\
  (\sin\theta)_{i+1}= & \sin\theta_i+\left(\omega_i\cos\theta_i
      +(\alpha_i\cos\theta_i-\omega_i^2\sin\theta_i)\frac{h}{2}\right)h \\
  \omega_{i+1} = & \omega_i+\alpha_i h \\
  \alpha_{\mathrm{dh}_{i+1}} = & \alpha_{\mathrm{dh}_i}
            +\left.\frac{d\alpha_\mathrm{dh}}{d\omega}\right|_i\alpha_i h
\end{align}
The history sum is
\[ \Sigma_{i+1} = \sum_{k=1}^N W_k \alpha_{\mathrm{dh}_{i+1-k}} \]
and the fractional derivative is, therefore, predicted to be
\[ \left(\mathcal{D}^\beta_t \alpha_\mathrm{dh} \right)_{i+1} =
           \frac{\alpha_{\mathrm{dh}_{i+1}}+\Sigma_{i+1}}{h^\beta}. \]
The angular acceleration results from equation
(\ref{eq:beginningfraceqofmot})
\begin{equation}
  \label{eq:predcorralpha}
  \alpha_{i+1} = -\Omega_0^2(\sin\theta)_{i+1} + \alpha_{\mathrm{dl}_{i+1}}
  + \left(\mathcal{D}^\beta_t \alpha_\mathrm{dh} \right)_{i+1}.
\end{equation}  
The method continues by correcting the predictions via the linear
acceleration approximation
\cite{Gavin2018}
\begin{align}
  \theta_{i+1} = & \theta_i+\left(\omega_i+
                   (2\alpha_i+\alpha_{i+1})\frac{h}{6}\right)h \\ 
  \omega_{i+1} = & \omega_i+(\alpha_{i+1}+\alpha_i)\frac{h}{2} 
\end{align}  
and finishes reapplying equation (\ref{eq:predcorralpha}).  

In order to avoid numerical ambiguities related with
initial conditions we assumed that the pendulum was placed or launched
at constant velocity, that is, we assumed null acceleration on the
whole unkown past
\cite{Heymans2006,Achar2007}.
Note that the Gr\"unwald-Letnikov
definition is, in this case, equivalent to the original
Riemman-Liouville definition 
\cite{Gladkina2017}. 

\section{Fractional model}\label{sec:fractionalmodel}

Given the unclear results of $C_a(l_1)$ in Figure
\ref{plotfig:fourparamsclassic} and the absence of negative damping in
results of both the classic model and the OPA model, we opted for
replacing \ainer, equation (\ref{eq:ainerdef}), with a fractional
derivative. The introduction of a fractional derivative made us try to
non-dimensionalize the equation of motion in order avoid changing units. 
This is achieved dividing by $(2\Omega_0)^2$.
A long trial-and-error process finally led us to the following
fractional model. 
\begin{equation}
  \label{eq:ourfinalmodel}
  \frac{d\mathcal{A}_\omega}{dt^\ast} +\frac{\sin\theta}{4} =  
  -\frac{G_1\sqrt[4]{|\mathcal{A}_\omega|}}{\signalop(\mathcal{A}_\omega)} 
-G_2 \FracD{\beta}\left(\mathcal{A}_\omega|\mathcal{A}_\omega|^{p_f-1}\right)
\end{equation}
where $t^\ast = 2\Omega_0 t$ is a non-dimensional time and
$\FracD{\beta}$ is the left Gr\"unwald-Letnikov fractional derivative
of order $\beta$ in $t^\ast$. Note that
\begin{equation}
  \label{eq:Aomegaoftast}
  \mathcal{A}_\omega= \frac{d\theta}{dt^\ast}=\FracD{1}\theta.
\end{equation}
Also note that $G_1\sqrt[4]{|\mathcal{A}_\omega|}$ is
similar to the friction proposed in
\cite{Threlfall1978}.

\subsection{Fitting models to data}
 
We used \texttt{amebsa} 
\cite{NumericalRecipes},
almost completely reimplemented in PASCAL language to fit the fractional
model \S\ref{sec:fractionalmodel} (with initial estimates guided by the
classic model). 

Fitting is in this case an iterative process and, therefore, depends
on an 
initial estimate of the model parameters. In order to calculate
parameter uncertainties one must generate a sample of best-fitting
parameter sets. We generated 150 best-fitting
parameter sets of the fractional model (\S\ref{sec:fractionalmodel}) for
each experimental data run. Each individual parameter set was obtained
from a single fitting procedure, each with a different initial
estimate selected randomly from a range set. A single range set was
defined 
by previous fitting trial-and-error for each experimental data
run. This trial-and-error means is-or-is-not in the basin of
attraction to the global best-fitting parameter set. Out of each sample
of 150 individual parameter sets, we identified the one corresponding
to the minimum of 
\begin{equation}
  \label{eq:leastsquarescostfunction}
  \Xi^2 = \frac{1}{N}\sum_{i=1}^N \left( 
  \left({\mathcal{A}_\theta}_\mathrm{mod}-{\mathcal{A}_\theta}_\mathrm{exp}\right)^2
  +
  \left({\mathcal{A}_\omega}_\mathrm{mod}-{\mathcal{A}_\omega}_\mathrm{exp}\right)^2
  \right)
\end{equation}
where $N$ is the number of experimental data points of the run.
This minimum is $\Xi^2_\mathrm{min}$. We then selected the parameter
sets having $\Xi^2\leq\frac{5}{4}\Xi^2_\mathrm{min}$. From this
selection we identified the maximum $\Psi_\mathrm{max}$ and the
minimum $\Psi_\mathrm{min}$ of each parameter $\Psi$. Finally, we
calculated the uncertainty as
\begin{equation}
  \label{eq:paramuncertain}
  u(\Psi)=\frac{\Psi_\mathrm{max}-\Psi_\mathrm{min}}{2}.
\end{equation}
The uncertainties thus calculated appear in figures 
\ref{plotfig:paramsfractional} and \ref{plotfig:fittexOmegaZer}.

\subsection{Fraccional model results}\label{sec:fractionalresults}

\begin{figure*}[p]
\centering
\includegraphics[width=0.85\textwidth]{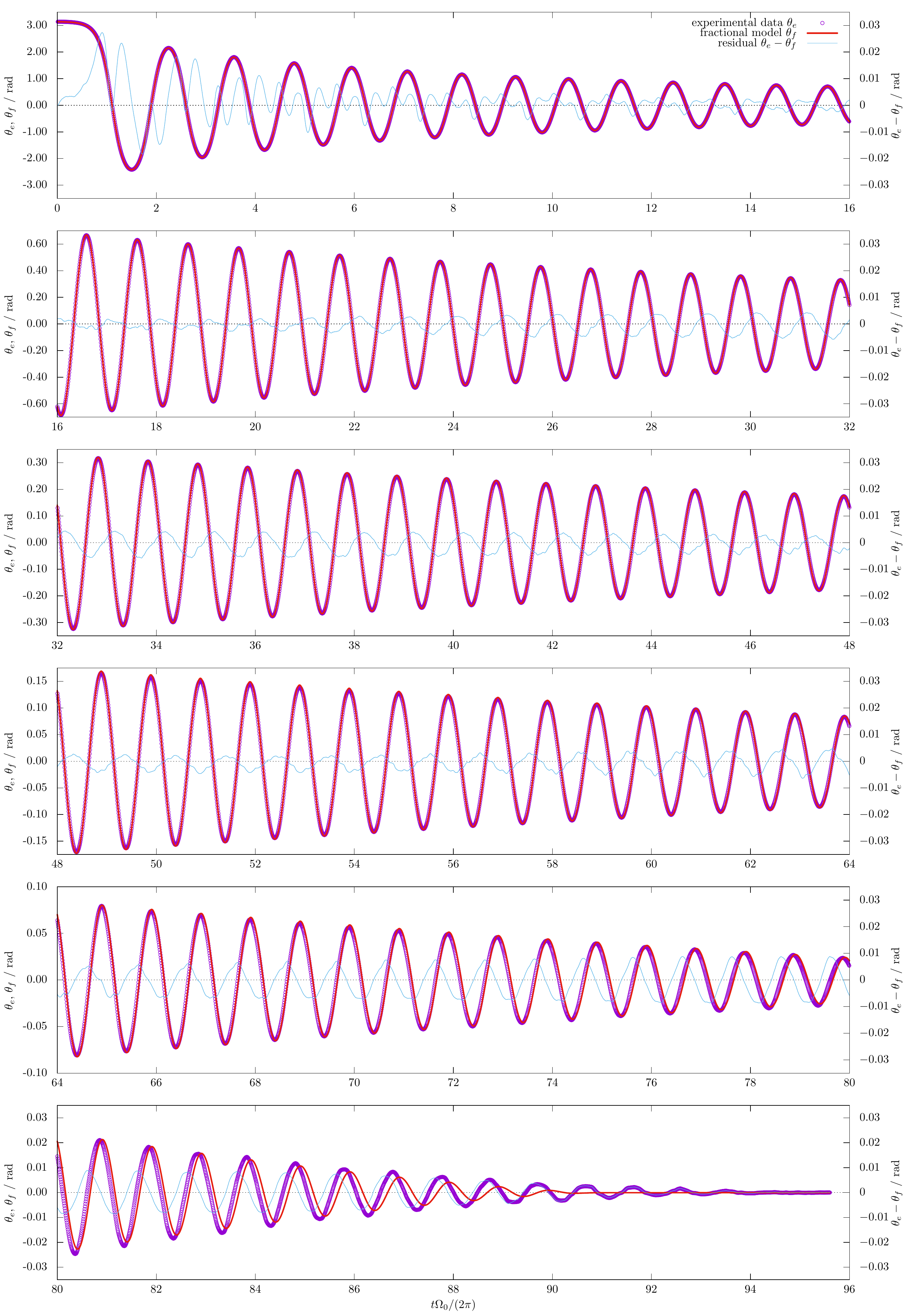}
\caption{Results for $\theta$ corresponding to run \texttt{03na1}
  fitted by the fractional model. Compare with Figure 
  \protect \ref{plotfig:spuriousosc}.}  
\label{plotfig:superfit}
\end{figure*}

\begin{figure}
\centering
\begin{tabular}{c}
  \includegraphics[width=53mm]{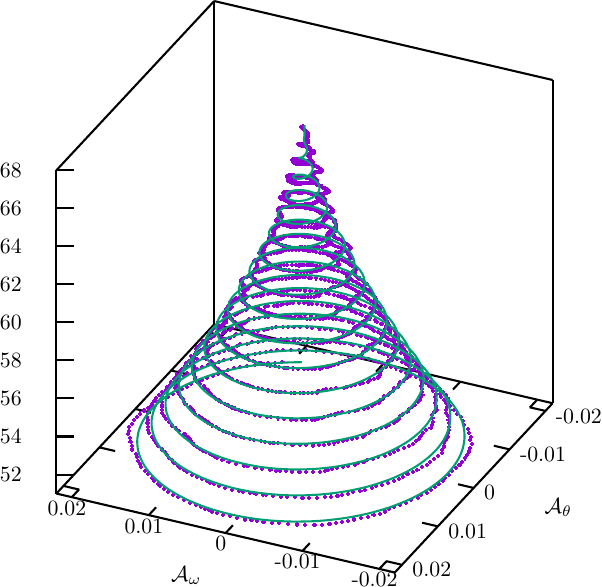} \\
  \includegraphics[width=53mm]{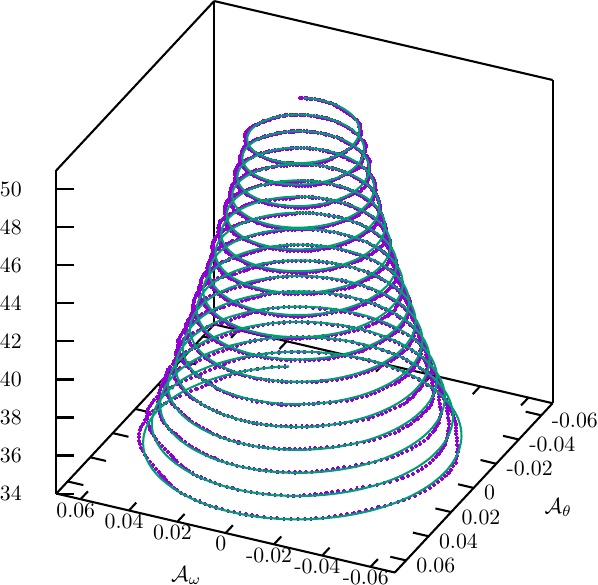} \\
  \includegraphics[width=53mm]{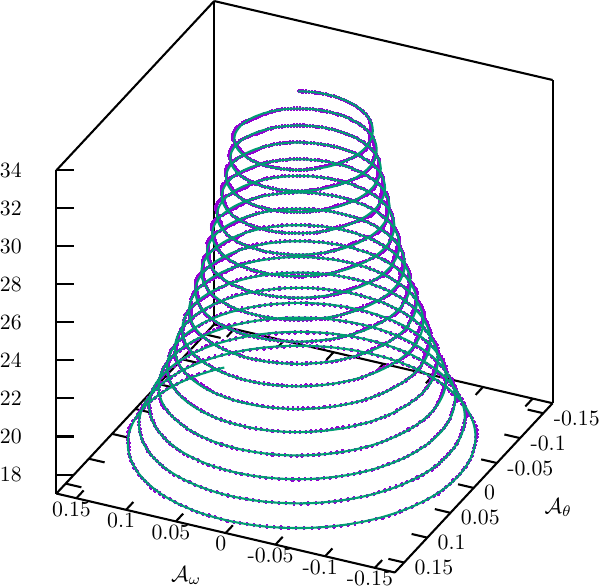} \\
  \includegraphics[width=53mm]{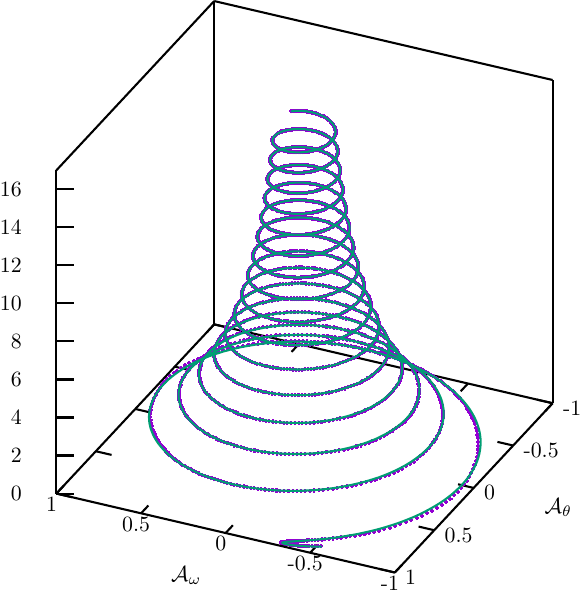}
\end{tabular}
\caption{Nyquist-like plots 
\cite{Magalas1996}
with both experimental data points and fractional model
corresponding to the fit of run \texttt{09ln1}. The vertical axis is
time in units of natural period ($t\Omega_0/(2\pi)$). } 
\label{plotfig:nyquist}
\end{figure}

Our main result is the long-time accurate fit of experimental data for
all runs. In Figure \ref{plotfig:superfit} we present the results
obtained for the best fit of run \texttt{03na1}, as an example,
showing that apparent mismatch between fit and data occurs only in the
last fifth of oscillation time.
An alternative 3D representation, corresponding to equations 
(\ref{eq:leastsquarescostfunction}), 
(\ref{eq:amplitudesobservable}) and
(\ref{eq:phasephi})
is presented in Figure
\ref{plotfig:nyquist}. This figure highlights a curious feature of
launched runs: phase velocity changes sign when
$\theta=\SI{\pi}{\radian}$.

\begin{figure}
\centering
\includegraphics[width=0.9\columnwidth]{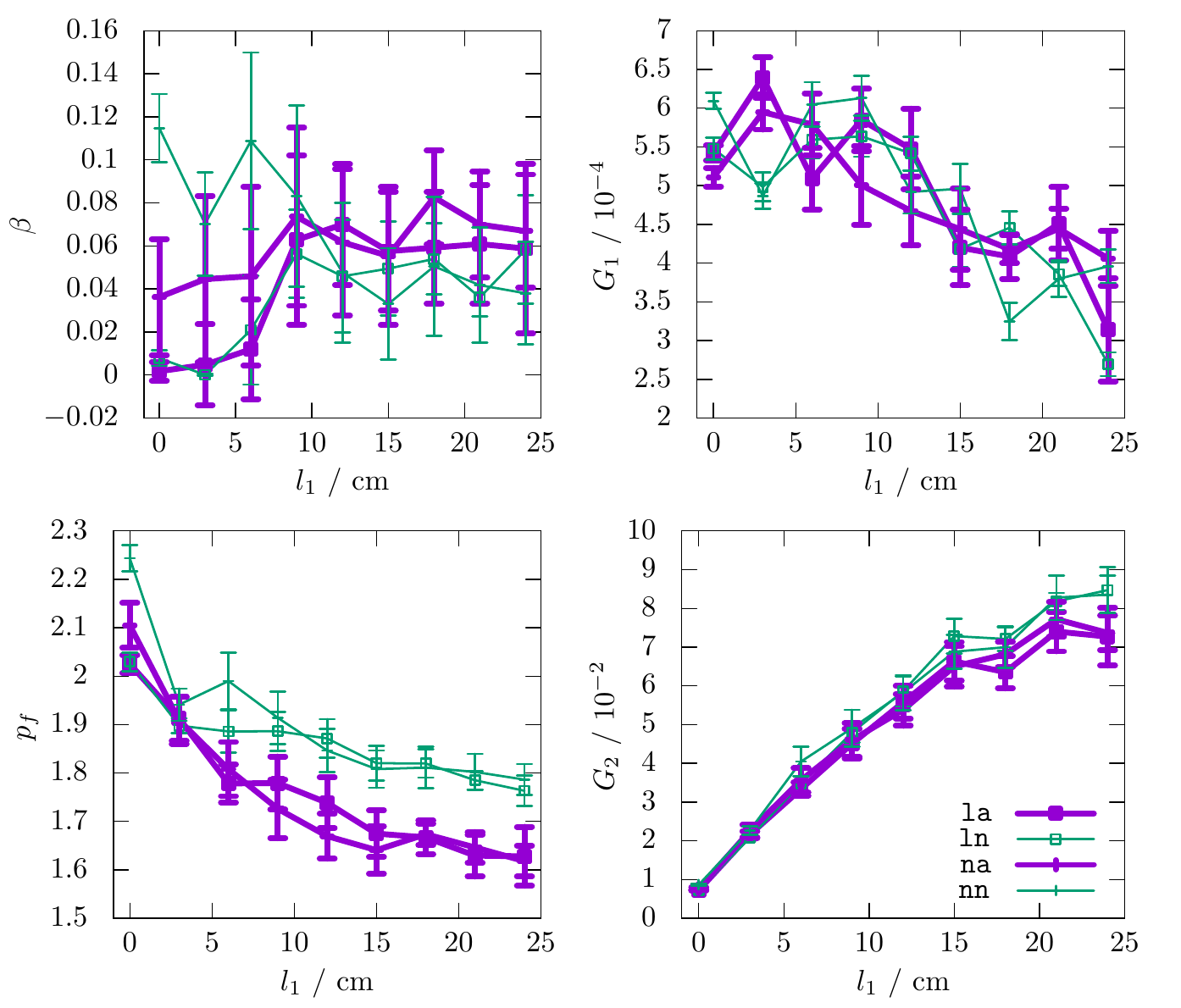}
\caption{Results from the fractional model for parameters
  $\beta$, $G_1$, $p_f$ and $G_2$. Compare with Figure 
  \protect \ref{plotfig:fourparamsclassic} noting that 
  $G_1\simeq\frac{C_0}{4\Omega_0^2}$ and 
  $G_2\simeq\frac{C_3}{2^{2-p}\Omega_0^2}$.} 
\label{plotfig:paramsfractional}
\end{figure}

The results for the fractional derivative order $\beta$ in Figure
\ref{plotfig:paramsfractional}  
allow an easier interpretation than those for $C_a$ in Figure
\ref{plotfig:fourparamsclassic}. The high uncertainties of $\beta$ do not
hide the fact that there are two behaviours: one for small tiles and
another for big tiles. Small tiles imply near zero memory effects
for launched runs and big tiles imply that the alley increases the
memory effects. Taken together these two behaviours say that memory
effects are enhanced by low speed.
As expected, the results for $G_1$, $G_2$ and $p_f$ in Figure
\ref{plotfig:paramsfractional} are 
qualitatively similar to the results for $C_0$, $C_3$ and $p$ in Figure
\ref{plotfig:fourparamsclassic}.
Note, however, that the derivative of a polynomial reduces its degree
and, therefore, it would be expected that 
\begin{equation}
  \label{eq:pubetapproxptwo}
  p_f-\beta\approx p.
\end{equation}

\begin{figure}
\centering
\includegraphics[width=0.9\columnwidth]{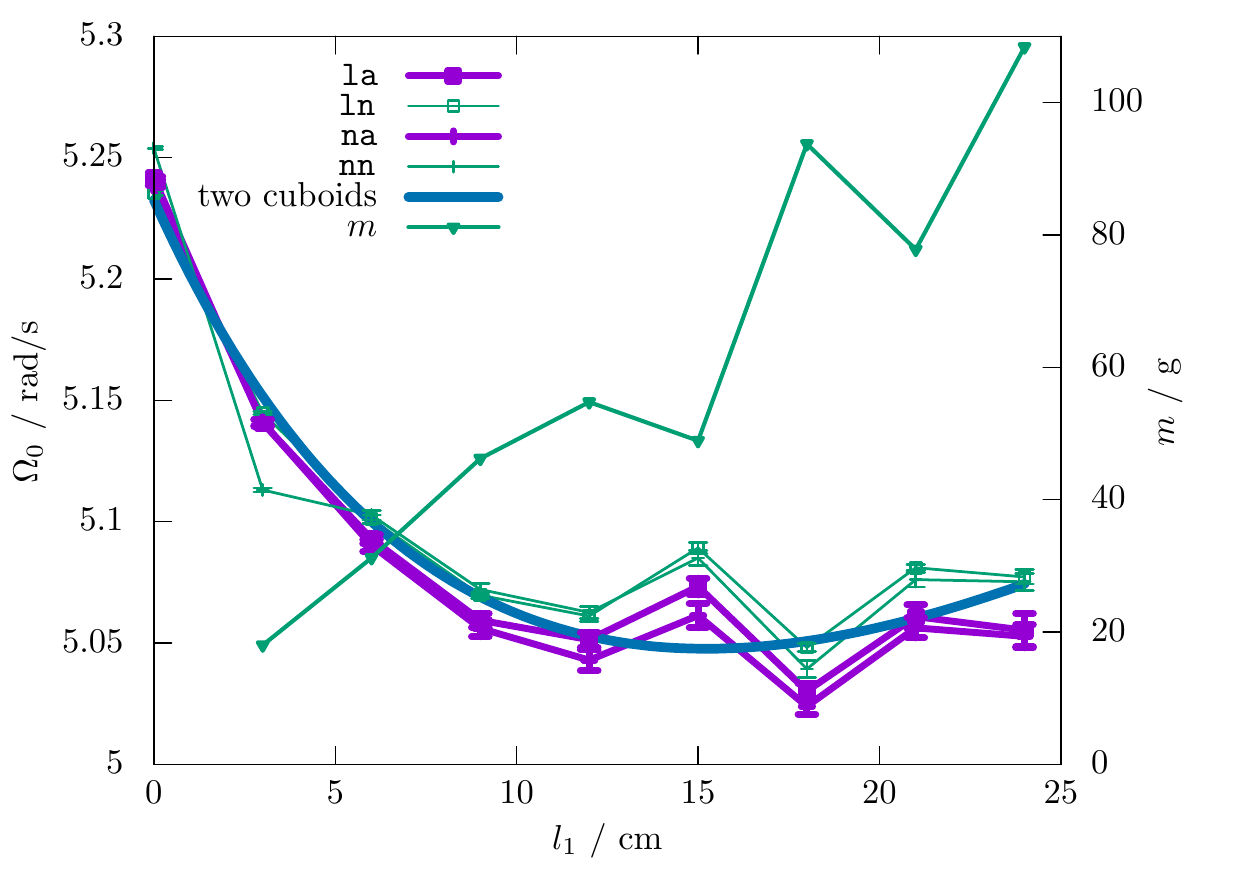}
\caption{Complete set of results from the fractional model for parameter
  $\Omega_0$. The ``two cuboids'' line is a fitting of equation 
  (\protect \ref{eq:consistencycheck}). The mass $m$ of each tile is
  also plotted.}
\label{plotfig:fittexOmegaZer}
\end{figure}

Finaly, the results for $\Omega_0$ in Figure
\ref{plotfig:fittexOmegaZer} demonstrate the high accuracy of the
fractional model both by the small relative uncertainties ($u_r<0.3\%$)
and by the good consistency with the independent theoretical model
explained below.
Figure \ref{plotfig:fittexOmegaZer} shows that $\Omega_0$ is slightly
lower for the alley runs. This is consistent with an augmented
system's inertia and reveals the effect of the coherent air flowing
along the alley lateral and bottom surfaces, specially in the small
angle oscillations.  

Assume that the pendulum and the tiles are perfect cuboids. 
Suppose that we assemble the pendulum of mass $M_0$ with one tile of
mass $m$.
We may now write
\begin{equation}
  \label{eq:omegazerobase}
  \Omega_0^2=\frac{(M_0l_\mathrm{com}+mR)g}{I_0
                                +\left(R^2+\frac{l_1^2}{12}\right)m}
\end{equation}
where $R=l_0-\frac{l_1}{2}$ and $I_0=M_0\frac{l_0^2}{3}$ (see Figure
\ref{basicscheme}). Considering 
that $M_0=\lambda_0 l_0$, $m=\lambda_1l_1$ and 
$l_\mathrm{com}=l_0/2$, where $\lambda$ is mass per unit length
assumed equal for all tiles, 
one arrives at 
\begin{equation}
  \label{eq:consistencycheck}
  \Omega_0=\sqrt{\frac{g}{l_0}}\sqrt{\frac{\frac{1}{2}X
                  +\frac{l_1}{l_0}
                  -\frac{1}{2}\left(\frac{l_1}{l_0}\right)^2}{
                  \frac{1}{3}X
                  +\frac{l_1}{l_0}
                  -\left(\frac{l_1}{l_0}\right)^2
                  +\frac{1}{3}\left(\frac{l_1}{l_0}\right)^3}}
\end{equation}
where $X=\lambda_0/\lambda_1$. We fitted this equation to the results
of $\Omega_0(l_1)$ in Figure \ref{plotfig:fittexOmegaZer} and obtained
$X\approx 1.48$ and $l_0\approx\SI{54}{\centi\metre}$. The actual
value is $\SI{52}{\centi\metre}$. The fitted and the actual $l_0$
values don't match exactly because the actual pendulum isn't a cuboid
and the tiles don't have equal mass per unit length. However, the
results for $\Omega_0$ in Figure \ref{plotfig:fittexOmegaZer} show
that when $\lambda_1=m/l_1$ increases, $\Omega_0$ consistently decreases.

\begin{figure*}
\centering
\includegraphics[width=0.9\textwidth]{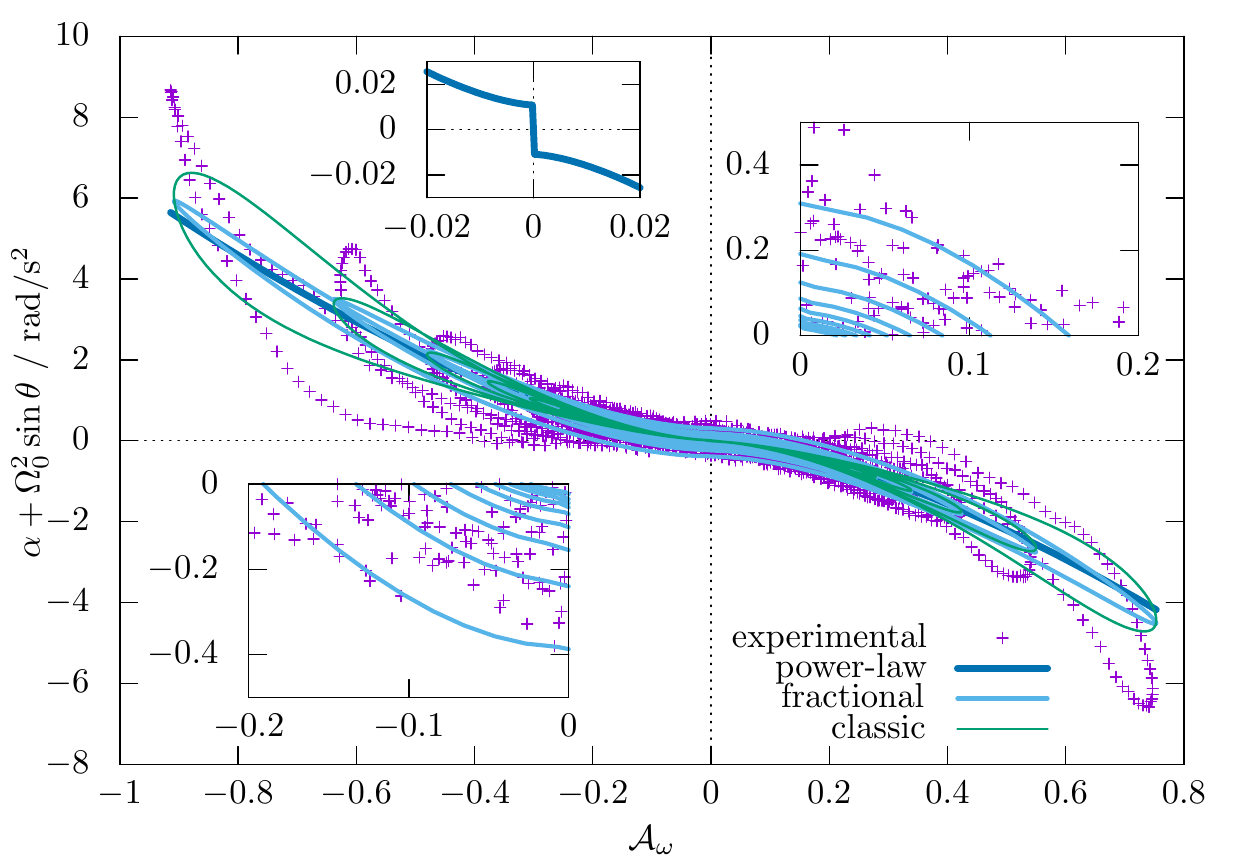}
\caption{Sum of the dissipative acceleration and the inertial
  acceleration for the initial ten seconds of 
  the \texttt{24la1} run. Two insets show where dissipative
  acceleration and velocity have the same sign (negative
  damping). The fitted classic model does not describe negative
  damping (see top-left insert). Nevertheless,  
  this plot shows that both the classic model and the fractional model
  reproduce a hysteretic dissipation. Equation 
  (\protect \ref{eq:classicmodel}) with $C_a=0$ is
  indicated as ``power-law''.}   
\label{plotfig:dissipativehysteresis}
\end{figure*}

The set of results from the fractional model for run \texttt{24la1}
allows the revisualization of Figure \ref{fig:lobolusdecaras} as in
Figure \ref{plotfig:dissipativehysteresis}.

\section{Conclusion}\label{sec:umdiscuss}

The physical pendulum is traditionally treated as a rigid body and a
one-body equation is used to describe a rotating center of mass. Only
gravitational and strictly dissipative torques are traditionally
considered but the experimental data collected in this experiment
clearly shows, for large amplitude oscillations, a hysteretic
behaviour. Also, the experimental results confirm that a fixed
structure unconnected to the pendulum may modify its motion. This
means that the pendulum damping depends on the flow of air surrounding
the pendulum. The pendulum equation of motion must, therefore, account
for a multitude of air flow consequences, such as non-constant moment
of inertia, recoverable air kinetic energy, automatic parametric
pumping, and the compound hysteretic behaviour. The special
consequence of non-null air flow when the strictly rigid pendulum
stops, imposes the consideration of memory effects that may be
adequately modelled by fractional derivatives.

Some attention was payed to the extraction of another classic rigid
body concept, the natural angular frequency of the linear harmonic 
oscillator.  
On the one hand, estimates of the natural angular frequency can be
obtained directly from both angle and acceleration experimental data
but, on the other hand, our model-based estimates establish an
interdependency between the use of an unconnected structure and
the natural angular frequency. 

Given the obtained results it is possible to expect not only that the 
proposed fractional model will be able to fit general pendular
phenomena including forcing, amplitude resonance, and rotatory
regime but also that the fractional derivative of a power-law can be
used as a generic model of air drag. 

The physical pendulum is, after all, 
the \emph{exponent} of classics. 

\begin{acknowledgments}
  We gratefully acknowledge enlightning discussions with Manuel
  Ortigueira, Arnaldo Baptista, Gr{\' e}goire Bonfait, Carlos Dias, Carlos
  Cruz and Mendanha Dias. We also acknowledge the american english
  language revision by Jeffrey Keefer. 
\end{acknowledgments}

\begin{thebibliography}{170}%
\makeatletter
\providecommand \@ifxundefined [1]{%
 \@ifx{#1\undefined}
}%
\providecommand \@ifnum [1]{%
 \ifnum #1\expandafter \@firstoftwo
 \else \expandafter \@secondoftwo
 \fi
}%
\providecommand \@ifx [1]{%
 \ifx #1\expandafter \@firstoftwo
 \else \expandafter \@secondoftwo
 \fi
}%
\providecommand \natexlab [1]{#1}%
\providecommand \enquote  [1]{``#1''}%
\providecommand \bibnamefont  [1]{#1}%
\providecommand \bibfnamefont [1]{#1}%
\providecommand \citenamefont [1]{#1}%
\providecommand \href@noop [0]{\@secondoftwo}%
\providecommand \href [0]{\begingroup \@sanitize@url \@href}%
\providecommand \@href[1]{\@@startlink{#1}\@@href}%
\providecommand \@@href[1]{\endgroup#1\@@endlink}%
\providecommand \@sanitize@url [0]{\catcode `\\12\catcode `\$12\catcode
  `\&12\catcode `\#12\catcode `\^12\catcode `\_12\catcode `\%12\relax}%
\providecommand \@@startlink[1]{}%
\providecommand \@@endlink[0]{}%
\providecommand \url  [0]{\begingroup\@sanitize@url \@url }%
\providecommand \@url [1]{\endgroup\@href {#1}{\urlprefix }}%
\providecommand \urlprefix  [0]{URL }%
\providecommand \Eprint [0]{\href }%
\providecommand \doibase [0]{http://dx.doi.org/}%
\providecommand \selectlanguage [0]{\@gobble}%
\providecommand \bibinfo  [0]{\@secondoftwo}%
\providecommand \bibfield  [0]{\@secondoftwo}%
\providecommand \translation [1]{[#1]}%
\providecommand \BibitemOpen [0]{}%
\providecommand \bibitemStop [0]{}%
\providecommand \bibitemNoStop [0]{.\EOS\space}%
\providecommand \EOS [0]{\spacefactor3000\relax}%
\providecommand \BibitemShut  [1]{\csname bibitem#1\endcsname}%
\let\auto@bib@innerbib\@empty
\bibitem [{\citenamefont {Erlichson}(1999)}]{Erlichson1999}%
  \BibitemOpen
  \bibfield  {author} {\bibinfo {author} {\bibfnamefont {H.}~\bibnamefont
  {Erlichson}},\ }\href {\doibase 10.1119/1.880380} {\bibfield  {journal}
  {\bibinfo  {journal} {The Physics Teacher}\ }\textbf {\bibinfo {volume}
  {37}},\ \bibinfo {pages} {478} (\bibinfo {year} {1999})}\BibitemShut
  {NoStop}%
\bibitem [{\citenamefont {Huygens}(1966)}]{Huygens1673}%
  \BibitemOpen
  \bibfield  {author} {\bibinfo {author} {\bibfnamefont {C.}~\bibnamefont
  {Huygens}},\ }\href@noop {} {\emph {\bibinfo {title} {Horologium
  oscillatorium: sive de motu pendulorum ad horologia aptato demonstrationes
  geometric\ae}}}\ (\bibinfo  {publisher} {F. Muguet},\ \bibinfo {year}
  {1966})\BibitemShut {NoStop}%
\bibitem [{\citenamefont {Kater}\ and\ \citenamefont
  {Young}(1818)}]{Kater1818}%
  \BibitemOpen
  \bibfield  {author} {\bibinfo {author} {\bibfnamefont {H.}~\bibnamefont
  {Kater}}\ and\ \bibinfo {author} {\bibfnamefont {T.}~\bibnamefont {Young}},\
  }\href {\doibase 10.1098/rstl.1818.0006} {\bibfield  {journal} {\bibinfo
  {journal} {Philosophical Transactions of the Royal Society of London}\
  }\textbf {\bibinfo {volume} {108}},\ \bibinfo {pages} {33} (\bibinfo {year}
  {1818})}\BibitemShut {NoStop}%
\bibitem [{\citenamefont {Jackson}(1961)}]{Jackson1961}%
  \BibitemOpen
  \bibfield  {author} {\bibinfo {author} {\bibfnamefont {J.~E.}\ \bibnamefont
  {Jackson}},\ }\href {\doibase 10.1111/j.1365-246X.1961.tb06826.x} {\bibfield
  {journal} {\bibinfo  {journal} {Geophysical Journal International}\ }\textbf
  {\bibinfo {volume} {4}},\ \bibinfo {pages} {375} (\bibinfo {year} {1961})},\
  \Eprint
  {http://arxiv.org/abs/http://oup.prod.sis.lan/gji/article-pdf/4/Supplement\_1/375/1833118/4-Supplement\_1-375.pdf}
  {http://oup.prod.sis.lan/gji/article-pdf/4/Supplement\_1/375/1833118/4-Supplement\_1-375.pdf}
  \BibitemShut {NoStop}%
\bibitem [{\citenamefont {Marson}\ and\ \citenamefont
  {Riccardi}(2012)}]{Marson2012}%
  \BibitemOpen
  \bibfield  {author} {\bibinfo {author} {\bibfnamefont {I.}~\bibnamefont
  {Marson}}\ and\ \bibinfo {author} {\bibfnamefont {U.}~\bibnamefont
  {Riccardi}},\ }\href {\doibase 10.1155/2012/687813} {\bibfield  {journal}
  {\bibinfo  {journal} {International Journal of Geophysics}\ } (\bibinfo
  {year} {2012}),\ 10.1155/2012/687813}\BibitemShut {NoStop}%
\bibitem [{\citenamefont {Blair}\ \emph {et~al.}(1993)\citenamefont {Blair},
  \citenamefont {Ju},\ and\ \citenamefont {Notcutt}}]{Blair1993}%
  \BibitemOpen
  \bibfield  {author} {\bibinfo {author} {\bibfnamefont {D.~G.}\ \bibnamefont
  {Blair}}, \bibinfo {author} {\bibfnamefont {L.}~\bibnamefont {Ju}}, \ and\
  \bibinfo {author} {\bibfnamefont {M.}~\bibnamefont {Notcutt}},\ }\href
  {\doibase 10.1063/1.1143974} {\bibfield  {journal} {\bibinfo  {journal}
  {Review of Scientific Instruments}\ }\textbf {\bibinfo {volume} {64}},\
  \bibinfo {pages} {1899} (\bibinfo {year} {1993})}\BibitemShut {NoStop}%
\bibitem [{\citenamefont {Mitrofanov}\ and\ \citenamefont
  {Styazhkina}(1999)}]{Mitrofanov1999}%
  \BibitemOpen
  \bibfield  {author} {\bibinfo {author} {\bibfnamefont {V.}~\bibnamefont
  {Mitrofanov}}\ and\ \bibinfo {author} {\bibfnamefont {N.}~\bibnamefont
  {Styazhkina}},\ }\href {\doibase
  https://doi.org/10.1016/S0375-9601(99)00244-3} {\bibfield  {journal}
  {\bibinfo  {journal} {Physics Letters A}\ }\textbf {\bibinfo {volume}
  {256}},\ \bibinfo {pages} {351 } (\bibinfo {year} {1999})}\BibitemShut
  {NoStop}%
\bibitem [{\citenamefont {Uchiyama}\ \emph {et~al.}(2000)\citenamefont
  {Uchiyama}, \citenamefont {Tomaru}, \citenamefont {Tatsumi}, \citenamefont
  {Miyoki}, \citenamefont {Ohashi}, \citenamefont {Kuroda}, \citenamefont
  {Suzuki}, \citenamefont {Yamamoto},\ and\ \citenamefont
  {Shintomi}}]{Uchiyama2000}%
  \BibitemOpen
  \bibfield  {author} {\bibinfo {author} {\bibfnamefont {T.}~\bibnamefont
  {Uchiyama}}, \bibinfo {author} {\bibfnamefont {T.}~\bibnamefont {Tomaru}},
  \bibinfo {author} {\bibfnamefont {D.}~\bibnamefont {Tatsumi}}, \bibinfo
  {author} {\bibfnamefont {S.}~\bibnamefont {Miyoki}}, \bibinfo {author}
  {\bibfnamefont {M.}~\bibnamefont {Ohashi}}, \bibinfo {author} {\bibfnamefont
  {K.}~\bibnamefont {Kuroda}}, \bibinfo {author} {\bibfnamefont
  {T.}~\bibnamefont {Suzuki}}, \bibinfo {author} {\bibfnamefont
  {A.}~\bibnamefont {Yamamoto}}, \ and\ \bibinfo {author} {\bibfnamefont
  {T.}~\bibnamefont {Shintomi}},\ }\href {\doibase
  https://doi.org/10.1016/S0375-9601(00)00514-4} {\bibfield  {journal}
  {\bibinfo  {journal} {Physics Letters A}\ }\textbf {\bibinfo {volume}
  {273}},\ \bibinfo {pages} {310 } (\bibinfo {year} {2000})}\BibitemShut
  {NoStop}%
\bibitem [{\citenamefont {Cagnoli}\ \emph {et~al.}(2000)\citenamefont
  {Cagnoli}, \citenamefont {Gammaitoni}, \citenamefont {Hough}, \citenamefont
  {Kovalik}, \citenamefont {McIntosh}, \citenamefont {Punturo},\ and\
  \citenamefont {Rowan}}]{Cagnoli2000}%
  \BibitemOpen
  \bibfield  {author} {\bibinfo {author} {\bibfnamefont {G.}~\bibnamefont
  {Cagnoli}}, \bibinfo {author} {\bibfnamefont {L.}~\bibnamefont {Gammaitoni}},
  \bibinfo {author} {\bibfnamefont {J.}~\bibnamefont {Hough}}, \bibinfo
  {author} {\bibfnamefont {J.}~\bibnamefont {Kovalik}}, \bibinfo {author}
  {\bibfnamefont {S.}~\bibnamefont {McIntosh}}, \bibinfo {author}
  {\bibfnamefont {M.}~\bibnamefont {Punturo}}, \ and\ \bibinfo {author}
  {\bibfnamefont {S.}~\bibnamefont {Rowan}},\ }\href {\doibase
  10.1103/PhysRevLett.85.2442} {\bibfield  {journal} {\bibinfo  {journal}
  {Phys. Rev. Lett.}\ }\textbf {\bibinfo {volume} {85}},\ \bibinfo {pages}
  {2442} (\bibinfo {year} {2000})}\BibitemShut {NoStop}%
\bibitem [{\citenamefont {Lima}(2008)}]{Lima2008}%
  \BibitemOpen
  \bibfield  {author} {\bibinfo {author} {\bibfnamefont {F.~M.~S.}\
  \bibnamefont {Lima}},\ }\href {\doibase 10.1088/0143-0807/29/5/021}
  {\bibfield  {journal} {\bibinfo  {journal} {European Journal of Physics}\
  }\textbf {\bibinfo {volume} {29}},\ \bibinfo {pages} {1091} (\bibinfo {year}
  {2008})}\BibitemShut {NoStop}%
\bibitem [{\citenamefont {Sanmartín}(1984)}]{Sanmartin1984}%
  \BibitemOpen
  \bibfield  {author} {\bibinfo {author} {\bibfnamefont {J.~R.}\ \bibnamefont
  {Sanmartín}},\ }\href {\doibase 10.1119/1.13798} {\bibfield  {journal}
  {\bibinfo  {journal} {American Journal of Physics}\ }\textbf {\bibinfo
  {volume} {52}},\ \bibinfo {pages} {937} (\bibinfo {year} {1984})}\BibitemShut
  {NoStop}%
\bibitem [{\citenamefont {Wirkus}\ \emph {et~al.}(1998)\citenamefont {Wirkus},
  \citenamefont {Rand},\ and\ \citenamefont {Ruina}}]{Wirkus1998}%
  \BibitemOpen
  \bibfield  {author} {\bibinfo {author} {\bibfnamefont {S.}~\bibnamefont
  {Wirkus}}, \bibinfo {author} {\bibfnamefont {R.}~\bibnamefont {Rand}}, \ and\
  \bibinfo {author} {\bibfnamefont {A.}~\bibnamefont {Ruina}},\ }\href
  {http://www.jstor.org/stable/2687680} {\bibfield  {journal} {\bibinfo
  {journal} {The College Mathematics Journal}\ }\textbf {\bibinfo {volume}
  {29}},\ \bibinfo {pages} {266} (\bibinfo {year} {1998})}\BibitemShut
  {NoStop}%
\bibitem [{\citenamefont {Stilling}\ and\ \citenamefont
  {Szyszkowski}(2002)}]{Stilling2002}%
  \BibitemOpen
  \bibfield  {author} {\bibinfo {author} {\bibfnamefont {D.~S.}\ \bibnamefont
  {Stilling}}\ and\ \bibinfo {author} {\bibfnamefont {W.}~\bibnamefont
  {Szyszkowski}},\ }\href {\doibase
  https://doi.org/10.1016/S0020-7462(00)00099-8} {\bibfield  {journal}
  {\bibinfo  {journal} {International Journal of Non-Linear Mechanics}\
  }\textbf {\bibinfo {volume} {37}},\ \bibinfo {pages} {89 } (\bibinfo {year}
  {2002})}\BibitemShut {NoStop}%
\bibitem [{\citenamefont {Post}\ \emph {et~al.}(2007)\citenamefont {Post},
  \citenamefont {de~Groot}, \citenamefont {Daffertshofer},\ and\ \citenamefont
  {Beek}}]{Post2007}%
  \BibitemOpen
  \bibfield  {author} {\bibinfo {author} {\bibfnamefont {A.~A.}\ \bibnamefont
  {Post}}, \bibinfo {author} {\bibfnamefont {G.}~\bibnamefont {de~Groot}},
  \bibinfo {author} {\bibfnamefont {A.}~\bibnamefont {Daffertshofer}}, \ and\
  \bibinfo {author} {\bibfnamefont {P.~J.}\ \bibnamefont {Beek}},\ }\href
  {https://journals.humankinetics.com/view/journals/mcj/11/2/article-p136.xml}
  {\bibfield  {journal} {\bibinfo  {journal} {Motor Control}\ }\textbf
  {\bibinfo {volume} {11}},\ \bibinfo {pages} {136 } (\bibinfo {year}
  {2007})}\BibitemShut {NoStop}%
\bibitem [{\citenamefont {Greenwood}\ \emph {et~al.}(2008)\citenamefont
  {Greenwood}, \citenamefont {Johnson}, \citenamefont {Choi},\ and\
  \citenamefont {Chaudhury}}]{Greenwood2009}%
  \BibitemOpen
  \bibfield  {author} {\bibinfo {author} {\bibfnamefont {J.~A.}\ \bibnamefont
  {Greenwood}}, \bibinfo {author} {\bibfnamefont {K.~L.}\ \bibnamefont
  {Johnson}}, \bibinfo {author} {\bibfnamefont {S.-H.}\ \bibnamefont {Choi}}, \
  and\ \bibinfo {author} {\bibfnamefont {M.~K.}\ \bibnamefont {Chaudhury}},\
  }\href {\doibase 10.1088/0022-3727/42/3/035301} {\bibfield  {journal}
  {\bibinfo  {journal} {Journal of Physics D: Applied Physics}\ }\textbf
  {\bibinfo {volume} {42}},\ \bibinfo {pages} {035301} (\bibinfo {year}
  {2008})}\BibitemShut {NoStop}%
\bibitem [{\citenamefont {Obligado}\ \emph {et~al.}(2013)\citenamefont
  {Obligado}, \citenamefont {Puy},\ and\ \citenamefont
  {Bourgoin}}]{Obligado2013}%
  \BibitemOpen
  \bibfield  {author} {\bibinfo {author} {\bibfnamefont {M.}~\bibnamefont
  {Obligado}}, \bibinfo {author} {\bibfnamefont {M.}~\bibnamefont {Puy}}, \
  and\ \bibinfo {author} {\bibfnamefont {M.}~\bibnamefont {Bourgoin}},\ }\href
  {\doibase 10.1017/jfm.2013.312} {\bibfield  {journal} {\bibinfo  {journal}
  {Journal of Fluid Mechanics}\ }\textbf {\bibinfo {volume} {728}},\ \bibinfo
  {pages} {R2} (\bibinfo {year} {2013})}\BibitemShut {NoStop}%
\bibitem [{\citenamefont {Cuerno}\ \emph {et~al.}(1992)\citenamefont {Cuerno},
  \citenamefont {Rañada},\ and\ \citenamefont {Ruiz‐Lorenzo}}]{Cuerno1992}%
  \BibitemOpen
  \bibfield  {author} {\bibinfo {author} {\bibfnamefont {R.}~\bibnamefont
  {Cuerno}}, \bibinfo {author} {\bibfnamefont {A.~F.}\ \bibnamefont {Rañada}},
  \ and\ \bibinfo {author} {\bibfnamefont {J.~J.}\ \bibnamefont
  {Ruiz‐Lorenzo}},\ }\href {\doibase 10.1119/1.17047} {\bibfield  {journal}
  {\bibinfo  {journal} {American Journal of Physics}\ }\textbf {\bibinfo
  {volume} {60}},\ \bibinfo {pages} {73} (\bibinfo {year} {1992})}\BibitemShut
  {NoStop}%
\bibitem [{\citenamefont {DeSerio}(2003)}]{Deserio2003}%
  \BibitemOpen
  \bibfield  {author} {\bibinfo {author} {\bibfnamefont {R.}~\bibnamefont
  {DeSerio}},\ }\href {\doibase 10.1119/1.1526465} {\bibfield  {journal}
  {\bibinfo  {journal} {American Journal of Physics}\ }\textbf {\bibinfo
  {volume} {71}},\ \bibinfo {pages} {250} (\bibinfo {year} {2003})}\BibitemShut
  {NoStop}%
\bibitem [{\citenamefont {Ya.~Azbel}\ and\ \citenamefont
  {Bak}(1984)}]{Azbel1984}%
  \BibitemOpen
  \bibfield  {author} {\bibinfo {author} {\bibfnamefont {M.}~\bibnamefont
  {Ya.~Azbel}}\ and\ \bibinfo {author} {\bibfnamefont {P.}~\bibnamefont
  {Bak}},\ }\href {\doibase 10.1103/PhysRevB.30.3722} {\bibfield  {journal}
  {\bibinfo  {journal} {Phys. Rev. B}\ }\textbf {\bibinfo {volume} {30}},\
  \bibinfo {pages} {3722} (\bibinfo {year} {1984})}\BibitemShut {NoStop}%
\bibitem [{\citenamefont {Romeiras}\ and\ \citenamefont
  {Ott}(1987)}]{Romeiras1987}%
  \BibitemOpen
  \bibfield  {author} {\bibinfo {author} {\bibfnamefont {F.~J.}\ \bibnamefont
  {Romeiras}}\ and\ \bibinfo {author} {\bibfnamefont {E.}~\bibnamefont {Ott}},\
  }\href {\doibase 10.1103/PhysRevA.35.4404} {\bibfield  {journal} {\bibinfo
  {journal} {Phys. Rev. A}\ }\textbf {\bibinfo {volume} {35}},\ \bibinfo
  {pages} {4404} (\bibinfo {year} {1987})}\BibitemShut {NoStop}%
\bibitem [{\citenamefont {Milburn}\ and\ \citenamefont
  {Walls}(1983)}]{Milburn1983}%
  \BibitemOpen
  \bibfield  {author} {\bibinfo {author} {\bibfnamefont {G.~J.}\ \bibnamefont
  {Milburn}}\ and\ \bibinfo {author} {\bibfnamefont {D.~F.}\ \bibnamefont
  {Walls}},\ }\href {\doibase 10.1119/1.13324} {\bibfield  {journal} {\bibinfo
  {journal} {American Journal of Physics}\ }\textbf {\bibinfo {volume} {51}},\
  \bibinfo {pages} {1134} (\bibinfo {year} {1983})}\BibitemShut {NoStop}%
\bibitem [{\citenamefont {Yurke}(1986)}]{Yurke1986}%
  \BibitemOpen
  \bibfield  {author} {\bibinfo {author} {\bibfnamefont {B.}~\bibnamefont
  {Yurke}},\ }\href {\doibase 10.1119/1.14730} {\bibfield  {journal} {\bibinfo
  {journal} {American Journal of Physics}\ }\textbf {\bibinfo {volume} {54}},\
  \bibinfo {pages} {1133} (\bibinfo {year} {1986})}\BibitemShut {NoStop}%
\bibitem [{\citenamefont {Doubochinski}\ and\ \citenamefont
  {Tennenbaum}(2007)}]{Doubochinski2007}%
  \BibitemOpen
  \bibfield  {author} {\bibinfo {author} {\bibfnamefont {D.}~\bibnamefont
  {Doubochinski}}\ and\ \bibinfo {author} {\bibfnamefont {J.}~\bibnamefont
  {Tennenbaum}},\ }\href@noop {} {\enquote {\bibinfo {title} {The macroscopic
  quantum effect in nonlinear oscillating systems: a possible bridge between
  classical and quantum physics},}\ } (\bibinfo {year} {2007}),\ \Eprint
  {http://arxiv.org/abs/0711.4892} {arXiv:0711.4892 [physics.gen-ph]}
  \BibitemShut {NoStop}%
\bibitem [{\citenamefont {Shumaev}\ and\ \citenamefont
  {Maizelis}(2017)}]{Shumaev2017}%
  \BibitemOpen
  \bibfield  {author} {\bibinfo {author} {\bibfnamefont {A.~I.}\ \bibnamefont
  {Shumaev}}\ and\ \bibinfo {author} {\bibfnamefont {Z.~A.}\ \bibnamefont
  {Maizelis}},\ }\href {\doibase 10.1063/1.4979800} {\bibfield  {journal}
  {\bibinfo  {journal} {Journal of Applied Physics}\ }\textbf {\bibinfo
  {volume} {121}},\ \bibinfo {pages} {154902} (\bibinfo {year}
  {2017})}\BibitemShut {NoStop}%
\bibitem [{\citenamefont {Pigneur}\ and\ \citenamefont
  {Schmiedmayer}(2018)}]{Pigneur2018}%
  \BibitemOpen
  \bibfield  {author} {\bibinfo {author} {\bibfnamefont {M.}~\bibnamefont
  {Pigneur}}\ and\ \bibinfo {author} {\bibfnamefont {J.}~\bibnamefont
  {Schmiedmayer}},\ }\href {\doibase 10.1103/PhysRevA.98.063632} {\bibfield
  {journal} {\bibinfo  {journal} {Phys. Rev. A}\ }\textbf {\bibinfo {volume}
  {98}},\ \bibinfo {pages} {063632} (\bibinfo {year} {2018})}\BibitemShut
  {NoStop}%
\bibitem [{\citenamefont {Naudts}(2005)}]{Naudts2005}%
  \BibitemOpen
  \bibfield  {author} {\bibinfo {author} {\bibfnamefont {J.}~\bibnamefont
  {Naudts}},\ }\href {\doibase 10.1209/epl/i2004-10413-1} {\bibfield  {journal}
  {\bibinfo  {journal} {Europhysics Letters ({EPL})}\ }\textbf {\bibinfo
  {volume} {69}},\ \bibinfo {pages} {719} (\bibinfo {year} {2005})}\BibitemShut
  {NoStop}%
\bibitem [{\citenamefont {Baeten}\ and\ \citenamefont
  {Naudts}(2011)}]{Baeten2011}%
  \BibitemOpen
  \bibfield  {author} {\bibinfo {author} {\bibfnamefont {M.}~\bibnamefont
  {Baeten}}\ and\ \bibinfo {author} {\bibfnamefont {J.}~\bibnamefont
  {Naudts}},\ }\href {https://doi.org/10.3390/e13061186} {\bibfield  {journal}
  {\bibinfo  {journal} {Entropy}\ }\textbf {\bibinfo {volume} {13}},\ \bibinfo
  {pages} {1186} (\bibinfo {year} {2011})}\BibitemShut {NoStop}%
\bibitem [{\citenamefont {Lukichev}(2014)}]{Lukichev2014}%
  \BibitemOpen
  \bibfield  {author} {\bibinfo {author} {\bibfnamefont {A.}~\bibnamefont
  {Lukichev}},\ }\href {\doibase
  https://doi.org/10.1016/j.chemphys.2013.10.021} {\bibfield  {journal}
  {\bibinfo  {journal} {Chemical Physics}\ }\textbf {\bibinfo {volume} {428}},\
  \bibinfo {pages} {29 } (\bibinfo {year} {2014})}\BibitemShut {NoStop}%
\bibitem [{\citenamefont {Lukichev}(2015)}]{Lukichev2015}%
  \BibitemOpen
  \bibfield  {author} {\bibinfo {author} {\bibfnamefont {A.}~\bibnamefont
  {Lukichev}},\ }\href {\doibase
  https://doi.org/10.1016/j.jnoncrysol.2015.04.012} {\bibfield  {journal}
  {\bibinfo  {journal} {Journal of Non-Crystalline Solids}\ }\textbf {\bibinfo
  {volume} {420}},\ \bibinfo {pages} {43 } (\bibinfo {year}
  {2015})}\BibitemShut {NoStop}%
\bibitem [{\citenamefont {Lukichev}(2016)}]{Lukichev2016}%
  \BibitemOpen
  \bibfield  {author} {\bibinfo {author} {\bibfnamefont {A.}~\bibnamefont
  {Lukichev}},\ }\href {\doibase
  https://doi.org/10.1016/j.jnoncrysol.2016.02.027} {\bibfield  {journal}
  {\bibinfo  {journal} {Journal of Non-Crystalline Solids}\ }\textbf {\bibinfo
  {volume} {442}},\ \bibinfo {pages} {17 } (\bibinfo {year}
  {2016})}\BibitemShut {NoStop}%
\bibitem [{\citenamefont {Lukichev}(2019)}]{Lukichev2019}%
  \BibitemOpen
  \bibfield  {author} {\bibinfo {author} {\bibfnamefont {A.}~\bibnamefont
  {Lukichev}},\ }\href {\doibase
  https://doi.org/10.1016/j.physleta.2019.06.029} {\bibfield  {journal}
  {\bibinfo  {journal} {Physics Letters A}\ }\textbf {\bibinfo {volume}
  {383}},\ \bibinfo {pages} {2983 } (\bibinfo {year} {2019})}\BibitemShut
  {NoStop}%
\bibitem [{\citenamefont {Rabinovich}()}]{Rabinovich2009}%
  \BibitemOpen
  \bibfield  {author} {\bibinfo {author} {\bibfnamefont {A.~B.}\ \bibnamefont
  {Rabinovich}},\ }\enquote {\bibinfo {title} {Seiches and harbor
  oscillations},}\ in\ \href {\doibase 10.1142/9789812819307\_0009} {\emph
  {\bibinfo {booktitle} {Handbook of Coastal and Ocean Engineering}}},\ pp.\
  \bibinfo {pages} {193--236}\BibitemShut {NoStop}%
\bibitem [{\citenamefont {{Fee}}(2002)}]{Fee2002}%
  \BibitemOpen
  \bibfield  {author} {\bibinfo {author} {\bibfnamefont {J.~W.}\ \bibnamefont
  {{Fee}}},\ }in\ \href {\doibase 10.1109/NEBC.2002.999451} {\emph {\bibinfo
  {booktitle} {Proceedings of the IEEE 28th Annual Northeast Bioengineering
  Conference (IEEE Cat. No.02CH37342)}}}\ (\bibinfo {year} {2002})\ pp.\
  \bibinfo {pages} {33--34}\BibitemShut {NoStop}%
\bibitem [{\citenamefont {Peters}(2003)}]{Peters2003}%
  \BibitemOpen
  \bibfield  {author} {\bibinfo {author} {\bibfnamefont {R.~D.}\ \bibnamefont
  {Peters}},\ }\href@noop {} {\enquote {\bibinfo {title} {Nonlinear damping of
  the 'linear' pendulum},}\ } (\bibinfo {year} {2003}),\ \Eprint
  {http://arxiv.org/abs/physics/0306081} {arXiv:physics/0306081
  [physics.class-ph]} \BibitemShut {NoStop}%
\bibitem [{\citenamefont {Fernandes}\ \emph {et~al.}(2017)\citenamefont
  {Fernandes}, \citenamefont {Sebasti{\~{a}}o}, \citenamefont
  {Gon{\c{c}}alves},\ and\ \citenamefont {Ferraz}}]{Fernandes2017}%
  \BibitemOpen
  \bibfield  {author} {\bibinfo {author} {\bibfnamefont {J.~C.}\ \bibnamefont
  {Fernandes}}, \bibinfo {author} {\bibfnamefont {P.~J.}\ \bibnamefont
  {Sebasti{\~{a}}o}}, \bibinfo {author} {\bibfnamefont {L.~N.}\ \bibnamefont
  {Gon{\c{c}}alves}}, \ and\ \bibinfo {author} {\bibfnamefont {A.}~\bibnamefont
  {Ferraz}},\ }\href {\doibase 10.1088/1361-6404/aa6c52} {\bibfield  {journal}
  {\bibinfo  {journal} {European Journal of Physics}\ }\textbf {\bibinfo
  {volume} {38}},\ \bibinfo {pages} {045004} (\bibinfo {year}
  {2017})}\BibitemShut {NoStop}%
\bibitem [{\citenamefont {Kavyanpoor}\ and\ \citenamefont
  {Shokrollahi}(2019)}]{Kavyanpoor2017}%
  \BibitemOpen
  \bibfield  {author} {\bibinfo {author} {\bibfnamefont {M.}~\bibnamefont
  {Kavyanpoor}}\ and\ \bibinfo {author} {\bibfnamefont {S.}~\bibnamefont
  {Shokrollahi}},\ }\href {\doibase
  https://doi.org/10.1016/j.jksus.2017.03.006} {\bibfield  {journal} {\bibinfo
  {journal} {Journal of King Saud University - Science}\ }\textbf {\bibinfo
  {volume} {31}},\ \bibinfo {pages} {14 } (\bibinfo {year} {2019})}\BibitemShut
  {NoStop}%
\bibitem [{\citenamefont {Amabili}(2019)}]{Amabili2018}%
  \BibitemOpen
  \bibfield  {author} {\bibinfo {author} {\bibfnamefont {M.}~\bibnamefont
  {Amabili}},\ }\href@noop {} {\bibfield  {journal} {\bibinfo  {journal}
  {Nonlinear Dynamics}\ }\textbf {\bibinfo {volume} {97}},\ \bibinfo {pages}
  {1785} (\bibinfo {year} {2019})}\BibitemShut {NoStop}%
\bibitem [{\citenamefont {Fulcher}\ and\ \citenamefont
  {Davis}(1976)}]{Fulcher1976}%
  \BibitemOpen
  \bibfield  {author} {\bibinfo {author} {\bibfnamefont {L.~P.}\ \bibnamefont
  {Fulcher}}\ and\ \bibinfo {author} {\bibfnamefont {B.~F.}\ \bibnamefont
  {Davis}},\ }\href {\doibase 10.1119/1.10137} {\bibfield  {journal} {\bibinfo
  {journal} {American Journal of Physics}\ }\textbf {\bibinfo {volume} {44}},\
  \bibinfo {pages} {51} (\bibinfo {year} {1976})}\BibitemShut {NoStop}%
\bibitem [{\citenamefont {Hall}\ and\ \citenamefont {Shea}(1977)}]{Hall1977}%
  \BibitemOpen
  \bibfield  {author} {\bibinfo {author} {\bibfnamefont {D.~E.}\ \bibnamefont
  {Hall}}\ and\ \bibinfo {author} {\bibfnamefont {M.~J.}\ \bibnamefont
  {Shea}},\ }\href {\doibase 10.1119/1.10621} {\bibfield  {journal} {\bibinfo
  {journal} {American Journal of Physics}\ }\textbf {\bibinfo {volume} {45}},\
  \bibinfo {pages} {355} (\bibinfo {year} {1977})}\BibitemShut {NoStop}%
\bibitem [{\citenamefont {Zilio}(1982)}]{Zilio1982}%
  \BibitemOpen
  \bibfield  {author} {\bibinfo {author} {\bibfnamefont {S.~C.}\ \bibnamefont
  {Zilio}},\ }\href {\doibase 10.1119/1.12832} {\bibfield  {journal} {\bibinfo
  {journal} {American Journal of Physics}\ }\textbf {\bibinfo {volume} {50}},\
  \bibinfo {pages} {450} (\bibinfo {year} {1982})}\BibitemShut {NoStop}%
\bibitem [{\citenamefont {Gil}\ \emph {et~al.}(2008)\citenamefont {Gil},
  \citenamefont {Legarreta},\ and\ \citenamefont {Di~Gregorio}}]{Gil2008}%
  \BibitemOpen
  \bibfield  {author} {\bibinfo {author} {\bibfnamefont {S.}~\bibnamefont
  {Gil}}, \bibinfo {author} {\bibfnamefont {A.~E.}\ \bibnamefont {Legarreta}},
  \ and\ \bibinfo {author} {\bibfnamefont {D.~E.}\ \bibnamefont
  {Di~Gregorio}},\ }\href {\doibase 10.1119/1.2908184} {\bibfield  {journal}
  {\bibinfo  {journal} {American Journal of Physics}\ }\textbf {\bibinfo
  {volume} {76}},\ \bibinfo {pages} {843} (\bibinfo {year} {2008})}\BibitemShut
  {NoStop}%
\bibitem [{\citenamefont {Squire}(1986)}]{Squire1986}%
  \BibitemOpen
  \bibfield  {author} {\bibinfo {author} {\bibfnamefont {P.~T.}\ \bibnamefont
  {Squire}},\ }\href {\doibase 10.1119/1.14838} {\bibfield  {journal} {\bibinfo
   {journal} {American Journal of Physics}\ }\textbf {\bibinfo {volume} {54}},\
  \bibinfo {pages} {984} (\bibinfo {year} {1986})}\BibitemShut {NoStop}%
\bibitem [{\citenamefont {Basano}\ and\ \citenamefont
  {Ottonello}(1991)}]{Basano1991}%
  \BibitemOpen
  \bibfield  {author} {\bibinfo {author} {\bibfnamefont {L.}~\bibnamefont
  {Basano}}\ and\ \bibinfo {author} {\bibfnamefont {P.}~\bibnamefont
  {Ottonello}},\ }\href {\doibase 10.1119/1.16639} {\bibfield  {journal}
  {\bibinfo  {journal} {American Journal of Physics}\ }\textbf {\bibinfo
  {volume} {59}},\ \bibinfo {pages} {1018} (\bibinfo {year}
  {1991})}\BibitemShut {NoStop}%
\bibitem [{\citenamefont {Zonetti}\ \emph {et~al.}(1999)\citenamefont
  {Zonetti}, \citenamefont {Camargo}, \citenamefont {Sartori}, \citenamefont
  {de~Sousa},\ and\ \citenamefont {Nunes}}]{Zonetti1999}%
  \BibitemOpen
  \bibfield  {author} {\bibinfo {author} {\bibfnamefont {L.~F.~C.}\
  \bibnamefont {Zonetti}}, \bibinfo {author} {\bibfnamefont {A.~S.~S.}\
  \bibnamefont {Camargo}}, \bibinfo {author} {\bibfnamefont {J.}~\bibnamefont
  {Sartori}}, \bibinfo {author} {\bibfnamefont {D.~F.}\ \bibnamefont
  {de~Sousa}}, \ and\ \bibinfo {author} {\bibfnamefont {L.~A.~O.}\ \bibnamefont
  {Nunes}},\ }\href {\doibase 10.1088/0143-0807/20/2/004} {\bibfield  {journal}
  {\bibinfo  {journal} {European Journal of Physics}\ }\textbf {\bibinfo
  {volume} {20}},\ \bibinfo {pages} {85} (\bibinfo {year} {1999})}\BibitemShut
  {NoStop}%
\bibitem [{\citenamefont {jun Wang}\ \emph {et~al.}(2002)\citenamefont {jun
  Wang}, \citenamefont {Schmitt},\ and\ \citenamefont {Payne}}]{Wang2002}%
  \BibitemOpen
  \bibfield  {author} {\bibinfo {author} {\bibfnamefont {X.}~\bibnamefont {jun
  Wang}}, \bibinfo {author} {\bibfnamefont {C.}~\bibnamefont {Schmitt}}, \ and\
  \bibinfo {author} {\bibfnamefont {M.}~\bibnamefont {Payne}},\ }\href
  {\doibase 10.1088/0143-0807/23/2/309} {\bibfield  {journal} {\bibinfo
  {journal} {European Journal of Physics}\ }\textbf {\bibinfo {volume} {23}},\
  \bibinfo {pages} {155} (\bibinfo {year} {2002})}\BibitemShut {NoStop}%
\bibitem [{\citenamefont {Bacon}\ and\ \citenamefont
  {Nguyen}(2005)}]{Bacon2005}%
  \BibitemOpen
  \bibfield  {author} {\bibinfo {author} {\bibfnamefont {M.~E.}\ \bibnamefont
  {Bacon}}\ and\ \bibinfo {author} {\bibfnamefont {D.~D.}\ \bibnamefont
  {Nguyen}},\ }\href {\doibase 10.1088/0143-0807/26/4/011} {\bibfield
  {journal} {\bibinfo  {journal} {European Journal of Physics}\ }\textbf
  {\bibinfo {volume} {26}},\ \bibinfo {pages} {651} (\bibinfo {year}
  {2005})}\BibitemShut {NoStop}%
\bibitem [{\citenamefont {Simbach}\ and\ \citenamefont
  {Priest}(2005)}]{Simbach2005}%
  \BibitemOpen
  \bibfield  {author} {\bibinfo {author} {\bibfnamefont {J.~C.}\ \bibnamefont
  {Simbach}}\ and\ \bibinfo {author} {\bibfnamefont {J.}~\bibnamefont
  {Priest}},\ }\href {\doibase 10.1119/1.1858488} {\bibfield  {journal}
  {\bibinfo  {journal} {American Journal of Physics}\ }\textbf {\bibinfo
  {volume} {73}},\ \bibinfo {pages} {1079} (\bibinfo {year}
  {2005})}\BibitemShut {NoStop}%
\bibitem [{\citenamefont {Smith}(2012)}]{Smith2012}%
  \BibitemOpen
  \bibfield  {author} {\bibinfo {author} {\bibfnamefont {B.~R.}\ \bibnamefont
  {Smith}},\ }\href {\doibase 10.1119/1.4729440} {\bibfield  {journal}
  {\bibinfo  {journal} {American Journal of Physics}\ }\textbf {\bibinfo
  {volume} {80}},\ \bibinfo {pages} {816} (\bibinfo {year} {2012})}\BibitemShut
  {NoStop}%
\bibitem [{\citenamefont {Mungan}\ and\ \citenamefont
  {Lipscombe}(2013)}]{Mungan2013}%
  \BibitemOpen
  \bibfield  {author} {\bibinfo {author} {\bibfnamefont {C.~E.}\ \bibnamefont
  {Mungan}}\ and\ \bibinfo {author} {\bibfnamefont {T.~C.}\ \bibnamefont
  {Lipscombe}},\ }\href {\doibase 10.1088/0143-0807/34/5/1243} {\bibfield
  {journal} {\bibinfo  {journal} {European Journal of Physics}\ }\textbf
  {\bibinfo {volume} {34}},\ \bibinfo {pages} {1243} (\bibinfo {year}
  {2013})}\BibitemShut {NoStop}%
\bibitem [{\citenamefont {Mathai}\ \emph {et~al.}(2019)\citenamefont {Mathai},
  \citenamefont {Loeffen}, \citenamefont {Chan},\ and\ \citenamefont
  {Wildeman}}]{Mathai2019}%
  \BibitemOpen
  \bibfield  {author} {\bibinfo {author} {\bibfnamefont {V.}~\bibnamefont
  {Mathai}}, \bibinfo {author} {\bibfnamefont {L.~A. W.~M.}\ \bibnamefont
  {Loeffen}}, \bibinfo {author} {\bibfnamefont {T.~T.~K.}\ \bibnamefont
  {Chan}}, \ and\ \bibinfo {author} {\bibfnamefont {S.}~\bibnamefont
  {Wildeman}},\ }\href {\doibase 10.1017/jfm.2018.867} {\bibfield  {journal}
  {\bibinfo  {journal} {Journal of Fluid Mechanics}\ }\textbf {\bibinfo
  {volume} {862}},\ \bibinfo {pages} {348–363} (\bibinfo {year}
  {2019})}\BibitemShut {NoStop}%
\bibitem [{\citenamefont {Eng}\ \emph {et~al.}(2008)\citenamefont {Eng},
  \citenamefont {Lau}, \citenamefont {Low},\ and\ \citenamefont
  {Seet}}]{Eng2008}%
  \BibitemOpen
  \bibfield  {author} {\bibinfo {author} {\bibfnamefont {Y.}~\bibnamefont
  {Eng}}, \bibinfo {author} {\bibfnamefont {W.}~\bibnamefont {Lau}}, \bibinfo
  {author} {\bibfnamefont {E.}~\bibnamefont {Low}}, \ and\ \bibinfo {author}
  {\bibfnamefont {G.}~\bibnamefont {Seet}},\ }in\ \href@noop {} {\emph
  {\bibinfo {booktitle} {International MultiConference of Engineers and
  Computer Scientists}}},\ Vol.~\bibinfo {volume} {2}\ (\bibinfo {year}
  {2008})\ pp.\ \bibinfo {pages} {423--430}\BibitemShut {NoStop}%
\bibitem [{\citenamefont {Bolster}\ \emph {et~al.}(2010)\citenamefont
  {Bolster}, \citenamefont {Hershberger},\ and\ \citenamefont
  {Donnelly}}]{Bolster2010}%
  \BibitemOpen
  \bibfield  {author} {\bibinfo {author} {\bibfnamefont {D.}~\bibnamefont
  {Bolster}}, \bibinfo {author} {\bibfnamefont {R.~E.}\ \bibnamefont
  {Hershberger}}, \ and\ \bibinfo {author} {\bibfnamefont {R.~J.}\ \bibnamefont
  {Donnelly}},\ }\href {\doibase 10.1103/PhysRevE.81.046317} {\bibfield
  {journal} {\bibinfo  {journal} {Phys. Rev. E}\ }\textbf {\bibinfo {volume}
  {81}},\ \bibinfo {pages} {046317} (\bibinfo {year} {2010})}\BibitemShut
  {NoStop}%
\bibitem [{\citenamefont {Whineray}(1991)}]{Whineray1991}%
  \BibitemOpen
  \bibfield  {author} {\bibinfo {author} {\bibfnamefont {S.}~\bibnamefont
  {Whineray}},\ }\href {\doibase 10.1088/0143-0807/12/2/008} {\bibfield
  {journal} {\bibinfo  {journal} {European Journal of Physics}\ }\textbf
  {\bibinfo {volume} {12}},\ \bibinfo {pages} {90} (\bibinfo {year}
  {1991})}\BibitemShut {NoStop}%
\bibitem [{\citenamefont {Hinrichsen}\ and\ \citenamefont
  {Larnder}(2018)}]{Hinrichsen2018}%
  \BibitemOpen
  \bibfield  {author} {\bibinfo {author} {\bibfnamefont {P.~F.}\ \bibnamefont
  {Hinrichsen}}\ and\ \bibinfo {author} {\bibfnamefont {C.~I.}\ \bibnamefont
  {Larnder}},\ }\href {\doibase 10.1119/1.5034345} {\bibfield  {journal}
  {\bibinfo  {journal} {American Journal of Physics}\ }\textbf {\bibinfo
  {volume} {86}},\ \bibinfo {pages} {577} (\bibinfo {year} {2018})}\BibitemShut
  {NoStop}%
\bibitem [{\citenamefont {Lorenceau}\ \emph {et~al.}(2002)\citenamefont
  {Lorenceau}, \citenamefont {Qu\'{e}r\'{e}}, \citenamefont {Ollitrault},\ and\
  \citenamefont {Clanet}}]{Lorenceau2002}%
  \BibitemOpen
  \bibfield  {author} {\bibinfo {author} {\bibfnamefont {E.}~\bibnamefont
  {Lorenceau}}, \bibinfo {author} {\bibfnamefont {D.}~\bibnamefont
  {Qu\'{e}r\'{e}}}, \bibinfo {author} {\bibfnamefont {J.-Y.}\ \bibnamefont
  {Ollitrault}}, \ and\ \bibinfo {author} {\bibfnamefont {C.}~\bibnamefont
  {Clanet}},\ }\href {\doibase 10.1063/1.1476670} {\bibfield  {journal}
  {\bibinfo  {journal} {Physics of Fluids}\ }\textbf {\bibinfo {volume} {14}},\
  \bibinfo {pages} {1985} (\bibinfo {year} {2002})}\BibitemShut {NoStop}%
\bibitem [{\citenamefont {Smith}\ and\ \citenamefont
  {Matlis}(2019)}]{Smith2019}%
  \BibitemOpen
  \bibfield  {author} {\bibinfo {author} {\bibfnamefont {R.~P.}\ \bibnamefont
  {Smith}}\ and\ \bibinfo {author} {\bibfnamefont {E.~H.}\ \bibnamefont
  {Matlis}},\ }\href {\doibase 10.1119/1.5095945} {\bibfield  {journal}
  {\bibinfo  {journal} {American Journal of Physics}\ }\textbf {\bibinfo
  {volume} {87}},\ \bibinfo {pages} {433} (\bibinfo {year} {2019})}\BibitemShut
  {NoStop}%
\bibitem [{\citenamefont {Basano}\ and\ \citenamefont
  {Ottonello}(1989)}]{Basano1989}%
  \BibitemOpen
  \bibfield  {author} {\bibinfo {author} {\bibfnamefont {L.}~\bibnamefont
  {Basano}}\ and\ \bibinfo {author} {\bibfnamefont {P.}~\bibnamefont
  {Ottonello}},\ }\href {\doibase 10.1119/1.15784} {\bibfield  {journal}
  {\bibinfo  {journal} {American Journal of Physics}\ }\textbf {\bibinfo
  {volume} {57}},\ \bibinfo {pages} {999} (\bibinfo {year} {1989})}\BibitemShut
  {NoStop}%
\bibitem [{\citenamefont {Alho}\ \emph {et~al.}(2018)\citenamefont {Alho},
  \citenamefont {Silva}, \citenamefont {Teodoro},\ and\ \citenamefont
  {Bonfait}}]{Alho2019}%
  \BibitemOpen
  \bibfield  {author} {\bibinfo {author} {\bibfnamefont {J.}~\bibnamefont
  {Alho}}, \bibinfo {author} {\bibfnamefont {H.}~\bibnamefont {Silva}},
  \bibinfo {author} {\bibfnamefont {V.}~\bibnamefont {Teodoro}}, \ and\
  \bibinfo {author} {\bibfnamefont {G.}~\bibnamefont {Bonfait}},\ }\href
  {\doibase 10.1088/1361-6552/aaea9d} {\bibfield  {journal} {\bibinfo
  {journal} {Physics Education}\ }\textbf {\bibinfo {volume} {54}},\ \bibinfo
  {pages} {015015} (\bibinfo {year} {2018})}\BibitemShut {NoStop}%
\bibitem [{\citenamefont {Larnder}(2019)}]{Larnder2019}%
  \BibitemOpen
  \bibfield  {author} {\bibinfo {author} {\bibfnamefont {C.~I.}\ \bibnamefont
  {Larnder}},\ }\href {\doibase 10.1119/1.5123455} {\bibfield  {journal}
  {\bibinfo  {journal} {American Journal of Physics}\ }\textbf {\bibinfo
  {volume} {87}},\ \bibinfo {pages} {784} (\bibinfo {year} {2019})}\BibitemShut
  {NoStop}%
\bibitem [{\citenamefont {Mann}\ and\ \citenamefont
  {Khasawneh}(2009)}]{Mann2009}%
  \BibitemOpen
  \bibfield  {author} {\bibinfo {author} {\bibfnamefont {B.}~\bibnamefont
  {Mann}}\ and\ \bibinfo {author} {\bibfnamefont {F.}~\bibnamefont
  {Khasawneh}},\ }\href {\doibase https://doi.org/10.1016/j.jsv.2008.09.036}
  {\bibfield  {journal} {\bibinfo  {journal} {Journal of Sound and Vibration}\
  }\textbf {\bibinfo {volume} {321}},\ \bibinfo {pages} {65 } (\bibinfo {year}
  {2009})}\BibitemShut {NoStop}%
\bibitem [{\citenamefont {Jakšić}(2011)}]{Jaksic2011}%
  \BibitemOpen
  \bibfield  {author} {\bibinfo {author} {\bibfnamefont {N.}~\bibnamefont
  {Jakšić}},\ }\href {\doibase https://doi.org/10.1016/j.jsv.2011.07.029}
  {\bibfield  {journal} {\bibinfo  {journal} {Journal of Sound and Vibration}\
  }\textbf {\bibinfo {volume} {330}},\ \bibinfo {pages} {5878 } (\bibinfo
  {year} {2011})}\BibitemShut {NoStop}%
\bibitem [{\citenamefont {Gon{\c{c}}alves}(2004)}]{featpost}%
  \BibitemOpen
  \bibfield  {author} {\bibinfo {author} {\bibfnamefont {L.~N.}\ \bibnamefont
  {Gon{\c{c}}alves}},\ }in\ \href@noop {} {\emph {\bibinfo {booktitle}
  {Internatinoal Conference on TEX, XML, and Digital Typography}}}\ (\bibinfo
  {organization} {Springer},\ \bibinfo {year} {2004})\ pp.\ \bibinfo {pages}
  {112--124}\BibitemShut {NoStop}%
\bibitem [{Note1()}]{Note1}%
  \BibitemOpen
  \bibinfo {note} {The color pink was chosen because it produced the best
  contrast.}\BibitemShut {Stop}%
\bibitem [{\citenamefont {Brown}()}]{Tracker}%
  \BibitemOpen
  \bibfield  {author} {\bibinfo {author} {\bibfnamefont {D.}~\bibnamefont
  {Brown}},\ }\href@noop {} {\enquote {\bibinfo {title} {Tracker video analysis
  and modeling tool},}\ }\bibinfo {howpublished}
  {\url{http://physlets.org/tracker/}}\BibitemShut {NoStop}%
\bibitem [{\citenamefont {Press}\ and\ \citenamefont
  {Teukolsky}(1990)}]{Press1990}%
  \BibitemOpen
  \bibfield  {author} {\bibinfo {author} {\bibfnamefont {W.~H.}\ \bibnamefont
  {Press}}\ and\ \bibinfo {author} {\bibfnamefont {S.~A.}\ \bibnamefont
  {Teukolsky}},\ }\href@noop {} {\bibfield  {journal} {\bibinfo  {journal}
  {Computers in Physics}\ }\textbf {\bibinfo {volume} {4}},\ \bibinfo {pages}
  {669} (\bibinfo {year} {1990})}\BibitemShut {NoStop}%
\bibitem [{\citenamefont {Ermanyuk}(2000)}]{Ermanyuk2000}%
  \BibitemOpen
  \bibfield  {author} {\bibinfo {author} {\bibfnamefont {E.}~\bibnamefont
  {Ermanyuk}},\ }\href@noop {} {\bibfield  {journal} {\bibinfo  {journal}
  {Experiments in fluids}\ }\textbf {\bibinfo {volume} {28}},\ \bibinfo {pages}
  {152} (\bibinfo {year} {2000})}\BibitemShut {NoStop}%
\bibitem [{\citenamefont {Neill}\ \emph {et~al.}(2007)\citenamefont {Neill},
  \citenamefont {Livelybrooks},\ and\ \citenamefont {Donnelly}}]{Neill2007}%
  \BibitemOpen
  \bibfield  {author} {\bibinfo {author} {\bibfnamefont {D.}~\bibnamefont
  {Neill}}, \bibinfo {author} {\bibfnamefont {D.}~\bibnamefont {Livelybrooks}},
  \ and\ \bibinfo {author} {\bibfnamefont {R.~J.}\ \bibnamefont {Donnelly}},\
  }\href {\doibase 10.1119/1.2360993} {\bibfield  {journal} {\bibinfo
  {journal} {American Journal of Physics}\ }\textbf {\bibinfo {volume} {75}},\
  \bibinfo {pages} {226} (\bibinfo {year} {2007})}\BibitemShut {NoStop}%
\bibitem [{\citenamefont {Messer}\ and\ \citenamefont
  {Pantaleone}(2010)}]{Messer2010}%
  \BibitemOpen
  \bibfield  {author} {\bibinfo {author} {\bibfnamefont {J.}~\bibnamefont
  {Messer}}\ and\ \bibinfo {author} {\bibfnamefont {J.}~\bibnamefont
  {Pantaleone}},\ }\href {\doibase 10.1119/1.3274365} {\bibfield  {journal}
  {\bibinfo  {journal} {The Physics Teacher}\ }\textbf {\bibinfo {volume}
  {48}},\ \bibinfo {pages} {52} (\bibinfo {year} {2010})}\BibitemShut {NoStop}%
\bibitem [{\citenamefont {Pantaleone}\ and\ \citenamefont
  {Messer}(2011)}]{Pantaleone2011}%
  \BibitemOpen
  \bibfield  {author} {\bibinfo {author} {\bibfnamefont {J.}~\bibnamefont
  {Pantaleone}}\ and\ \bibinfo {author} {\bibfnamefont {J.}~\bibnamefont
  {Messer}},\ }\href {\doibase 10.1119/1.3644334} {\bibfield  {journal}
  {\bibinfo  {journal} {American Journal of Physics}\ }\textbf {\bibinfo
  {volume} {79}},\ \bibinfo {pages} {1202} (\bibinfo {year}
  {2011})}\BibitemShut {NoStop}%
\bibitem [{\citenamefont {{Raza}}\ \emph {et~al.}(2012)\citenamefont {{Raza}},
  \citenamefont {{Mehmood}}, \citenamefont {{Rafiuddin}},\ and\ \citenamefont
  {{Rafique}}}]{Raza2012}%
  \BibitemOpen
  \bibfield  {author} {\bibinfo {author} {\bibfnamefont {N.}~\bibnamefont
  {{Raza}}}, \bibinfo {author} {\bibfnamefont {I.}~\bibnamefont {{Mehmood}}},
  \bibinfo {author} {\bibfnamefont {H.}~\bibnamefont {{Rafiuddin}}}, \ and\
  \bibinfo {author} {\bibfnamefont {M.}~\bibnamefont {{Rafique}}},\ }in\ \href
  {\doibase 10.1109/IBCAST.2012.6177564} {\emph {\bibinfo {booktitle}
  {Proceedings of 2012 9th International Bhurban Conference on Applied Sciences
  Technology (IBCAST)}}}\ (\bibinfo {year} {2012})\ pp.\ \bibinfo {pages}
  {270--273}\BibitemShut {NoStop}%
\bibitem [{\citenamefont {Konstantinidis}(2013)}]{Konstantinidis2013}%
  \BibitemOpen
  \bibfield  {author} {\bibinfo {author} {\bibfnamefont {E.}~\bibnamefont
  {Konstantinidis}},\ }\href {\doibase 10.1098/rspa.2013.0135} {\bibfield
  {journal} {\bibinfo  {journal} {Proceedings of the Royal Society A:
  Mathematical, Physical and Engineering Sciences}\ }\textbf {\bibinfo {volume}
  {469}},\ \bibinfo {pages} {20130135} (\bibinfo {year} {2013})}\BibitemShut
  {NoStop}%
\bibitem [{\citenamefont {Peters}(2004)}]{Peters2004}%
  \BibitemOpen
  \bibfield  {author} {\bibinfo {author} {\bibfnamefont {R.}~\bibnamefont
  {Peters}},\ }\href@noop {} {\bibfield  {journal} {\bibinfo  {journal}
  {Contemporary Physics}\ }\textbf {\bibinfo {volume} {45}},\ \bibinfo {pages}
  {475} (\bibinfo {year} {2004})}\BibitemShut {NoStop}%
\bibitem [{\citenamefont {Penchina}(1978)}]{Penchina1978}%
  \BibitemOpen
  \bibfield  {author} {\bibinfo {author} {\bibfnamefont {C.~M.}\ \bibnamefont
  {Penchina}},\ }\href {\doibase 10.1119/1.11359} {\bibfield  {journal}
  {\bibinfo  {journal} {American Journal of Physics}\ }\textbf {\bibinfo
  {volume} {46}},\ \bibinfo {pages} {295} (\bibinfo {year} {1978})}\BibitemShut
  {NoStop}%
\bibitem [{\citenamefont {Copeland}(1982)}]{Copeland1982}%
  \BibitemOpen
  \bibfield  {author} {\bibinfo {author} {\bibfnamefont {J.}~\bibnamefont
  {Copeland}},\ }\href {\doibase 10.1119/1.12785} {\bibfield  {journal}
  {\bibinfo  {journal} {American Journal of Physics}\ }\textbf {\bibinfo
  {volume} {50}},\ \bibinfo {pages} {599} (\bibinfo {year} {1982})}\BibitemShut
  {NoStop}%
\bibitem [{\citenamefont {Sherwood}(1983)}]{Sherwood1983}%
  \BibitemOpen
  \bibfield  {author} {\bibinfo {author} {\bibfnamefont {B.~A.}\ \bibnamefont
  {Sherwood}},\ }\href {\doibase 10.1119/1.13173} {\bibfield  {journal}
  {\bibinfo  {journal} {American Journal of Physics}\ }\textbf {\bibinfo
  {volume} {51}},\ \bibinfo {pages} {597} (\bibinfo {year} {1983})}\BibitemShut
  {NoStop}%
\bibitem [{\citenamefont {Arons}(1999)}]{Arons1999}%
  \BibitemOpen
  \bibfield  {author} {\bibinfo {author} {\bibfnamefont {A.~B.}\ \bibnamefont
  {Arons}},\ }\href {\doibase 10.1119/1.19182} {\bibfield  {journal} {\bibinfo
  {journal} {American Journal of Physics}\ }\textbf {\bibinfo {volume} {67}},\
  \bibinfo {pages} {1063} (\bibinfo {year} {1999})}\BibitemShut {NoStop}%
\bibitem [{\citenamefont {Gü{\'{e}}mez}\ and\ \citenamefont
  {Fiolhais}(2013)}]{Guemez2013}%
  \BibitemOpen
  \bibfield  {author} {\bibinfo {author} {\bibfnamefont {J.}~\bibnamefont
  {Gü{\'{e}}mez}}\ and\ \bibinfo {author} {\bibfnamefont {M.}~\bibnamefont
  {Fiolhais}},\ }\href {\doibase 10.1088/0143-0807/34/2/345} {\bibfield
  {journal} {\bibinfo  {journal} {European Journal of Physics}\ }\textbf
  {\bibinfo {volume} {34}},\ \bibinfo {pages} {345} (\bibinfo {year}
  {2013})}\BibitemShut {NoStop}%
\bibitem [{\citenamefont {Gü{\'{e}}mez}\ and\ \citenamefont
  {Fiolhais}(2016)}]{Guemez2016}%
  \BibitemOpen
  \bibfield  {author} {\bibinfo {author} {\bibfnamefont {J.}~\bibnamefont
  {Gü{\'{e}}mez}}\ and\ \bibinfo {author} {\bibfnamefont {M.}~\bibnamefont
  {Fiolhais}},\ }\href {\doibase 10.1088/0143-0807/37/4/045101} {\bibfield
  {journal} {\bibinfo  {journal} {European Journal of Physics}\ }\textbf
  {\bibinfo {volume} {37}},\ \bibinfo {pages} {045101} (\bibinfo {year}
  {2016})}\BibitemShut {NoStop}%
\bibitem [{\citenamefont {Gü{\'{e}}mez}\ and\ \citenamefont
  {Fiolhais}(2018)}]{Guemez2018}%
  \BibitemOpen
  \bibfield  {author} {\bibinfo {author} {\bibfnamefont {J.}~\bibnamefont
  {Gü{\'{e}}mez}}\ and\ \bibinfo {author} {\bibfnamefont {M.}~\bibnamefont
  {Fiolhais}},\ }\href {\doibase 10.1088/1361-6404/aabbfb} {\bibfield
  {journal} {\bibinfo  {journal} {European Journal of Physics}\ }\textbf
  {\bibinfo {volume} {39}},\ \bibinfo {pages} {045010} (\bibinfo {year}
  {2018})}\BibitemShut {NoStop}%
\bibitem [{\citenamefont {Lima}(2010)}]{Lima2010}%
  \BibitemOpen
  \bibfield  {author} {\bibinfo {author} {\bibfnamefont {F.~M.~S.}\
  \bibnamefont {Lima}},\ }\href {\doibase 10.1119/1.3442472} {\bibfield
  {journal} {\bibinfo  {journal} {American Journal of Physics}\ }\textbf
  {\bibinfo {volume} {78}},\ \bibinfo {pages} {1146} (\bibinfo {year}
  {2010})}\BibitemShut {NoStop}%
\bibitem [{\citenamefont {Nelson}\ and\ \citenamefont
  {Olsson}(1986)}]{Nelson1986}%
  \BibitemOpen
  \bibfield  {author} {\bibinfo {author} {\bibfnamefont {R.~A.}\ \bibnamefont
  {Nelson}}\ and\ \bibinfo {author} {\bibfnamefont {M.~G.}\ \bibnamefont
  {Olsson}},\ }\href {\doibase 10.1119/1.14703} {\bibfield  {journal} {\bibinfo
   {journal} {American Journal of Physics}\ }\textbf {\bibinfo {volume} {54}},\
  \bibinfo {pages} {112} (\bibinfo {year} {1986})}\BibitemShut {NoStop}%
\bibitem [{\citenamefont {Takahashi}\ and\ \citenamefont
  {Thompson}(1999)}]{Takahashi1999}%
  \BibitemOpen
  \bibfield  {author} {\bibinfo {author} {\bibfnamefont {K.}~\bibnamefont
  {Takahashi}}\ and\ \bibinfo {author} {\bibfnamefont {D.}~\bibnamefont
  {Thompson}},\ }\href {https://doi.org/10.1119/1.19356} {\bibfield  {journal}
  {\bibinfo  {journal} {American Journal of Physics}\ }\textbf {\bibinfo
  {volume} {67}},\ \bibinfo {pages} {709} (\bibinfo {year} {1999})}\BibitemShut
  {NoStop}%
\bibitem [{\citenamefont {Arora}\ \emph {et~al.}(2006)\citenamefont {Arora},
  \citenamefont {Rawat}, \citenamefont {Kaur},\ and\ \citenamefont
  {Arun}}]{Arora2006}%
  \BibitemOpen
  \bibfield  {author} {\bibinfo {author} {\bibfnamefont {A.}~\bibnamefont
  {Arora}}, \bibinfo {author} {\bibfnamefont {R.}~\bibnamefont {Rawat}},
  \bibinfo {author} {\bibfnamefont {S.}~\bibnamefont {Kaur}}, \ and\ \bibinfo
  {author} {\bibfnamefont {P.}~\bibnamefont {Arun}},\ }\href
  {https://arxiv.org/abs/physics/0608071} {\enquote {\bibinfo {title} {{Study
  of the Damped Pendulum}},}\ } (\bibinfo {year} {2006}),\ \Eprint
  {http://arxiv.org/abs/physics/0608071} {arXiv:physics/0608071 [physics]}
  \BibitemShut {NoStop}%
\bibitem [{\citenamefont {Guo}(2011)}]{Guo2011}%
  \BibitemOpen
  \bibfield  {author} {\bibinfo {author} {\bibfnamefont {J.}~\bibnamefont
  {Guo}},\ }\href {\doibase 10.1080/00221686.2010.538572} {\bibfield  {journal}
  {\bibinfo  {journal} {Journal of Hydraulic Research}\ }\textbf {\bibinfo
  {volume} {49}},\ \bibinfo {pages} {32} (\bibinfo {year} {2011})}\BibitemShut
  {NoStop}%
\bibitem [{\citenamefont {Dahmen}(2014)}]{Dahmen2014}%
  \BibitemOpen
  \bibfield  {author} {\bibinfo {author} {\bibfnamefont {S.~R.}\ \bibnamefont
  {Dahmen}},\ }\href {\doibase 10.1140/epjh/e2015-50054-8} {\  (\bibinfo {year}
  {2014}),\ 10.1140/epjh/e2015-50054-8},\ \Eprint
  {http://arxiv.org/abs/1409.7446} {arXiv:1409.7446} \BibitemShut {NoStop}%
\bibitem [{\citenamefont {Klein}\ \emph {et~al.}(2017)\citenamefont {Klein},
  \citenamefont {M{\"{u}}ller}, \citenamefont {Gr{\"{o}}ber}, \citenamefont
  {Molz},\ and\ \citenamefont {Kuhn}}]{Klein2017}%
  \BibitemOpen
  \bibfield  {author} {\bibinfo {author} {\bibfnamefont {P.}~\bibnamefont
  {Klein}}, \bibinfo {author} {\bibfnamefont {A.}~\bibnamefont {M{\"{u}}ller}},
  \bibinfo {author} {\bibfnamefont {S.}~\bibnamefont {Gr{\"{o}}ber}}, \bibinfo
  {author} {\bibfnamefont {A.}~\bibnamefont {Molz}}, \ and\ \bibinfo {author}
  {\bibfnamefont {J.}~\bibnamefont {Kuhn}},\ }\href {\doibase
  10.1119/1.4964134} {\bibfield  {journal} {\bibinfo  {journal} {American
  Journal of Physics}\ }\textbf {\bibinfo {volume} {85}},\ \bibinfo {pages}
  {30} (\bibinfo {year} {2017})}\BibitemShut {NoStop}%
\bibitem [{\citenamefont {Graef}(1972)}]{Graef1972}%
  \BibitemOpen
  \bibfield  {author} {\bibinfo {author} {\bibfnamefont {J.~R.}\ \bibnamefont
  {Graef}},\ }\href {\doibase 10.1016/0022-0396(72)90004-6} {\bibfield
  {journal} {\bibinfo  {journal} {Journal of Differential Equations}\ }\textbf
  {\bibinfo {volume} {12}},\ \bibinfo {pages} {34} (\bibinfo {year}
  {1972})}\BibitemShut {NoStop}%
\bibitem [{\citenamefont {Fulcher}\ \emph {et~al.}(2006)\citenamefont
  {Fulcher}, \citenamefont {Scherer}, \citenamefont {Melnykov}, \citenamefont
  {Gateva},\ and\ \citenamefont {Limes}}]{Fulcher2006}%
  \BibitemOpen
  \bibfield  {author} {\bibinfo {author} {\bibfnamefont {L.~P.}\ \bibnamefont
  {Fulcher}}, \bibinfo {author} {\bibfnamefont {R.~C.}\ \bibnamefont
  {Scherer}}, \bibinfo {author} {\bibfnamefont {A.}~\bibnamefont {Melnykov}},
  \bibinfo {author} {\bibfnamefont {V.}~\bibnamefont {Gateva}}, \ and\ \bibinfo
  {author} {\bibfnamefont {M.~E.}\ \bibnamefont {Limes}},\ }\href {\doibase
  10.1119/1.2173272} {\bibfield  {journal} {\bibinfo  {journal} {American
  Journal of Physics}\ }\textbf {\bibinfo {volume} {74}},\ \bibinfo {pages}
  {386} (\bibinfo {year} {2006})}\BibitemShut {NoStop}%
\bibitem [{\citenamefont {Stoop}\ \emph {et~al.}(2006)\citenamefont {Stoop},
  \citenamefont {Kern}, \citenamefont {G{\"{o}}pfert}, \citenamefont {Smirnov},
  \citenamefont {Dikanev},\ and\ \citenamefont {Bezrucko}}]{Stoop2006}%
  \BibitemOpen
  \bibfield  {author} {\bibinfo {author} {\bibfnamefont {R.}~\bibnamefont
  {Stoop}}, \bibinfo {author} {\bibfnamefont {A.}~\bibnamefont {Kern}},
  \bibinfo {author} {\bibfnamefont {M.~C.}\ \bibnamefont {G{\"{o}}pfert}},
  \bibinfo {author} {\bibfnamefont {D.~A.}\ \bibnamefont {Smirnov}}, \bibinfo
  {author} {\bibfnamefont {T.~V.}\ \bibnamefont {Dikanev}}, \ and\ \bibinfo
  {author} {\bibfnamefont {B.~P.}\ \bibnamefont {Bezrucko}},\ }\href {\doibase
  10.1007/s00249-006-0059-5} {\bibfield  {journal} {\bibinfo  {journal}
  {European Biophysics Journal}\ }\textbf {\bibinfo {volume} {35}},\ \bibinfo
  {pages} {511} (\bibinfo {year} {2006})}\BibitemShut {NoStop}%
\bibitem [{\citenamefont {Jenkins}(2013)}]{Jenkins2013}%
  \BibitemOpen
  \bibfield  {author} {\bibinfo {author} {\bibfnamefont {A.}~\bibnamefont
  {Jenkins}},\ }\href {\doibase https://doi.org/10.1016/j.physrep.2012.10.007}
  {\bibfield  {journal} {\bibinfo  {journal} {Physics Reports}\ }\textbf
  {\bibinfo {volume} {525}},\ \bibinfo {pages} {167 } (\bibinfo {year}
  {2013})}\BibitemShut {NoStop}%
\bibitem [{\citenamefont {Kostov}\ \emph {et~al.}(2008)\citenamefont {Kostov},
  \citenamefont {Morshed}, \citenamefont {Höling}, \citenamefont {Chen},\ and\
  \citenamefont {Siegel}}]{Kostov2008}%
  \BibitemOpen
  \bibfield  {author} {\bibinfo {author} {\bibfnamefont {Y.}~\bibnamefont
  {Kostov}}, \bibinfo {author} {\bibfnamefont {R.}~\bibnamefont {Morshed}},
  \bibinfo {author} {\bibfnamefont {B.}~\bibnamefont {Höling}}, \bibinfo
  {author} {\bibfnamefont {R.}~\bibnamefont {Chen}}, \ and\ \bibinfo {author}
  {\bibfnamefont {P.~B.}\ \bibnamefont {Siegel}},\ }\href {\doibase
  10.1119/1.2937897} {\bibfield  {journal} {\bibinfo  {journal} {American
  Journal of Physics}\ }\textbf {\bibinfo {volume} {76}},\ \bibinfo {pages}
  {956} (\bibinfo {year} {2008})}\BibitemShut {NoStop}%
\bibitem [{\citenamefont {Crawford}(1975)}]{Crawford1975}%
  \BibitemOpen
  \bibfield  {author} {\bibinfo {author} {\bibfnamefont {F.~S.}\ \bibnamefont
  {Crawford}},\ }\href {\doibase 10.1119/1.10073} {\bibfield  {journal}
  {\bibinfo  {journal} {American Journal of Physics}\ }\textbf {\bibinfo
  {volume} {43}},\ \bibinfo {pages} {276} (\bibinfo {year} {1975})}\BibitemShut
  {NoStop}%
\bibitem [{\citenamefont {Ravindra}\ and\ \citenamefont
  {Mallik}(1994)}]{Ravindra1994}%
  \BibitemOpen
  \bibfield  {author} {\bibinfo {author} {\bibfnamefont {B.}~\bibnamefont
  {Ravindra}}\ and\ \bibinfo {author} {\bibfnamefont {A.}~\bibnamefont
  {Mallik}},\ }\href {\doibase https://doi.org/10.1006/jsvi.1994.1066}
  {\bibfield  {journal} {\bibinfo  {journal} {Journal of Sound and Vibration}\
  }\textbf {\bibinfo {volume} {170}},\ \bibinfo {pages} {325 } (\bibinfo {year}
  {1994})}\BibitemShut {NoStop}%
\bibitem [{\citenamefont {Baltanás}\ \emph {et~al.}(2001)\citenamefont
  {Baltanás}, \citenamefont {Trueba},\ and\ \citenamefont
  {Sanjuán}}]{Baltanas2001}%
  \BibitemOpen
  \bibfield  {author} {\bibinfo {author} {\bibfnamefont {J.~P.}\ \bibnamefont
  {Baltanás}}, \bibinfo {author} {\bibfnamefont {J.~L.}\ \bibnamefont
  {Trueba}}, \ and\ \bibinfo {author} {\bibfnamefont {M.~A.}\ \bibnamefont
  {Sanjuán}},\ }\href {\doibase https://doi.org/10.1016/S0167-2789(01)00329-3}
  {\bibfield  {journal} {\bibinfo  {journal} {Physica D: Nonlinear Phenomena}\
  }\textbf {\bibinfo {volume} {159}},\ \bibinfo {pages} {22 } (\bibinfo {year}
  {2001})}\BibitemShut {NoStop}%
\bibitem [{\citenamefont {Mickens}(2003)}]{Mickens2003}%
  \BibitemOpen
  \bibfield  {author} {\bibinfo {author} {\bibfnamefont {R.}~\bibnamefont
  {Mickens}},\ }\href {\doibase 10.1016/S0022-460X(02)01510-9} {\bibfield
  {journal} {\bibinfo  {journal} {Journal of Sound and Vibration}\ }\textbf
  {\bibinfo {volume} {264}},\ \bibinfo {pages} {1195} (\bibinfo {year}
  {2003})}\BibitemShut {NoStop}%
\bibitem [{\citenamefont {Elliott}\ \emph {et~al.}(2015)\citenamefont
  {Elliott}, \citenamefont {Tehrani},\ and\ \citenamefont
  {Langley}}]{Elliott2015}%
  \BibitemOpen
  \bibfield  {author} {\bibinfo {author} {\bibfnamefont {S.~J.}\ \bibnamefont
  {Elliott}}, \bibinfo {author} {\bibfnamefont {M.~G.}\ \bibnamefont
  {Tehrani}}, \ and\ \bibinfo {author} {\bibfnamefont {R.~S.}\ \bibnamefont
  {Langley}},\ }\href {\doibase 10.1098/rsta.2014.0402} {\bibfield  {journal}
  {\bibinfo  {journal} {Philosophical Transactions of the Royal Society A:
  Mathematical, Physical and Engineering Sciences}\ }\textbf {\bibinfo {volume}
  {373}},\ \bibinfo {pages} {20140402} (\bibinfo {year} {2015})}\BibitemShut
  {NoStop}%
\bibitem [{\citenamefont {Plastino}\ \emph {et~al.}(2018)\citenamefont
  {Plastino}, \citenamefont {Wedemann}, \citenamefont {Curado}, \citenamefont
  {Nobre},\ and\ \citenamefont {Tsallis}}]{Plastino2018b}%
  \BibitemOpen
  \bibfield  {author} {\bibinfo {author} {\bibfnamefont {A.~R.}\ \bibnamefont
  {Plastino}}, \bibinfo {author} {\bibfnamefont {R.~S.}\ \bibnamefont
  {Wedemann}}, \bibinfo {author} {\bibfnamefont {E.~M.~F.}\ \bibnamefont
  {Curado}}, \bibinfo {author} {\bibfnamefont {F.~D.}\ \bibnamefont {Nobre}}, \
  and\ \bibinfo {author} {\bibfnamefont {C.}~\bibnamefont {Tsallis}},\ }\href
  {\doibase 10.1103/PhysRevE.98.012129} {\bibfield  {journal} {\bibinfo
  {journal} {Phys. Rev. E}\ }\textbf {\bibinfo {volume} {98}},\ \bibinfo
  {pages} {012129} (\bibinfo {year} {2018})}\BibitemShut {NoStop}%
\bibitem [{\citenamefont {Flores}\ \emph {et~al.}(2008)\citenamefont {Flores},
  \citenamefont {Ambr{\'o}sio}, \citenamefont {Claro},\ and\ \citenamefont
  {Lankarani}}]{Flores2008}%
  \BibitemOpen
  \bibfield  {author} {\bibinfo {author} {\bibfnamefont {P.}~\bibnamefont
  {Flores}}, \bibinfo {author} {\bibfnamefont {J.}~\bibnamefont
  {Ambr{\'o}sio}}, \bibinfo {author} {\bibfnamefont {J.~P.}\ \bibnamefont
  {Claro}}, \ and\ \bibinfo {author} {\bibfnamefont {H.~M.}\ \bibnamefont
  {Lankarani}},\ }\enquote {\bibinfo {title} {Contact-impact force models for
  mechanical systems},}\ in\ \href {\doibase 10.1007/978-3-540-74361-3_3}
  {\emph {\bibinfo {booktitle} {Kinematics and Dynamics of Multibody Systems
  with Imperfect Joints: Models and Case Studies}}}\ (\bibinfo  {publisher}
  {Springer Berlin Heidelberg},\ \bibinfo {address} {Berlin, Heidelberg},\
  \bibinfo {year} {2008})\ pp.\ \bibinfo {pages} {47--66}\BibitemShut {NoStop}%
\bibitem [{\citenamefont {Muvengei}\ \emph {et~al.}(2014)\citenamefont
  {Muvengei}, \citenamefont {Kihiu},\ and\ \citenamefont
  {Ikua}}]{Muvengei2012}%
  \BibitemOpen
  \bibfield  {author} {\bibinfo {author} {\bibfnamefont {O.}~\bibnamefont
  {Muvengei}}, \bibinfo {author} {\bibfnamefont {J.}~\bibnamefont {Kihiu}}, \
  and\ \bibinfo {author} {\bibfnamefont {B.}~\bibnamefont {Ikua}},\ }\href
  {http://sri.jkuat.ac.ke/ojs/index.php/proceedings/article/view/195}
  {\bibfield  {journal} {\bibinfo  {journal} {Proceedings of Sustainable
  Research and Innovation Conference}\ }\textbf {\bibinfo {volume} {0}},\
  \bibinfo {pages} {99} (\bibinfo {year} {2014})}\BibitemShut {NoStop}%
\bibitem [{\citenamefont {Sebasti{\~{a}}o}(2013)}]{Sebastiao2013}%
  \BibitemOpen
  \bibfield  {author} {\bibinfo {author} {\bibfnamefont {P.~J.}\ \bibnamefont
  {Sebasti{\~{a}}o}},\ }\href {\doibase 10.1088/0143-0807/35/1/015017}
  {\bibfield  {journal} {\bibinfo  {journal} {European Journal of Physics}\
  }\textbf {\bibinfo {volume} {35}},\ \bibinfo {pages} {015017} (\bibinfo
  {year} {2013})}\BibitemShut {NoStop}%
\bibitem [{\citenamefont {Grosse}(2014)}]{Grosse2014}%
  \BibitemOpen
  \bibfield  {author} {\bibinfo {author} {\bibfnamefont {C.}~\bibnamefont
  {Grosse}},\ }\href {\doibase https://doi.org/10.1016/j.jcis.2013.12.031}
  {\bibfield  {journal} {\bibinfo  {journal} {Journal of Colloid and Interface
  Science}\ }\textbf {\bibinfo {volume} {419}},\ \bibinfo {pages} {102 }
  (\bibinfo {year} {2014})}\BibitemShut {NoStop}%
\bibitem [{\citenamefont {Minorsky}(1942)}]{Minorsky1942}%
  \BibitemOpen
  \bibfield  {author} {\bibinfo {author} {\bibfnamefont {N.}~\bibnamefont
  {Minorsky}},\ }\href
  {https://books.google.pt/books?hl=en&lr=&id=6to5DwAAQBAJ&oi=fnd&pg=PA141&dq=oscillation+derivative+of+phase#v=onepage&q=oscillation
  derivative of phase&f=false} {\bibfield  {journal} {\bibinfo  {journal}
  {Journal of Applied Mechanics}\ }\textbf {\bibinfo {volume} {9}},\ \bibinfo
  {pages} {65} (\bibinfo {year} {1942})}\BibitemShut {NoStop}%
\bibitem [{\citenamefont {Cerullo}\ and\ \citenamefont
  {De~Silvestri}(2003)}]{Cerullo2003}%
  \BibitemOpen
  \bibfield  {author} {\bibinfo {author} {\bibfnamefont {G.}~\bibnamefont
  {Cerullo}}\ and\ \bibinfo {author} {\bibfnamefont {S.}~\bibnamefont
  {De~Silvestri}},\ }\href {\doibase 10.1063/1.1523642} {\bibfield  {journal}
  {\bibinfo  {journal} {Review of Scientific Instruments}\ }\textbf {\bibinfo
  {volume} {74}},\ \bibinfo {pages} {1} (\bibinfo {year} {2003})}\BibitemShut
  {NoStop}%
\bibitem [{\citenamefont {Shen}(2006)}]{Shen2006}%
  \BibitemOpen
  \bibfield  {author} {\bibinfo {author} {\bibfnamefont {Y.}~\bibnamefont
  {Shen}},\ }\href {https://books.google.pt/books?id=-4iQDAAAQBAJ} {\emph
  {\bibinfo {title} {Nonlinear Infrared Generation}}},\ Topics in Applied
  Physics\ (\bibinfo  {publisher} {Springer Berlin Heidelberg},\ \bibinfo
  {year} {2006})\BibitemShut {NoStop}%
\bibitem [{\citenamefont {Boyd}\ and\ \citenamefont {Prato}(2008)}]{Boyd2008}%
  \BibitemOpen
  \bibfield  {author} {\bibinfo {author} {\bibfnamefont {R.}~\bibnamefont
  {Boyd}}\ and\ \bibinfo {author} {\bibfnamefont {D.}~\bibnamefont {Prato}},\
  }\href {https://books.google.pt/books?id=uoRUi1Yb7ooC} {\emph {\bibinfo
  {title} {Nonlinear Optics}}}\ (\bibinfo  {publisher} {Elsevier Science},\
  \bibinfo {year} {2008})\BibitemShut {NoStop}%
\bibitem [{\citenamefont {Lehmann}\ \emph {et~al.}(2013)\citenamefont
  {Lehmann}, \citenamefont {Spatschek},\ and\ \citenamefont
  {Sewell}}]{Lehmann2013}%
  \BibitemOpen
  \bibfield  {author} {\bibinfo {author} {\bibfnamefont {G.}~\bibnamefont
  {Lehmann}}, \bibinfo {author} {\bibfnamefont {K.~H.}\ \bibnamefont
  {Spatschek}}, \ and\ \bibinfo {author} {\bibfnamefont {G.}~\bibnamefont
  {Sewell}},\ }\href {\doibase 10.1103/PhysRevE.87.063107} {\bibfield
  {journal} {\bibinfo  {journal} {Phys. Rev. E}\ }\textbf {\bibinfo {volume}
  {87}},\ \bibinfo {pages} {063107} (\bibinfo {year} {2013})}\BibitemShut
  {NoStop}%
\bibitem [{\citenamefont {Schluck}\ \emph {et~al.}(2015)\citenamefont
  {Schluck}, \citenamefont {Lehmann},\ and\ \citenamefont
  {Spatschek}}]{Schluck2015}%
  \BibitemOpen
  \bibfield  {author} {\bibinfo {author} {\bibfnamefont {F.}~\bibnamefont
  {Schluck}}, \bibinfo {author} {\bibfnamefont {G.}~\bibnamefont {Lehmann}}, \
  and\ \bibinfo {author} {\bibfnamefont {K.~H.}\ \bibnamefont {Spatschek}},\
  }\href {\doibase 10.1063/1.4929859} {\bibfield  {journal} {\bibinfo
  {journal} {Physics of Plasmas}\ }\textbf {\bibinfo {volume} {22}},\ \bibinfo
  {pages} {093104} (\bibinfo {year} {2015})}\BibitemShut {NoStop}%
\bibitem [{\citenamefont {Christian}(2017)}]{Christian2017}%
  \BibitemOpen
  \bibfield  {author} {\bibinfo {author} {\bibfnamefont {J.~M.}\ \bibnamefont
  {Christian}},\ }\href {\doibase 10.1088/1361-6404/aa7cbe} {\bibfield
  {journal} {\bibinfo  {journal} {European Journal of Physics}\ }\textbf
  {\bibinfo {volume} {38}},\ \bibinfo {pages} {055002} (\bibinfo {year}
  {2017})}\BibitemShut {NoStop}%
\bibitem [{\citenamefont {George}\ and\ \citenamefont
  {Harris}(1983)}]{George1983}%
  \BibitemOpen
  \bibfield  {author} {\bibinfo {author} {\bibfnamefont {S.~M.}\ \bibnamefont
  {George}}\ and\ \bibinfo {author} {\bibfnamefont {C.~B.}\ \bibnamefont
  {Harris}},\ }\href {\doibase 10.1103/PhysRevA.28.863} {\bibfield  {journal}
  {\bibinfo  {journal} {Phys. Rev. A}\ }\textbf {\bibinfo {volume} {28}},\
  \bibinfo {pages} {863} (\bibinfo {year} {1983})}\BibitemShut {NoStop}%
\bibitem [{\citenamefont {{Boscolo}}\ \emph {et~al.}(2014)\citenamefont
  {{Boscolo}}, \citenamefont {{Castelli}}, \citenamefont {{Stellato}},\ and\
  \citenamefont {{Vercellati}}}]{Boscolo2014}%
  \BibitemOpen
  \bibfield  {author} {\bibinfo {author} {\bibfnamefont {I.}~\bibnamefont
  {{Boscolo}}}, \bibinfo {author} {\bibfnamefont {F.}~\bibnamefont
  {{Castelli}}}, \bibinfo {author} {\bibfnamefont {M.}~\bibnamefont
  {{Stellato}}}, \ and\ \bibinfo {author} {\bibfnamefont {S.}~\bibnamefont
  {{Vercellati}}},\ }\href@noop {} {\bibfield  {journal} {\bibinfo  {journal}
  {arXiv e-prints}\ ,\ \bibinfo {eid} {arXiv:1402.5318}} (\bibinfo {year}
  {2014})},\ \Eprint {http://arxiv.org/abs/1402.5318} {arXiv:1402.5318
  [physics.ed-ph]} \BibitemShut {NoStop}%
\bibitem [{\citenamefont {Basset}(1888)}]{Basset1888}%
  \BibitemOpen
  \bibfield  {author} {\bibinfo {author} {\bibfnamefont {A.~B.}\ \bibnamefont
  {Basset}},\ }\href@noop {} {\emph {\bibinfo {title} {A treatise on
  hydrodynamics: with numerous examples}}},\ Vol.~\bibinfo {volume} {2}\
  (\bibinfo  {publisher} {Deighton, Bell and Company},\ \bibinfo {year}
  {1888})\BibitemShut {NoStop}%
\bibitem [{\citenamefont {Hamilton}()}]{Hamilton1973}%
  \BibitemOpen
  \bibfield  {author} {\bibinfo {author} {\bibfnamefont {W.~S.}\ \bibnamefont
  {Hamilton}},\ }\enquote {\bibinfo {title} {Fluid force on accelerating
  bodies},}\ in\ \href {\doibase 10.1061/9780872620490.102} {\emph {\bibinfo
  {booktitle} {Coastal Engineering 1972}}},\ pp.\ \bibinfo {pages}
  {1767--1782}\BibitemShut {NoStop}%
\bibitem [{\citenamefont {Herringe}(1976)}]{Herringe1976}%
  \BibitemOpen
  \bibfield  {author} {\bibinfo {author} {\bibfnamefont {R.~A.}\ \bibnamefont
  {Herringe}},\ }\href {\doibase 10.1016/S0300-9467(76)80030-5} {\bibfield
  {journal} {\bibinfo  {journal} {The Chemical Engineering Journal}\ }\textbf
  {\bibinfo {volume} {11}},\ \bibinfo {pages} {89} (\bibinfo {year}
  {1976})}\BibitemShut {NoStop}%
\bibitem [{\citenamefont {Thomas}\ and\ \citenamefont
  {Thomasa}(1992)}]{Thomas1992}%
  \BibitemOpen
  \bibfield  {author} {\bibinfo {author} {\bibfnamefont {P.~J.}\ \bibnamefont
  {Thomas}}\ and\ \bibinfo {author} {\bibfnamefont {P.~J.}\ \bibnamefont
  {Thomasa}},\ }\href {\doibase 10.1063/1.858379} {\ \textbf {\bibinfo {volume}
  {2090}} (\bibinfo {year} {1992}),\ 10.1063/1.858379}\BibitemShut {NoStop}%
\bibitem [{\citenamefont {Mainardi}\ \emph {et~al.}(1995)\citenamefont
  {Mainardi}, \citenamefont {Pironi},\ and\ \citenamefont
  {Tampieri}}]{Mainardi1995}%
  \BibitemOpen
  \bibfield  {author} {\bibinfo {author} {\bibfnamefont {F.}~\bibnamefont
  {Mainardi}}, \bibinfo {author} {\bibfnamefont {P.}~\bibnamefont {Pironi}}, \
  and\ \bibinfo {author} {\bibfnamefont {F.}~\bibnamefont {Tampieri}},\
  }\href@noop {} {\bibfield  {journal} {\bibinfo  {journal} {Proceedings
  CANCAM}\ }\textbf {\bibinfo {volume} {95}},\ \bibinfo {pages} {836} (\bibinfo
  {year} {1995})}\BibitemShut {NoStop}%
\bibitem [{\citenamefont {Chang}\ and\ \citenamefont {Yen}(1998)}]{Chang1998}%
  \BibitemOpen
  \bibfield  {author} {\bibinfo {author} {\bibfnamefont {T.-J.}\ \bibnamefont
  {Chang}}\ and\ \bibinfo {author} {\bibfnamefont {B.~C.}\ \bibnamefont
  {Yen}},\ }\href {\doibase 10.1061/(ASCE)0733-9399(1998)124:11(1193)}
  {\bibfield  {journal} {\bibinfo  {journal} {Journal of Engineering
  Mechanics}\ }\textbf {\bibinfo {volume} {124}},\ \bibinfo {pages} {1193}
  (\bibinfo {year} {1998})}\BibitemShut {NoStop}%
\bibitem [{\citenamefont {Candelier}\ \emph {et~al.}(2004)\citenamefont
  {Candelier}, \citenamefont {Angilella},\ and\ \citenamefont
  {Souhar}}]{Candelier2004}%
  \BibitemOpen
  \bibfield  {author} {\bibinfo {author} {\bibfnamefont {F.}~\bibnamefont
  {Candelier}}, \bibinfo {author} {\bibfnamefont {J.~R.}\ \bibnamefont
  {Angilella}}, \ and\ \bibinfo {author} {\bibfnamefont {M.}~\bibnamefont
  {Souhar}},\ }\href {\doibase 10.1063/1.1689970} {\bibfield  {journal}
  {\bibinfo  {journal} {Physics of Fluids}\ }\textbf {\bibinfo {volume} {16}},\
  \bibinfo {pages} {1765} (\bibinfo {year} {2004})}\BibitemShut {NoStop}%
\bibitem [{\citenamefont {van Hinsberg}\ \emph {et~al.}(2011)\citenamefont {van
  Hinsberg}, \citenamefont {{ten Thije Boonkkamp}},\ and\ \citenamefont
  {Clercx}}]{Hinsberg2011}%
  \BibitemOpen
  \bibfield  {author} {\bibinfo {author} {\bibfnamefont {M.~A.}\ \bibnamefont
  {van Hinsberg}}, \bibinfo {author} {\bibfnamefont {J.~H.}\ \bibnamefont {{ten
  Thije Boonkkamp}}}, \ and\ \bibinfo {author} {\bibfnamefont {H.~J.}\
  \bibnamefont {Clercx}},\ }\href {\doibase 10.1016/j.jcp.2010.11.014}
  {\bibfield  {journal} {\bibinfo  {journal} {Journal of Computational
  Physics}\ }\textbf {\bibinfo {volume} {230}},\ \bibinfo {pages} {1465}
  (\bibinfo {year} {2011})},\ \Eprint {http://arxiv.org/abs/1008.0833}
  {arXiv:1008.0833} \BibitemShut {NoStop}%
\bibitem [{\citenamefont {Baleanu}\ \emph {et~al.}(2013)\citenamefont
  {Baleanu}, \citenamefont {Garra},\ and\ \citenamefont
  {Petras}}]{Baleanu2013}%
  \BibitemOpen
  \bibfield  {author} {\bibinfo {author} {\bibfnamefont {D.}~\bibnamefont
  {Baleanu}}, \bibinfo {author} {\bibfnamefont {R.}~\bibnamefont {Garra}}, \
  and\ \bibinfo {author} {\bibfnamefont {I.}~\bibnamefont {Petras}},\ }\href
  {\doibase 10.1016/S0034-4877(14)60004-5} {\bibfield  {journal} {\bibinfo
  {journal} {Reports on Mathematical Physics}\ }\textbf {\bibinfo {volume}
  {72}},\ \bibinfo {pages} {57} (\bibinfo {year} {2013})}\BibitemShut {NoStop}%
\bibitem [{\citenamefont {Daitche}(2015)}]{Daitche2015}%
  \BibitemOpen
  \bibfield  {author} {\bibinfo {author} {\bibfnamefont {A.}~\bibnamefont
  {Daitche}},\ }\href {\doibase 10.1017/jfm.2015.551} {\bibfield  {journal}
  {\bibinfo  {journal} {Journal of Fluid Mechanics}\ }\textbf {\bibinfo
  {volume} {782}},\ \bibinfo {pages} {567} (\bibinfo {year} {2015})},\ \Eprint
  {http://arxiv.org/abs/1501.04770} {arXiv:1501.04770} \BibitemShut {NoStop}%
\bibitem [{\citenamefont {Annamalai}\ and\ \citenamefont
  {Balachandar}(2017)}]{Annamalai2017}%
  \BibitemOpen
  \bibfield  {author} {\bibinfo {author} {\bibfnamefont {S.}~\bibnamefont
  {Annamalai}}\ and\ \bibinfo {author} {\bibfnamefont {S.}~\bibnamefont
  {Balachandar}},\ }\href {\doibase 10.1017/jfm.2017.77} {\bibfield  {journal}
  {\bibinfo  {journal} {Journal of Fluid Mechanics}\ }\textbf {\bibinfo
  {volume} {816}},\ \bibinfo {pages} {381} (\bibinfo {year}
  {2017})}\BibitemShut {NoStop}%
\bibitem [{\citenamefont {Maris}(2019)}]{Maris2019}%
  \BibitemOpen
  \bibfield  {author} {\bibinfo {author} {\bibfnamefont {H.~J.}\ \bibnamefont
  {Maris}},\ }\href {\doibase 10.1119/1.5100939} {\bibfield  {journal}
  {\bibinfo  {journal} {American Journal of Physics}\ }\textbf {\bibinfo
  {volume} {87}},\ \bibinfo {pages} {643} (\bibinfo {year} {2019})}\BibitemShut
  {NoStop}%
\bibitem [{\citenamefont {Tatom}(1988)}]{Tatom1988}%
  \BibitemOpen
  \bibfield  {author} {\bibinfo {author} {\bibfnamefont {F.}~\bibnamefont
  {Tatom}},\ }\href@noop {} {\bibfield  {journal} {\bibinfo  {journal} {Applied
  Scientific Research}\ }\textbf {\bibinfo {volume} {45}},\ \bibinfo {pages}
  {283} (\bibinfo {year} {1988})}\BibitemShut {NoStop}%
\bibitem [{\citenamefont {Mainardi}(1997)}]{Mainardi1997}%
  \BibitemOpen
  \bibfield  {author} {\bibinfo {author} {\bibfnamefont {F.}~\bibnamefont
  {Mainardi}},\ }\href@noop {} {\enquote {\bibinfo {title} {Fractional
  calculus. some basic problems in continuum and statistical mechanics,[in:] a.
  carpinteri, f. mainardi (eds.), fractals and fractional calculus in continuum
  mechanics},}\ } (\bibinfo {year} {1997})\BibitemShut {NoStop}%
\bibitem [{\citenamefont {Gonz{\'{a}}les}\ \emph {et~al.}(2008)\citenamefont
  {Gonz{\'{a}}les}, \citenamefont {Bombardelli},\ and\ \citenamefont
  {Ni{\~{n}}o}}]{Bombardelli2008}%
  \BibitemOpen
  \bibfield  {author} {\bibinfo {author} {\bibfnamefont {A.~E.}\ \bibnamefont
  {Gonz{\'{a}}les}}, \bibinfo {author} {\bibfnamefont {F.~A.}\ \bibnamefont
  {Bombardelli}}, \ and\ \bibinfo {author} {\bibfnamefont {Y.~I.}\ \bibnamefont
  {Ni{\~{n}}o}},\ }\href {\doibase 10.1061/(ASCE)0733-9429(2008)134:10(1513)}
  {\bibfield  {journal} {\bibinfo  {journal} {Journal of Hydraulic
  Engineering}\ }\textbf {\bibinfo {volume} {134}},\ \bibinfo {pages} {1513}
  (\bibinfo {year} {2008})}\BibitemShut {NoStop}%
\bibitem [{\citenamefont {Lukerchenko}(2010)}]{Lukerchenko2010}%
  \BibitemOpen
  \bibfield  {author} {\bibinfo {author} {\bibfnamefont {N.}~\bibnamefont
  {Lukerchenko}},\ }\href {\doibase 10.1061/(ASCE)HY.1943-7900.0000140}
  {\bibfield  {journal} {\bibinfo  {journal} {Journal of Hydraulic
  Engineering}\ }\textbf {\bibinfo {volume} {136}},\ \bibinfo {pages} {853}
  (\bibinfo {year} {2010})}\BibitemShut {NoStop}%
\bibitem [{\citenamefont {Du}\ \emph {et~al.}(2013)\citenamefont {Du},
  \citenamefont {Wang},\ and\ \citenamefont {Hu}}]{Du2013}%
  \BibitemOpen
  \bibfield  {author} {\bibinfo {author} {\bibfnamefont {M.}~\bibnamefont
  {Du}}, \bibinfo {author} {\bibfnamefont {Z.}~\bibnamefont {Wang}}, \ and\
  \bibinfo {author} {\bibfnamefont {H.}~\bibnamefont {Hu}},\ }\href {\doibase
  10.1038/srep03431} {\bibfield  {journal} {\bibinfo  {journal} {Scientific
  Reports}\ }\textbf {\bibinfo {volume} {3}},\ \bibinfo {pages} {1} (\bibinfo
  {year} {2013})}\BibitemShut {NoStop}%
\bibitem [{\citenamefont {{\"O}zgen}(2013)}]{Ozgen2013}%
  \BibitemOpen
  \bibfield  {author} {\bibinfo {author} {\bibfnamefont {O.}~\bibnamefont
  {{\"O}zgen}},\ }\emph {\bibinfo {title} {Physics-Based Animation Models Using
  Fractional Calculus}},\ \href@noop {} {Ph.D. thesis},\ \bibinfo  {school} {UC
  Merced} (\bibinfo {year} {2013})\BibitemShut {NoStop}%
\bibitem [{\citenamefont {Olejnik}\ and\ \citenamefont
  {Awrejcewicz}(2018)}]{Olejnik2018}%
  \BibitemOpen
  \bibfield  {author} {\bibinfo {author} {\bibfnamefont {P.}~\bibnamefont
  {Olejnik}}\ and\ \bibinfo {author} {\bibfnamefont {J.}~\bibnamefont
  {Awrejcewicz}},\ }\href {\doibase 10.1016/j.ymssp.2017.04.037} {\bibfield
  {journal} {\bibinfo  {journal} {Mechanical Systems and Signal Processing}\
  }\textbf {\bibinfo {volume} {98}},\ \bibinfo {pages} {91} (\bibinfo {year}
  {2018})}\BibitemShut {NoStop}%
\bibitem [{\citenamefont {Sakakibara}(1997)}]{Sakakibara1997}%
  \BibitemOpen
  \bibfield  {author} {\bibinfo {author} {\bibfnamefont {S.}~\bibnamefont
  {Sakakibara}},\ }\href {\doibase 10.1299/jsmec.40.393} {\bibfield  {journal}
  {\bibinfo  {journal} {JSME International Journal Series C}\ }\textbf
  {\bibinfo {volume} {40}},\ \bibinfo {pages} {393} (\bibinfo {year}
  {1997})}\BibitemShut {NoStop}%
\bibitem [{\citenamefont {Seredy{\'{n}}ska}\ and\ \citenamefont
  {Hanyga}(2005)}]{Seredynska2005}%
  \BibitemOpen
  \bibfield  {author} {\bibinfo {author} {\bibfnamefont {M.}~\bibnamefont
  {Seredy{\'{n}}ska}}\ and\ \bibinfo {author} {\bibfnamefont {A.}~\bibnamefont
  {Hanyga}},\ }\href {\doibase 10.1007/s00707-005-0220-8} {\bibfield  {journal}
  {\bibinfo  {journal} {Acta Mechanica}\ }\textbf {\bibinfo {volume} {176}},\
  \bibinfo {pages} {169} (\bibinfo {year} {2005})}\BibitemShut {NoStop}%
\bibitem [{\citenamefont {Yin}\ \emph {et~al.}(2007)\citenamefont {Yin},
  \citenamefont {Liu},\ and\ \citenamefont {Anh}}]{Yin}%
  \BibitemOpen
  \bibfield  {author} {\bibinfo {author} {\bibfnamefont {C.}~\bibnamefont
  {Yin}}, \bibinfo {author} {\bibfnamefont {F.}~\bibnamefont {Liu}}, \ and\
  \bibinfo {author} {\bibfnamefont {V.}~\bibnamefont {Anh}},\ }\href {\doibase
  10.1260/174830107783133888} {\bibfield  {journal} {\bibinfo  {journal}
  {Journal of Algorithms \& Computational Technology}\ }\textbf {\bibinfo
  {volume} {1}},\ \bibinfo {pages} {427} (\bibinfo {year} {2007})}\BibitemShut
  {NoStop}%
\bibitem [{\citenamefont {Bagley}\ and\ \citenamefont
  {Torvik}(1983{\natexlab{a}})}]{Bagley1983}%
  \BibitemOpen
  \bibfield  {author} {\bibinfo {author} {\bibfnamefont {R.~L.}\ \bibnamefont
  {Bagley}}\ and\ \bibinfo {author} {\bibfnamefont {P.~J.}\ \bibnamefont
  {Torvik}},\ }\href {\doibase 10.1122/1.549724} {\bibfield  {journal}
  {\bibinfo  {journal} {Journal of Rheology}\ }\textbf {\bibinfo {volume}
  {27}},\ \bibinfo {pages} {201} (\bibinfo {year}
  {1983}{\natexlab{a}})}\BibitemShut {NoStop}%
\bibitem [{\citenamefont {Torvik}\ and\ \citenamefont
  {Bagley}(1984)}]{Torvik1984}%
  \BibitemOpen
  \bibfield  {author} {\bibinfo {author} {\bibfnamefont {P.~J.}\ \bibnamefont
  {Torvik}}\ and\ \bibinfo {author} {\bibfnamefont {R.~L.}\ \bibnamefont
  {Bagley}},\ }\href {\doibase 10.1115/1.3167615} {\bibfield  {journal}
  {\bibinfo  {journal} {Journal of Applied Mechanics}\ }\textbf {\bibinfo
  {volume} {51}},\ \bibinfo {pages} {294} (\bibinfo {year} {1984})}\BibitemShut
  {NoStop}%
\bibitem [{\citenamefont {Gaul}\ \emph {et~al.}(1991)\citenamefont {Gaul},
  \citenamefont {Klein},\ and\ \citenamefont {Kemple}}]{Gaul1991}%
  \BibitemOpen
  \bibfield  {author} {\bibinfo {author} {\bibfnamefont {L.}~\bibnamefont
  {Gaul}}, \bibinfo {author} {\bibfnamefont {P.}~\bibnamefont {Klein}}, \ and\
  \bibinfo {author} {\bibfnamefont {S.}~\bibnamefont {Kemple}},\ }\href
  {\doibase 10.1016/0888-3270(91)90016-X} {\bibfield  {journal} {\bibinfo
  {journal} {Mechanical Systems and Signal Processing}\ }\textbf {\bibinfo
  {volume} {5}},\ \bibinfo {pages} {81} (\bibinfo {year} {1991})}\BibitemShut
  {NoStop}%
\bibitem [{\citenamefont {Metzler}\ and\ \citenamefont
  {Nonnenmacher}(2003)}]{Metzler2003}%
  \BibitemOpen
  \bibfield  {author} {\bibinfo {author} {\bibfnamefont {R.}~\bibnamefont
  {Metzler}}\ and\ \bibinfo {author} {\bibfnamefont {T.~F.}\ \bibnamefont
  {Nonnenmacher}},\ }\href {\doibase
  https://doi.org/10.1016/S0749-6419(02)00087-6} {\bibfield  {journal}
  {\bibinfo  {journal} {International Journal of Plasticity}\ }\textbf
  {\bibinfo {volume} {19}},\ \bibinfo {pages} {941 } (\bibinfo {year}
  {2003})}\BibitemShut {NoStop}%
\bibitem [{\citenamefont {Aribi}\ \emph {et~al.}(2014)\citenamefont {Aribi},
  \citenamefont {Farges}, \citenamefont {Aoun}, \citenamefont {Melchior},
  \citenamefont {Najar},\ and\ \citenamefont {Abdelkrim}}]{Aribi2014}%
  \BibitemOpen
  \bibfield  {author} {\bibinfo {author} {\bibfnamefont {A.}~\bibnamefont
  {Aribi}}, \bibinfo {author} {\bibfnamefont {C.}~\bibnamefont {Farges}},
  \bibinfo {author} {\bibfnamefont {M.}~\bibnamefont {Aoun}}, \bibinfo {author}
  {\bibfnamefont {P.}~\bibnamefont {Melchior}}, \bibinfo {author}
  {\bibfnamefont {S.}~\bibnamefont {Najar}}, \ and\ \bibinfo {author}
  {\bibfnamefont {M.~N.}\ \bibnamefont {Abdelkrim}},\ }\href {\doibase
  10.1016/j.cnsns.2014.03.006} {\bibfield  {journal} {\bibinfo  {journal}
  {Communications in Nonlinear Science and Numerical Simulation}\ }\textbf
  {\bibinfo {volume} {19}},\ \bibinfo {pages} {3679} (\bibinfo {year}
  {2014})}\BibitemShut {NoStop}%
\bibitem [{\citenamefont {Falaize}\ and\ \citenamefont
  {H{\'e}lie}(2014)}]{Falaize2014}%
  \BibitemOpen
  \bibfield  {author} {\bibinfo {author} {\bibfnamefont {A.}~\bibnamefont
  {Falaize}}\ and\ \bibinfo {author} {\bibfnamefont {T.}~\bibnamefont
  {H{\'e}lie}},\ }in\ \href {https://hal.archives-ouvertes.fr/hal-01161071}
  {\emph {\bibinfo {booktitle} {{International Symposium on Musical
  Acoustics}}}}\ (\bibinfo {address} {Le Mans, France},\ \bibinfo {year}
  {2014})\ pp.\ \bibinfo {pages} {1--5},\ \bibinfo {note} {cote interne IRCAM:
  Falaize14d}\BibitemShut {NoStop}%
\bibitem [{\citenamefont {Quintana}\ \emph {et~al.}(2006)\citenamefont
  {Quintana}, \citenamefont {Ramos},\ and\ \citenamefont
  {Nuez}}]{Quintana2006}%
  \BibitemOpen
  \bibfield  {author} {\bibinfo {author} {\bibfnamefont {J.}~\bibnamefont
  {Quintana}}, \bibinfo {author} {\bibfnamefont {A.}~\bibnamefont {Ramos}}, \
  and\ \bibinfo {author} {\bibfnamefont {I.}~\bibnamefont {Nuez}},\ }\href
  {\doibase https://doi.org/10.3182/20060719-3-PT-4902.00073} {\bibfield
  {journal} {\bibinfo  {journal} {IFAC Proceedings Volumes}\ }\textbf {\bibinfo
  {volume} {39}},\ \bibinfo {pages} {432 } (\bibinfo {year} {2006})},\ \bibinfo
  {note} {2nd IFAC Workshop on Fractional Differentiation and its
  Applications}\BibitemShut {NoStop}%
\bibitem [{\citenamefont {Magin}(2010)}]{Magin2010}%
  \BibitemOpen
  \bibfield  {author} {\bibinfo {author} {\bibfnamefont {R.~L.}\ \bibnamefont
  {Magin}},\ }\href {\doibase 10.1016/j.camwa.2009.08.039} {\bibfield
  {journal} {\bibinfo  {journal} {Computers and Mathematics with Applications}\
  }\textbf {\bibinfo {volume} {59}},\ \bibinfo {pages} {1586} (\bibinfo {year}
  {2010})}\BibitemShut {NoStop}%
\bibitem [{\citenamefont {Mainardi}(2018)}]{Mainardi2018}%
  \BibitemOpen
  \bibfield  {author} {\bibinfo {author} {\bibfnamefont {F.}~\bibnamefont
  {Mainardi}},\ }\href {\doibase 10.3390/math6010008} {\bibfield  {journal}
  {\bibinfo  {journal} {Mathematics}\ }\textbf {\bibinfo {volume} {6}},\
  \bibinfo {pages} {4} (\bibinfo {year} {2018})}\BibitemShut {NoStop}%
\bibitem [{\citenamefont {Li}(2018)}]{Li2018}%
  \BibitemOpen
  \bibfield  {author} {\bibinfo {author} {\bibfnamefont {M.}~\bibnamefont
  {Li}},\ }\href {\doibase 10.3390/sym10020040} {\bibfield  {journal} {\bibinfo
   {journal} {Symmetry}\ }\textbf {\bibinfo {volume} {10}} (\bibinfo {year}
  {2018}),\ 10.3390/sym10020040}\BibitemShut {NoStop}%
\bibitem [{\citenamefont {Odar}\ and\ \citenamefont
  {Hamilton}(1964)}]{Odar1964}%
  \BibitemOpen
  \bibfield  {author} {\bibinfo {author} {\bibfnamefont {F.}~\bibnamefont
  {Odar}}\ and\ \bibinfo {author} {\bibfnamefont {W.~S.}\ \bibnamefont
  {Hamilton}},\ }\href {\doibase 10.1017/S0022112064000210} {\bibfield
  {journal} {\bibinfo  {journal} {Journal of Fluid Mechanics}\ }\textbf
  {\bibinfo {volume} {18}},\ \bibinfo {pages} {302} (\bibinfo {year}
  {1964})}\BibitemShut {NoStop}%
\bibitem [{\citenamefont {Catalano}(1985)}]{Catalano1985}%
  \BibitemOpen
  \bibfield  {author} {\bibinfo {author} {\bibfnamefont {G.~D.}\ \bibnamefont
  {Catalano}},\ }\href {\doibase 10.2514/3.9134} {\bibfield  {journal}
  {\bibinfo  {journal} {AIAA Journal}\ }\textbf {\bibinfo {volume} {23}},\
  \bibinfo {pages} {1627} (\bibinfo {year} {1985})}\BibitemShut {NoStop}%
\bibitem [{\citenamefont {Parmar}\ \emph {et~al.}(2012)\citenamefont {Parmar},
  \citenamefont {Haselbacher},\ and\ \citenamefont {Balachandar}}]{Parmar2012}%
  \BibitemOpen
  \bibfield  {author} {\bibinfo {author} {\bibfnamefont {M.}~\bibnamefont
  {Parmar}}, \bibinfo {author} {\bibfnamefont {A.}~\bibnamefont {Haselbacher}},
  \ and\ \bibinfo {author} {\bibfnamefont {S.}~\bibnamefont {Balachandar}},\
  }\href {\doibase 10.1017/jfm.2012.109} {\bibfield  {journal} {\bibinfo
  {journal} {Journal of Fluid Mechanics}\ }\textbf {\bibinfo {volume} {699}},\
  \bibinfo {pages} {352} (\bibinfo {year} {2012})}\BibitemShut {NoStop}%
\bibitem [{\citenamefont {Lambertz}()}]{Lambertz2012}%
  \BibitemOpen
  \bibfield  {author} {\bibinfo {author} {\bibfnamefont {S.}~\bibnamefont
  {Lambertz}},\ }\emph {\bibinfo {title} {{Experimentelle Untersuchung der
  Basset Ged{\"{a}}chtniskraft auf eine starre Kugel in instation{\"{a}}rer
  Bewegung} --- Experimental measurement of the history forces on a rigid
  sphere in unsteady motion}},\ \href@noop {} {Ph.D. thesis}\BibitemShut
  {NoStop}%
\bibitem [{\citenamefont {Parmar}\ \emph {et~al.}(2011)\citenamefont {Parmar},
  \citenamefont {Haselbacher},\ and\ \citenamefont {Balachandar}}]{Parmar2011}%
  \BibitemOpen
  \bibfield  {author} {\bibinfo {author} {\bibfnamefont {M.}~\bibnamefont
  {Parmar}}, \bibinfo {author} {\bibfnamefont {A.}~\bibnamefont {Haselbacher}},
  \ and\ \bibinfo {author} {\bibfnamefont {S.}~\bibnamefont {Balachandar}},\
  }\href {\doibase 10.1103/PhysRevLett.106.084501} {\bibfield  {journal}
  {\bibinfo  {journal} {Phys. Rev. Lett.}\ }\textbf {\bibinfo {volume} {106}},\
  \bibinfo {pages} {084501} (\bibinfo {year} {2011})}\BibitemShut {NoStop}%
\bibitem [{\citenamefont {Lawrence}\ and\ \citenamefont
  {Weinbaum}(1986)}]{Lawrence1986}%
  \BibitemOpen
  \bibfield  {author} {\bibinfo {author} {\bibfnamefont {C.~J.}\ \bibnamefont
  {Lawrence}}\ and\ \bibinfo {author} {\bibfnamefont {S.}~\bibnamefont
  {Weinbaum}},\ }\href {\doibase 10.1017/S0022112086001428} {\bibfield
  {journal} {\bibinfo  {journal} {Journal of Fluid Mechanics}\ }\textbf
  {\bibinfo {volume} {171}},\ \bibinfo {pages} {209–218} (\bibinfo {year}
  {1986})}\BibitemShut {NoStop}%
\bibitem [{\citenamefont {Abbad}\ and\ \citenamefont
  {Souhar}(2004)}]{Abbad2004}%
  \BibitemOpen
  \bibfield  {author} {\bibinfo {author} {\bibfnamefont {M.}~\bibnamefont
  {Abbad}}\ and\ \bibinfo {author} {\bibfnamefont {M.}~\bibnamefont {Souhar}},\
  }\href {\doibase 10.1007/s00348-003-0759-x} {\bibfield  {journal} {\bibinfo
  {journal} {Experiments in Fluids}\ }\textbf {\bibinfo {volume} {36}},\
  \bibinfo {pages} {775} (\bibinfo {year} {2004})}\BibitemShut {NoStop}%
\bibitem [{\citenamefont {Coimbra}(2003)}]{Coimbra2003}%
  \BibitemOpen
  \bibfield  {author} {\bibinfo {author} {\bibfnamefont {C.}~\bibnamefont
  {Coimbra}},\ }\href {\doibase 10.1002/andp.200310032} {\bibfield  {journal}
  {\bibinfo  {journal} {Annalen der Physik}\ }\textbf {\bibinfo {volume}
  {12}},\ \bibinfo {pages} {692} (\bibinfo {year} {2003})}\BibitemShut
  {NoStop}%
\bibitem [{\citenamefont {Pedro}\ \emph {et~al.}(2005)\citenamefont {Pedro},
  \citenamefont {Pereira}, \citenamefont {Kobayashi},\ and\ \citenamefont
  {Coimbra}}]{Pedro2005}%
  \BibitemOpen
  \bibfield  {author} {\bibinfo {author} {\bibfnamefont {H.}~\bibnamefont
  {Pedro}}, \bibinfo {author} {\bibfnamefont {J.}~\bibnamefont {Pereira}},
  \bibinfo {author} {\bibfnamefont {M.}~\bibnamefont {Kobayashi}}, \ and\
  \bibinfo {author} {\bibfnamefont {C.}~\bibnamefont {Coimbra}},\ }\enquote
  {\bibinfo {title} {History forces in oscillating convective flow past a fixed
  particle},}\ in\ \href@noop {} {\emph {\bibinfo {booktitle} {43rd AIAA
  Aerospace Sciences Meeting and Exhibit}}}\ (\bibinfo {year} {2005})\ p.\
  \bibinfo {pages} {1393}\BibitemShut {NoStop}%
\bibitem [{\citenamefont {Bagley}\ and\ \citenamefont
  {Torvik}(1983{\natexlab{b}})}]{Bagley1983b}%
  \BibitemOpen
  \bibfield  {author} {\bibinfo {author} {\bibfnamefont {R.~L.}\ \bibnamefont
  {Bagley}}\ and\ \bibinfo {author} {\bibfnamefont {P.~J.}\ \bibnamefont
  {Torvik}},\ }\href {\doibase 10.1122/1.549724} {\bibfield  {journal}
  {\bibinfo  {journal} {Journal of Rheology}\ }\textbf {\bibinfo {volume}
  {27}},\ \bibinfo {pages} {201} (\bibinfo {year}
  {1983}{\natexlab{b}})}\BibitemShut {NoStop}%
\bibitem [{\citenamefont {Moshrefi-Torbati}\ and\ \citenamefont
  {Hammond}(1998)}]{Moshrefi-Torbati1998}%
  \BibitemOpen
  \bibfield  {author} {\bibinfo {author} {\bibfnamefont {M.}~\bibnamefont
  {Moshrefi-Torbati}}\ and\ \bibinfo {author} {\bibfnamefont {J.~K.}\
  \bibnamefont {Hammond}},\ }\href {\doibase 10.1016/s0016-0032(97)00048-3}
  {\bibfield  {journal} {\bibinfo  {journal} {Journal of the Franklin
  Institute}\ }\textbf {\bibinfo {volume} {335}},\ \bibinfo {pages} {1077}
  (\bibinfo {year} {1998})}\BibitemShut {NoStop}%
\bibitem [{\citenamefont {Podlubny}(2008)}]{Podlubny2008}%
  \BibitemOpen
  \bibfield  {author} {\bibinfo {author} {\bibfnamefont {I.}~\bibnamefont
  {Podlubny}},\ }\href@noop {} {\ ,\ \bibinfo {pages} {1} (\bibinfo {year}
  {2008})},\ \Eprint {http://arxiv.org/abs/0110241v1} {arXiv:0110241v1
  [arXiv:math]} \BibitemShut {NoStop}%
\bibitem [{\citenamefont {G{\'{o}}mez-Aguilar}\ \emph
  {et~al.}(2018)\citenamefont {G{\'{o}}mez-Aguilar}, \citenamefont
  {Escobar-Jim{\'{e}}nez}, \citenamefont {L{\'{o}}pez-L{\'{o}}pez},\ and\
  \citenamefont {Alvarado-Mart{\'{i}}nez}}]{GomezAguilar2018}%
  \BibitemOpen
  \bibfield  {author} {\bibinfo {author} {\bibfnamefont {J.~F.}\ \bibnamefont
  {G{\'{o}}mez-Aguilar}}, \bibinfo {author} {\bibfnamefont {R.~F.}\
  \bibnamefont {Escobar-Jim{\'{e}}nez}}, \bibinfo {author} {\bibfnamefont
  {M.~G.}\ \bibnamefont {L{\'{o}}pez-L{\'{o}}pez}}, \ and\ \bibinfo {author}
  {\bibfnamefont {V.~M.}\ \bibnamefont {Alvarado-Mart{\'{i}}nez}},\ }\href
  {\doibase 10.1140/epjp/i2018-11924-1} {\bibfield  {journal} {\bibinfo
  {journal} {European Physical Journal Plus}\ }\textbf {\bibinfo {volume}
  {133}} (\bibinfo {year} {2018}),\ 10.1140/epjp/i2018-11924-1}\BibitemShut
  {NoStop}%
\bibitem [{\citenamefont {Ebaid}(2011)}]{Ebaid2011}%
  \BibitemOpen
  \bibfield  {author} {\bibinfo {author} {\bibfnamefont {A.}~\bibnamefont
  {Ebaid}},\ }\href {\doibase 10.1016/j.apm.2010.08.010} {\bibfield  {journal}
  {\bibinfo  {journal} {Applied Mathematical Modelling}\ }\textbf {\bibinfo
  {volume} {35}},\ \bibinfo {pages} {1231} (\bibinfo {year}
  {2011})}\BibitemShut {NoStop}%
\bibitem [{\citenamefont {Rekhviashvili}\ \emph {et~al.}(2019)\citenamefont
  {Rekhviashvili}, \citenamefont {Pskhu}, \citenamefont {Agarwal},\ and\
  \citenamefont {Jain}}]{Rekhviashvili2019}%
  \BibitemOpen
  \bibfield  {author} {\bibinfo {author} {\bibfnamefont {S.}~\bibnamefont
  {Rekhviashvili}}, \bibinfo {author} {\bibfnamefont {A.}~\bibnamefont
  {Pskhu}}, \bibinfo {author} {\bibfnamefont {P.}~\bibnamefont {Agarwal}}, \
  and\ \bibinfo {author} {\bibfnamefont {S.}~\bibnamefont {Jain}},\ }\href
  {\doibase 10.3906/fiz-1811-16} {\bibfield  {journal} {\bibinfo  {journal}
  {Turkish Journal of Physics}\ }\textbf {\bibinfo {volume} {43}},\ \bibinfo
  {pages} {236} (\bibinfo {year} {2019})}\BibitemShut {NoStop}%
\bibitem [{\citenamefont {Khubalkar}\ \emph {et~al.}(2018)\citenamefont
  {Khubalkar}, \citenamefont {Junghare}, \citenamefont {Aware},\ and\
  \citenamefont {Das}}]{Khubalkar2018}%
  \BibitemOpen
  \bibfield  {author} {\bibinfo {author} {\bibfnamefont {S.}~\bibnamefont
  {Khubalkar}}, \bibinfo {author} {\bibfnamefont {A.}~\bibnamefont {Junghare}},
  \bibinfo {author} {\bibfnamefont {M.}~\bibnamefont {Aware}}, \ and\ \bibinfo
  {author} {\bibfnamefont {S.}~\bibnamefont {Das}},\ }\href {\doibase
  10.1177/0020720918799509} {\bibfield  {journal} {\bibinfo  {journal}
  {International Journal of Electrical Engineering Education}\ ,\ \bibinfo
  {pages} {1}} (\bibinfo {year} {2018})}\BibitemShut {NoStop}%
\bibitem [{\citenamefont {Richard}(2014)}]{Herrmann2014}%
  \BibitemOpen
  \bibfield  {author} {\bibinfo {author} {\bibfnamefont {H.}~\bibnamefont
  {Richard}},\ }\href@noop {} {\emph {\bibinfo {title} {Fractional calculus: an
  introduction for physicists}}}\ (\bibinfo  {publisher} {World Scientific},\
  \bibinfo {year} {2014})\BibitemShut {NoStop}%
\bibitem [{\citenamefont {Ortigueira}\ and\ \citenamefont {{Tenreiro
  Machado}}(2015)}]{Ortigueira2015c}%
  \BibitemOpen
  \bibfield  {author} {\bibinfo {author} {\bibfnamefont {M.~D.}\ \bibnamefont
  {Ortigueira}}\ and\ \bibinfo {author} {\bibfnamefont {J.~A.}\ \bibnamefont
  {{Tenreiro Machado}}},\ }\href {\doibase 10.1016/j.jcp.2014.07.019}
  {\bibfield  {journal} {\bibinfo  {journal} {Journal of Computational
  Physics}\ }\textbf {\bibinfo {volume} {293}},\ \bibinfo {pages} {4} (\bibinfo
  {year} {2015})}\BibitemShut {NoStop}%
\bibitem [{\citenamefont {Seredy{\' n}ska}\ and\ \citenamefont
  {Hanyga}(2000)}]{Seredynska2000}%
  \BibitemOpen
  \bibfield  {author} {\bibinfo {author} {\bibfnamefont {M.}~\bibnamefont
  {Seredy{\' n}ska}}\ and\ \bibinfo {author} {\bibfnamefont {A.}~\bibnamefont
  {Hanyga}},\ }\href {\doibase 10.1063/1.533231} {\bibfield  {journal}
  {\bibinfo  {journal} {Journal of Mathematical Physics}\ }\textbf {\bibinfo
  {volume} {41}},\ \bibinfo {pages} {2135} (\bibinfo {year}
  {2000})}\BibitemShut {NoStop}%
\bibitem [{\citenamefont {Spanos}\ and\ \citenamefont
  {Evangelatos}(2010)}]{Spanos2010}%
  \BibitemOpen
  \bibfield  {author} {\bibinfo {author} {\bibfnamefont {P.~D.}\ \bibnamefont
  {Spanos}}\ and\ \bibinfo {author} {\bibfnamefont {G.~I.}\ \bibnamefont
  {Evangelatos}},\ }\href {\doibase 10.1016/j.soildyn.2010.01.013} {\bibfield
  {journal} {\bibinfo  {journal} {Soil Dynamics and Earthquake Engineering}\
  }\textbf {\bibinfo {volume} {30}},\ \bibinfo {pages} {811} (\bibinfo {year}
  {2010})}\BibitemShut {NoStop}%
\bibitem [{\citenamefont {Diethelm}\ \emph {et~al.}(2002)\citenamefont
  {Diethelm}, \citenamefont {Ford},\ and\ \citenamefont
  {Freed}}]{Diethelm2002}%
  \BibitemOpen
  \bibfield  {author} {\bibinfo {author} {\bibfnamefont {K.}~\bibnamefont
  {Diethelm}}, \bibinfo {author} {\bibfnamefont {N.~J.}\ \bibnamefont {Ford}},
  \ and\ \bibinfo {author} {\bibfnamefont {A.~D.}\ \bibnamefont {Freed}},\
  }\href {\doibase 10.1023/A:1016592219341} {\bibfield  {journal} {\bibinfo
  {journal} {Nonlinear Dynamics}\ }\textbf {\bibinfo {volume} {29}},\ \bibinfo
  {pages} {3} (\bibinfo {year} {2002})}\BibitemShut {NoStop}%
\bibitem [{\citenamefont {Gavin}(2001)}]{Gavin2018}%
  \BibitemOpen
  \bibfield  {author} {\bibinfo {author} {\bibfnamefont {H.}~\bibnamefont
  {Gavin}},\ }\href@noop {} {\enquote {\bibinfo {title} {Numerical integration
  for structural dynamics},}\ }\bibinfo {howpublished}
  {\url{https://pdfs.semanticscholar.org/4907/eb47e4924ff9d32a643b8acc9dd0666c94c4.pdf}}
  (\bibinfo {year} {2001})\BibitemShut {NoStop}%
\bibitem [{\citenamefont {Heymans}\ and\ \citenamefont
  {Podlubny}(2006)}]{Heymans2006}%
  \BibitemOpen
  \bibfield  {author} {\bibinfo {author} {\bibfnamefont {N.}~\bibnamefont
  {Heymans}}\ and\ \bibinfo {author} {\bibfnamefont {I.}~\bibnamefont
  {Podlubny}},\ }\href {\doibase 10.1007/s00397-005-0043-5} {\bibfield
  {journal} {\bibinfo  {journal} {Rheologica Acta}\ }\textbf {\bibinfo {volume}
  {45}},\ \bibinfo {pages} {765} (\bibinfo {year} {2006})}\BibitemShut
  {NoStop}%
\bibitem [{\citenamefont {Achar}\ \emph {et~al.}(2007)\citenamefont {Achar},
  \citenamefont {Lorenzo},\ and\ \citenamefont {Hartley}}]{Achar2007}%
  \BibitemOpen
  \bibfield  {author} {\bibinfo {author} {\bibfnamefont {B.~N.~N.}\
  \bibnamefont {Achar}}, \bibinfo {author} {\bibfnamefont {C.~F.}\ \bibnamefont
  {Lorenzo}}, \ and\ \bibinfo {author} {\bibfnamefont {T.~T.}\ \bibnamefont
  {Hartley}},\ }\enquote {\bibinfo {title} {The caputo fractional derivative:
  Initialization issues relative to fractional differential equation},}\ in\
  \href {\doibase 10.1007/978-1-4020-6042-7_3} {\emph {\bibinfo {booktitle}
  {Advances in Fractional Calculus: Theoretical Developments and Applications
  in Physics and Engineering}}},\ \bibinfo {editor} {edited by\ \bibinfo
  {editor} {\bibfnamefont {J.}~\bibnamefont {Sabatier}}, \bibinfo {editor}
  {\bibfnamefont {O.~P.}\ \bibnamefont {Agrawal}}, \ and\ \bibinfo {editor}
  {\bibfnamefont {J.~A.~T.}\ \bibnamefont {Machado}}}\ (\bibinfo  {publisher}
  {Springer Netherlands},\ \bibinfo {address} {Dordrecht},\ \bibinfo {year}
  {2007})\ pp.\ \bibinfo {pages} {27--42}\BibitemShut {NoStop}%
\bibitem [{\citenamefont {Gladkina}\ \emph {et~al.}(2017)\citenamefont
  {Gladkina}, \citenamefont {Shchedrin}, \citenamefont {Khawaja},\ and\
  \citenamefont {Carr}}]{Gladkina2017}%
  \BibitemOpen
  \bibfield  {author} {\bibinfo {author} {\bibfnamefont {A.}~\bibnamefont
  {Gladkina}}, \bibinfo {author} {\bibfnamefont {G.}~\bibnamefont {Shchedrin}},
  \bibinfo {author} {\bibfnamefont {U.~A.}\ \bibnamefont {Khawaja}}, \ and\
  \bibinfo {author} {\bibfnamefont {L.~D.}\ \bibnamefont {Carr}},\ }\href@noop
  {} {\enquote {\bibinfo {title} {Expansion of fractional derivatives in terms
  of an integer derivative series: physical and numerical applications},}\ }
  (\bibinfo {year} {2017}),\ \Eprint {http://arxiv.org/abs/1710.06297}
  {arXiv:1710.06297 [math.NA]} \BibitemShut {NoStop}%
\bibitem [{\citenamefont {Threlfall}(1978)}]{Threlfall1978}%
  \BibitemOpen
  \bibfield  {author} {\bibinfo {author} {\bibfnamefont {D.}~\bibnamefont
  {Threlfall}},\ }\href {\doibase https://doi.org/10.1016/0094-114X(78)90020-4}
  {\bibfield  {journal} {\bibinfo  {journal} {Mechanism and Machine Theory}\
  }\textbf {\bibinfo {volume} {13}},\ \bibinfo {pages} {475 } (\bibinfo {year}
  {1978})}\BibitemShut {NoStop}%
\bibitem [{\citenamefont {Press}\ \emph {et~al.}(2007)\citenamefont {Press},
  \citenamefont {Teukolsky}, \citenamefont {Vetterling},\ and\ \citenamefont
  {Flannery}}]{NumericalRecipes}%
  \BibitemOpen
  \bibfield  {author} {\bibinfo {author} {\bibfnamefont {W.~H.}\ \bibnamefont
  {Press}}, \bibinfo {author} {\bibfnamefont {S.~A.}\ \bibnamefont
  {Teukolsky}}, \bibinfo {author} {\bibfnamefont {W.~T.}\ \bibnamefont
  {Vetterling}}, \ and\ \bibinfo {author} {\bibfnamefont {B.~P.}\ \bibnamefont
  {Flannery}},\ }\href@noop {} {\emph {\bibinfo {title} {Numerical Recipes 3rd
  Edition: The Art of Scientific Computing}}},\ \bibinfo {edition} {3rd}\ ed.\
  (\bibinfo  {publisher} {Cambridge University Press},\ \bibinfo {address} {New
  York, NY, USA},\ \bibinfo {year} {2007})\BibitemShut {NoStop}%
\bibitem [{\citenamefont {Magalas}(1996)}]{Magalas1996}%
  \BibitemOpen
  \bibfield  {author} {\bibinfo {author} {\bibfnamefont {L.~B.}\ \bibnamefont
  {Magalas}},\ }\href {\doibase 10.1051/jp4:1996834} {\bibfield  {journal}
  {\bibinfo  {journal} {Journal de Physique IV Colloque}\ }\textbf {\bibinfo
  {volume} {6}},\ \bibinfo {pages} {17} (\bibinfo {year} {1996})}\BibitemShut
  {NoStop}%
\end{thebibliography}
\end{document}